\documentclass[twoside,12pt]{article}
\usepackage[dvipdfmx]{graphicx}
\usepackage{amsmath,amssymb,bm}

\begin{document}

\title{ \vspace{1cm}
Fluctuations of conserved charges in relativistic heavy ion collisions:
An introduction
}
\author{
Masayuki\ Asakawa$^1$ and Masakiyo\ Kitazawa$^1$\\
$^1$Department of Physics, Osaka University, 
Toyonaka, Osaka 560-0043, Japan
}
\maketitle

\begin{abstract}

  Bulk fluctuations of conserved charges measured by event-by-event
  analysis in relativistic heavy ion collisions are observables which 
  are believed to carry significant amount of information on
  the hot medium created by the collisions. 
  Active studies have been done recently
  experimentally, theoretically, and on the lattice.
  In particular, non-Gaussianity of the fluctuations has acquired 
  much attention recently.
  In this review, we give a pedagogical introduction to these issues,
  and survey recent developments in this field of research.
  Starting from the definition of cumulants, 
  basic concepts in fluctuation physics, such as 
  thermal fluctuations in statistical mechanics and time evolution
  of fluctuations in diffusive systems, are described.
  Phenomena which are expected to occur in finite temperature and/or
  density QCD matter and their measurement by event-by-event analyses are also
  elucidated.
 
\end{abstract}

\newpage

\section{Introduction}

\subsection{Background}

The medium described by quantum chromodynamics (QCD) is 
expected to have various phase transitions with variations of 
external thermodynamic parameters such as temperature $T$.
Although the basic degrees of freedom of QCD, quarks and gluons, 
are confined into hadrons in the vacuum, 
they are expected to be liberated at extremely high temperature 
and form a new state of the matter called the quark-gluon 
plasma (QGP).
It is also known that the chiral symmetry, which is 
spontaneously broken in vacuum, is restored at extremely hot and/or
dense environment.
These phase transitions at vanishing baryon chemical potential
($\mu_{\rm B}$) are investigated with lattice QCD
Monte Carlo simulations.
The numerical analyses show that the phase transition is 
a smooth crossover \cite{Cheng:2006qk,Aoki:2006we}.
On the other hand, various models predict that there exists
a discontinuous first order phase transition at nonzero $\mu_{\rm B}$.
The existence of the endpoint of the first order transition,
the QCD critical point \cite{Asakawa:1989bq}, and possibly
multiple critical points \cite{Kitazawa:2002bc}, are 
anticipated in the phase diagram of QCD on $T$-$\mu_{\rm B}$ plane
\cite{Stephanov:2007fk,Fukushima:2010bq,Philipsen:2011zx}.

After the advent of the relativistic heavy ion collisions,
the quark-gluon plasma has come to be created and investigated
on the Earth.
At the Relativistic Heavy Ion Collider (RHIC) \cite{RHIC} and 
the Large Hadron Collider (LHC) \cite{LHC}, active experimental 
studies on the QGP have been being performed.
The discovery of the strongly-coupled property of the QGP near the 
crossover region \cite{RHIC,LHC} is one of the highlights 
of these experiments.
With the top RHIC energy $\sqrt{s_{\rm NN}}= 200$ GeV and 
LHC energy $\sqrt{s_{\rm NN}}= 2.76$ TeV, 
hot medium with almost vanishing $\mu_{\rm B}$ is 
created \cite{Andronic:2008gu,Kumar:2012fb}.
On the other hand, the chemical freezeout picture for 
particle abundances \cite{BraunMunzinger:2003zd} suggests that 
the net-baryon number density 
and $\mu_{\rm B}$ of the hot medium increase as $\sqrt{s_{\rm NN}}$ 
is lowered 
down to $\sqrt{s_{\rm NN}}\simeq 5 - 10$ GeV 
\cite{Kumar:2012fb,Randrup:2006nr}.
The relativistic heavy ion collisions, therefore, can 
investigate various regions of the QCD phase diagram on 
$T$-$\mu_{\rm B}$ plane by changing the collision energy $\sqrt{s_{\rm NN}}$.
Such an experimental program is now ongoing at RHIC, which is 
called the beam-energy scan (BES) program \cite{BES-I}.
The upgraded stage of the BES called the BES-II is planned
to start in 2019 \cite{BES-II}.
The future experiments prepared at FAIR \cite{FAIR}, NICA \cite{NICA} 
and J-PARC will also contribute to the study of the medium 
with large $\mu_{\rm B}$.
The searches of the QCD critical point \cite{Stephanov:1998dy} and
the first-order phase transition are among the most interesting
subjects in this program.

In relativistic heavy ion collisions, after the formation of the 
QGP the medium undergoes confinement transition
before they arrive at the detector.
During rescatterings in the hadronic stage, the signals formed in 
the QGP tend to be blurred.
In order to study the properties of QGP in these experiments, 
therefore, it is important to choose observables which are 
sensitive to the medium property in the early stage.

Recently, as unique hadronic observables which reflect 
thermal property of the primordial medium created by 
relativistic heavy ion collisions, the bulk fluctuations
have acquired much attentions \cite{Kitazawa:2014nja}.
Although these observables are hadronic ones, it is believed
that they reflect the thermal property in the 
early stage \cite{Koch:2008ia}.
They are believed to be good observables
in investigating the deconfinement transition 
\cite{Asakawa:2000wh,Jeon:2000wg,Ejiri:2005wq}
and finding the location of the QCD critical point 
\cite{Stephanov:1998dy,Stephanov:1999zu,Hatta:2003wn}.
Active experimental studies have been carried out 
\cite{STAR-fluc,ALICE,Adamczyk:2013dal,Anticic:2013htn,
Adamczyk:2014fia,PHENIX-fluc,Luo:2015ewa,Adare:2015aqk}
as well as analyses on the numerical simulations 
on the lattice \cite{Ding:2015ona,Borsanyi:2015axp}.
In particular, fluctuations of conserved charges and 
their higher order cumulants representing non-Gaussianity 
\cite{Ejiri:2005wq,Stephanov:2008qz,Asakawa:2009aj}
are actively studied recently.
The purpose of this review is to give a basic introduction
to the physics of fluctuations in relativistic 
heavy ion collisions, and give an overview of the recent experimental 
and theoretical progress in this field of research.

\subsection{Fluctuations}

Before starting the discussion of relativistic heavy ion collisions,
we first give a general review on fluctuations briefly.
When one measures an observable in some system, 
the result of the measurements would take different values 
for different measurements, even if the measurement is performed 
with an ideal detector with an infinitesimal resolution.
This distribution of the result of measurements is referred to as 
fluctuations.
In typical thermal systems, the fluctuations are predominantly 
attributed to thermal effects, which can be calculated in statistical
mechanics. Quantum effects also give rise to fluctuations.

In contrast to standard observables, fluctuations are 
sometimes regarded as the noise associated with the 
measurement and thus are obstacles.
As expressed by Landauer as ``The noise is the signal'' \cite{Landauer},
however, the fluctuations sometimes can become
invaluable physical observables in spite of their obstacle characters.
Here, in order to spur the motivations of the readers 
we list three examples of the physics in which fluctuations 
play a crucial role.

\begin{enumerate}
\item
{\bf Brownian motion:}
The first example is a historical one on Brownian motion.
As first discovered by Brown in 1827, small objects, such as pollens, 
floating on water show a quick and random motion.
Due to this motion, the position of the pollen after several
time duration fluctuates even if the initial position is fixed.
The origin of this motion was first revealed by Einstein.
In his historical paper in 1905 \cite{Einstein}, Einstein pointed 
out that the Brownian motion is attributed to the thermal motion
of water molecules.
This prediction was confirmed by Perrin, who calculated 
the Avogadro constant based on this picture \cite{Perrin}.
In this era, the existence of atoms had not been confirmed, yet.
These studies served as a piece of the earliest evidence for 
the existence of molecules and atoms.
In other words, human beings saw atoms for the 
first time behind fluctuations.

This example tells us that fluctuations are powerful
tools to diagnose microscopic physics.
One century after Einstein's era, now that we know 
substructures of atoms, hadrons, and quarks and gluons,
it seems a natural idea to utilize fluctuations in 
relativistic heavy ion collisions in exploring subnuclear
physics.
In this review, a Brownian particle model for diffusion of 
fluctuations will be discussed in Sec.~\ref{sec:diffusion:DME}.

\item
{\bf Cosmic microwave background:}
The second example is found in cosmology.
As a remnant of Big Bang and as a result of transparent to radiation,
our Universe has $2.7$~K thermal radiation called cosmic 
microwave background (CMB) \cite{Ade:2013sjv}.
The temperature of this radiation is almost uniform in all 
directions in the Universe, but has a tiny fluctuation at 
different angles.
This fluctuation is now considered as the remnant of quantum 
fluctuations in the primordial Universe.
With this picture the power spectrum of this fluctuation 
tells us various properties of our Universe \cite{Baumann:2009ds};
for example, our Universe has
started with an inflational expansion $13.8$ billion years ago. 
In other words, we can see the hot primordial Universe behind
the fluctuation of CMB.

This example tells us that fluctuations can be powerful
tools to trace back the history of a system. 
It thus seems a natural idea to utilize fluctuations
to investigate the early stage of the ``little bang''
created by relativistic heavy ion collisions.
A common feature in the study of fluctuations in CMB
and heavy ion collisions is that 
the non-Gaussian fluctuations acquire attentions.
In fact, the non-Gaussianity of the CMB has been one of the 
hot topics in this community \cite{Maldacena:2002vr,Bartolo:2004if},
although the Planck spacecraft has not succeeded in the 
measurement of statistically significant non-Gaussianity thus far
\cite{Ade:2013ydc}.

\item
{\bf Shot noise:}
The final example is the fluctuations of the electric current
in an electric circuit called the shot noise.
The electric current at a resistor $R$ is generally fluctuating.
Even without an applied voltage, the current has thermal noise
proportional to $T/R$, which is called the Johnson-Nyquist noise
\cite{Johnson,Nyquist}.
On the other hand, there is a contribution of the noise 
which takes place when a voltage is applied and is proportional to
the average current $\langle I\rangle$.
When a circuit has a potential barrier, variance of such a noise 
tends to be proportional to $e^*\langle I\rangle$,
where $e^*$ is the electric charge of the elementary degrees of freedom
carrying electric current.
(This proportionality comes from the Poisson nature of the noise 
as will be clarified in Sec.~\ref{sec:equil:shot}.)
This noise is called the shot noise \cite{Schottky}.
Because of this proportionality, this fluctuation can be used to 
investigate the quasi-particle property.
When the material undergoes the phase transition to superconductivity,
for example, electrons are ``confined'' into Cooper pairs
and the electric charge carried by the elementary degrees of freedom
is doubled.
This behavior is in fact observed in the measurement of the shot noise
\cite{Jehl}.
More surprisingly, in the materials in which the fractional 
quantum Hall effect is realized, the shot noise behaves as 
if there were excitations having fractional charges \cite{FQHE}.

This example tells us that the fluctuations are powerful tools to 
investigate elementary degrees of freedom in the system although 
they are macroscopic observables.
It thus seems a natural idea to utilize this property of 
fluctuations to explore the confinement/deconfinement property
of quarks in relativistic heavy ion collisions.
In fact, this is a relatively old idea in heavy ion community
\cite{Asakawa:2000wh,Jeon:2000wg,Ejiri:2005wq}.
It is also notable that the non-Gaussianity of the shot noise
has been observed in mesoscopic systems \cite{Gustavssona}.

\end{enumerate}

\subsection{Bulk fluctuations in relativistic heavy ion collisions}

In this review, among various fluctuations 
we focus on the bulk fluctuations of conserved charges.
When one measures a charge in a phase space in some system,
the amount of the charge, $Q$, fluctuates
measurement by measurement.
We refer to the distribution of $Q$ as the bulk fluctuation
(or, simply fluctuation).
When we perform this measurement in a spatial volume in a thermal 
system, this fluctuation is called the thermal fluctuation.

The bulk fluctuations are closely related to 
correlation functions.
The total charge $Q$ in a phase space $V$ is given by the integral 
of the density of the charge $n(x)$ as 
\begin{align}
Q= \int_V dx n(x),
\end{align}
where $x$ is the coordinate in the phase space.
The variance of $Q$ thus is given by 
\begin{align}
\langle \delta Q^2 \rangle_V 
= \langle ( Q - \langle Q \rangle_V )^2 \rangle_V 
= \int_V dx_1 dx_2 \langle \delta n(x_1) \delta n(x_2) \rangle ,
\label{eq:<dQ^2>=nn}
\end{align}
where $\delta n(x) = n(x)-\langle n \rangle$.
In this equation, the left-hand side is the quantity 
that we call (second-order) bulk fluctuation, while 
the integrand on the right-hand side is called 
correlation function.
Equaiton~(\ref{eq:<dQ^2>=nn}) shows that 
$\langle \delta Q^2 \rangle_V $ can be obtained from the 
correlation function by taking the integral.
If one knows the value of $\langle \delta Q^2 \rangle_V $ for 
all $V$ the correlation function can also be constructed 
from $\langle \delta Q^2 \rangle_V $.
In this sense, the correlation function carries the same 
physical information as the bulk fluctuation, and the choice of 
observables, bulk fluctuation or correlation function, is a matter 
of taste for the second-order fluctuation $\langle \delta Q^2 \rangle_V $.
(For higher orders, correlation functions contain more
information than bulk fluctuations.)
Phenomenological studies on the correlation functions of 
conserved charges in relativistic heavy ion collisions are widely 
performed, especially in terms of the so-called balance function 
\cite{Pratt:2012dz}.
The relation between the correlation functions and bulk fluctuations
are also discussed in the
literature \cite{Bass:2000az,Jeon:2001ue,Ling:2013ksb}.
In this review, however, we basically stick to bulk fluctuations
in our discussion.

In relativistic heavy ion collisions, 
the bulk fluctuations are observed by the event-by-event analyses. 
In these analyses, the number of some charge or a species of particle 
observed by the detector is counted event by event.
The distribution of the numbers counted in this way is called 
event-by-event fluctuation.
As we will discuss in detail in Sec.~\ref{sec:e-v-e}, the 
fluctuations observed in this way are believed to carry 
information on the thermal fluctuation in the primordial 
stage.
The event-by-event fluctuations, however, are not the thermal
fluctuations themselves.
The hot media created by collisions are dynamical systems, and 
the detector can only measure their final states.
Moreover, fluctuations other than thermal fluctuations contribute to
event-by-event fluctuations.
Careful treatment and interpretation, therefore, are 
required in comparing the event-by-event fluctuations with 
theoretical analysis on thermal fluctuations.
As will be discussed in Sec.~\ref{sec:e-v-e}, however, 
there are sufficient reasons to expect that
event-by-event fluctuations can be compared with 
thermal fluctuations with an appropriate treatment.

\begin{figure}
\begin{center}
\includegraphics*[width=8cm]{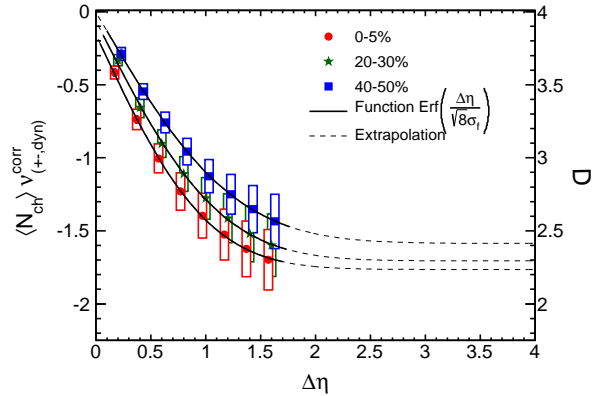}
\caption{
Rapidity window ($\Delta\eta$) dependence of 
net-electric charge fluctuation measured by ALICE 
collaboration at LHC \cite{ALICE}.
The right vertical axis is the D-measure $D$, 
the magnitude of net-electric charge fluctuation in a 
normalization that the value should become $3 \sim 4$ 
in an equilibrated hadronic medium \cite{Jeon:2000wg}.
}
\label{fig:ALICEfluc}
\end{center}
\end{figure}

\begin{figure}
\begin{center}
\includegraphics*[width=5.5cm]{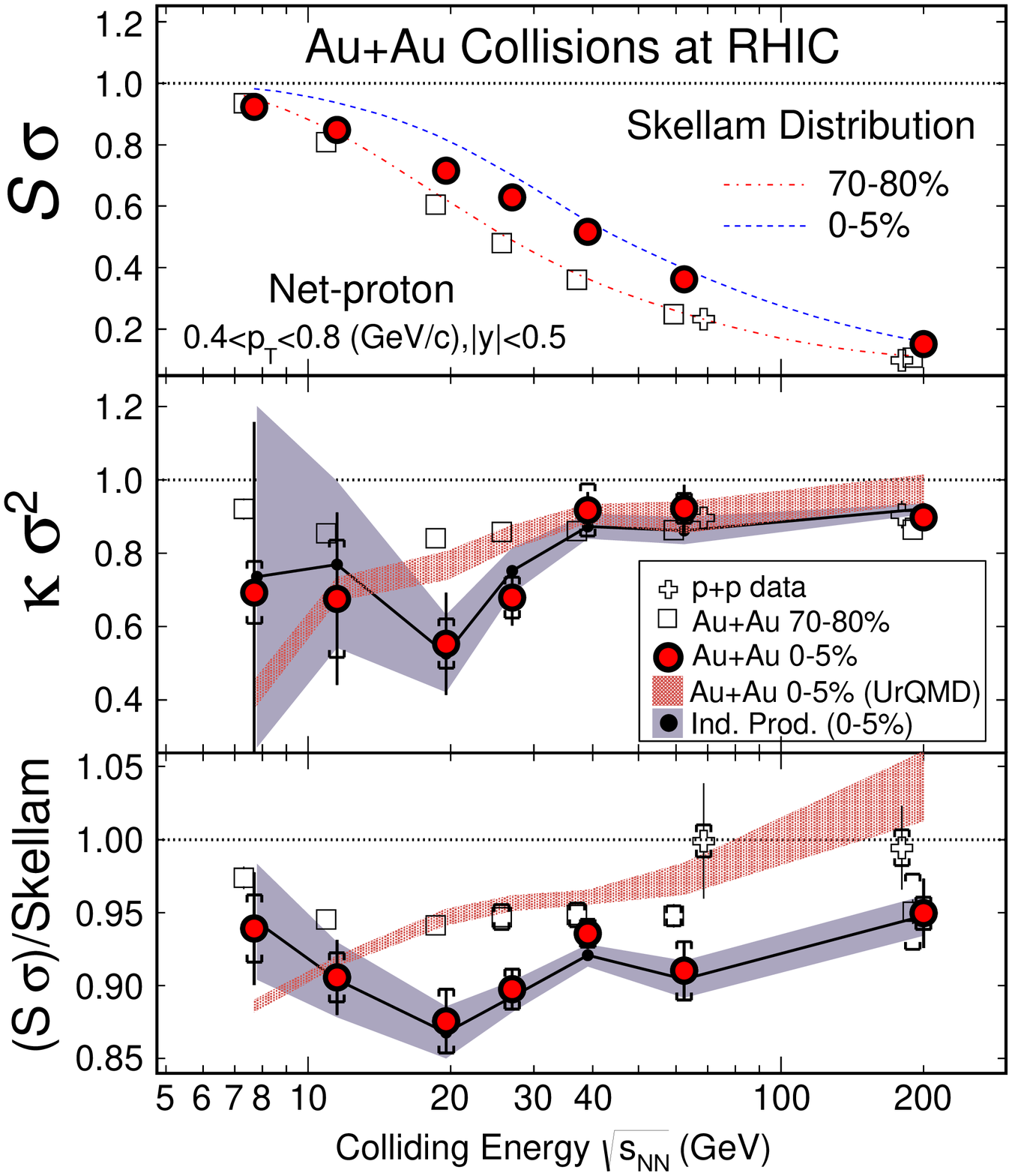}
\hspace{1.cm}
\includegraphics*[width=8cm]{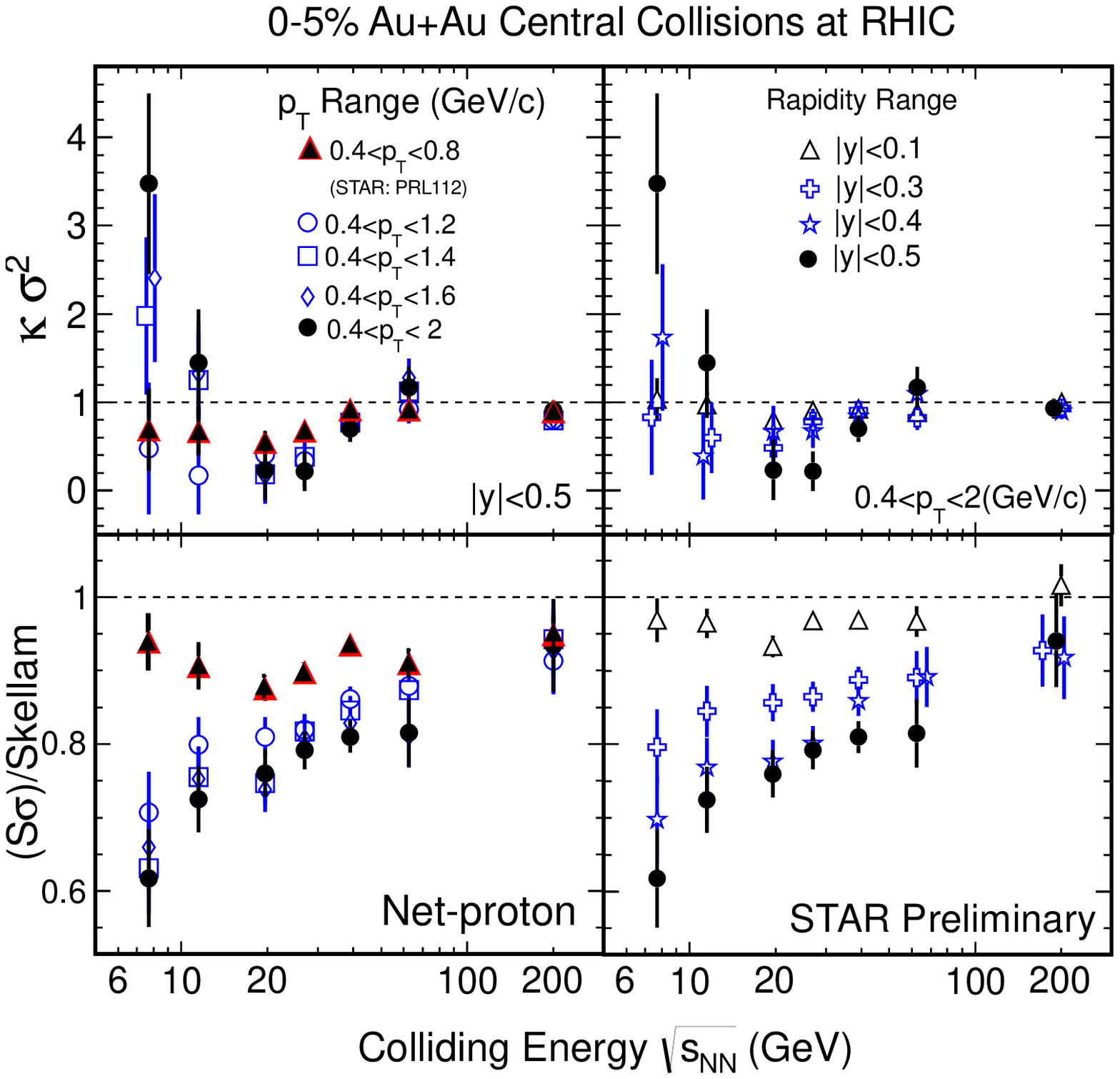}
\caption{
Ratios of net-proton number cumulants 
measured by STAR Collaboration at RHIC \cite{Adamczyk:2013dal,Luo:2015ewa}.
The right panel \cite{Luo:2015ewa} is the updated version
of the left panel \cite{Adamczyk:2013dal}.
}
\label{fig:STARfluc}
\end{center}
\end{figure}

In Figs.~\ref{fig:ALICEfluc} and \ref{fig:STARfluc}, we show two 
examples of recent experimental results on event-by-event analyses
of bulk fluctuations.
Figure~\ref{fig:ALICEfluc} shows the experimental result obtained 
by ALICE Collaboration at LHC \cite{ALICE}.
This figure shows the variance of the net-electric charge; 
the right vertical axis shows the quantity called D-measure, i.e. 
the variance normalized in such a way that the value in the 
equilibrated hadronic medium becomes $3 \sim 4$ \cite{Jeon:2000wg}.
The horizontal axis is the rapidity window $\Delta\eta$ to count 
the particle number.
The figure shows that the experimental result has 
a nontrivial suppression, which cannot be described by
the equilibrated hadronic degrees of freedom.
The result thus suggests that the net-electric charge fluctuation 
at the LHC energy contains non-hadronic or non-thermal physics
in the primordial medium.
The origin of the suppression in Fig.~\ref{fig:ALICEfluc} will be 
discussed in Sec.~\ref{sec:equil:QCD}.
The experimental result also shows that the fluctuation is 
more suppressed for larger $\Delta\eta$.
This behavior will be discussed in detail in Sec.~\ref{sec:diffusion:SDE}.

In Fig.~\ref{fig:STARfluc}, we show the experimental results on 
the non-Gaussian fluctuations measured by STAR Collaboration at RHIC
\cite{Adamczyk:2013dal,Luo:2015ewa}.
The two panels show the same quantity, the ratios of the net-proton
number cumulants, as a function of the collision energy $\sqrt{s_{\rm NN}}$;
since the baryon chemical potential $\mu_{\rm B}$ of the hot medium 
becomes smaller as 
$\sqrt{s_{\rm NN}}$ increases \cite{Kumar:2012fb,Randrup:2006nr},
these plots can be interpreted as the $\mu_{\rm B}$ dependence of 
the cumulants.
The right panel \cite{Luo:2015ewa} is the updated version of the 
left panel \cite{Adamczyk:2013dal}; the quality of the experimental 
analysis is ever-improving in this field \cite{Luo:2015ewa}.
In the right panel, the vertical axes are quantities which is 
expected to take unity in the equilibrated hadronic medium.
The panel shows that these quantities 
take values which are close to the hadronic one
but have statistically-significant deviation from those values.
These deviations are believed to be important
observables to explore the QCD phase structure.
In this review, we will elucidate the meanings of the vertical
axes in Fig.~\ref{fig:STARfluc} and the reason why these quantities are
widely discussed.

\subsection{Contents of this review}

In this review, we give a pedagogical introduction
to the physics of fluctuations in relativistic heavy ion
collisions.
In particular, one of the objectives of the introductory part
is to understand the meanings of Figs.~\ref{fig:ALICEfluc} and 
\ref{fig:STARfluc}.
We aimed at answering, for example, the following questions
in this review:
\begin{itemize}
\item
What are the {\it cumulants}?
Why should we focus on these quantities in the discussion 
of non-Gaussian fluctuations?
Why are the cumulants sometimes called susceptibility, and 
what are the relation of cumulants with moments, skewness and
kurtosis?

\item
Meanings of the vertical axes of
Figs.~\ref{fig:ALICEfluc} and \ref{fig:STARfluc}.
How to understand these experimental data?

\item
What happens in the fluctuation observables in heavy ion 
collisions if the hot medium undergoes a phase transition to 
deconfinement transition, or passes near the QCD critical point?

\item
Why can the event-by-event fluctuation be compared with 
thermal fluctuations?
Why should one {\it not} directly compare the event-by-event 
fluctuations with thermal fluctuations?

\item
The concept of ``equilibration of the fluctuation of conserved charges.''
What is the difference of this concept from 
the ``local equilibration,'' and why should we distinguish
them?

\end{itemize}
In addition to the answers to these questions, 
we have tried to describe recent progress in this field of research.

The outline of this review is as follows.
In the next section, 
we give a pedagogical review on the probability distribution
function, which is a basic quantity describing fluctuations.
The cumulants are introduced here, and their properties,
especially the extensive nature, are discussed.
In Sec.~\ref{sec:equil}, 
we discuss the thermal fluctuations, i.e. the fluctuations 
in an equilibrated medium.
The behaviors of cumulants in the QCD phase diagram is also
considered.
Sections~\ref{sec:e-v-e} -- \ref{sec:binomial} are devoted to 
a review on the event-by-event fluctuations in experimental 
analyses.
In Sec.~\ref{sec:e-v-e}, we summarize general properties of the 
event-by-event fluctuations. Various cautions in the interpretation
of these quantities are given in this section.
In Sec.~\ref{sec:diffusion}, we focus on the non-equilibrium
property of the event-by-event fluctuations, by describing 
the time evolution of fluctuations 
using stochastic formalisms.
In Sec.~\ref{sec:binomial}, we consider a model for probability
distribution functions which treats the efficiency problem
in the observation of fluctuations. The difference between
net-baryon and net-proton number cumulants are also discussed here.
We then give a short summary in Sec.~\ref{sec:concl}.

\section{General introduction to probability distribution function}
\label{sec:basic}

Because fluctuation is a distribution of some observable,
it is mathematically represented by probability distribution 
functions. 
For example, if one repeats a measurement of an observable in an 
equilibrated medium many times, the result of the measurement 
would fluctuate measurement by measurement.
The distribution of the result of the measurement is 
represented by a histogram.
After accumulating the results of many measurements, the 
histogram with an appropriate normalization can be regarded as 
the probability distribution function.
This distribution is nothing other than the fluctuation.
In many contexts, the width of the distribution is particularly
called fluctuations. 

In this section, as preliminaries of later sections we give 
a pedagogical introduction to basic concepts in probability.
We introduce moments and cumulants as quantities 
characterizing probability distribution functions.
Advantages of cumulants, especially for the description of 
non-Gaussianity of distribution functions, are elucidated.
We also discuss properties of specific distribution functions,
Poisson, Skellam, binomial and Gauss distributions.
The properties of these distribution functions play important 
roles in later sections for interpreting fluctuation 
observables in relativistic heavy ion collisions.

\subsection{Moments and cumulants}
\label{sec:basic:def}

We start from a probability distribution function $P(m)$
satisfying $\sum_m P(m)=1$
for an integer stochastic variable $m$.
One of the set of quantities which characterizes $P(m)$
is the moments.
The $n$-th order moment is defined by 
\begin{align}
\langle m^n \rangle = \sum_m m^n P(m),
\end{align}
where the bracket on the left-hand side represents the 
statistical average with $P(m)$.
If the moments for all $n>0$ exists,
they carry all information encoded in $P(m)$.
For a probability distribution function $P(x)$ for 
a continuous stochastic variable $x$, 
the moments are defined by
\begin{align}
\langle x^n \rangle = \int dx x^n P(x),
\label{eq:<x^n>}
\end{align}
where the integral is taken over the range of $x$.

To calculate the moments for a given probability distribution $P(m)$,
it is convenient to introduce the moment generating function,
\begin{align}
G(\theta) = \sum_m e^{m\theta} P(m) = \langle e^{m\theta} \rangle.
\label{eq:G(theta)}
\end{align}
Moments are then given by the derivatives of $G(\theta)$ as 
\begin{align}
\langle m^n \rangle =
\left. \frac{d^n}{d\theta^n} G(\theta) \right |_{\theta =0} .
\label{eq:moment}
\end{align}
For the continuous case the generating function is defined by
\begin{align}
G(\theta) = \int dx  e^{ x \theta} P(x).
\end{align}

For many practical purposes, it is more convenient to use 
{\it cumulants} rather than moments for characterizing a 
probability distribution.
To define the cumulants, we start from the cumulant 
generating function,
\begin{align}
K(\theta) = \ln G(\theta).
\label{eq:K(theta)}
\end{align}
The cumulants of $P(m)$ are then defined by 
\begin{align}
\langle m^n \rangle_{\rm c}
= \left. \frac{d^n}{d\theta^n} K(\theta) \right |_{\theta=0} .
\label{eq:cumulant}
\end{align}
As we will see below, cumulants have several useful features
for describing fluctuations, especially 
their non-Gaussianity.

Before discussing the advantages of cumulants, 
let us clarify the relation between moments and cumulants.
These relations are obtained straightforwardly from 
their definitions Eqs.~(\ref{eq:moment}) and (\ref{eq:cumulant}).
For example, to write cumulants in terms of 
moments, we calculate as follows:
\begin{align}
\langle m \rangle_{\rm c} 
&= \left. \frac{d}{d\theta} \ln G(\theta) \right |_{\theta=0} 
= \frac{G^{(1)}(0)}{G(0)} = \langle m \rangle ,
\label{eq:<m1>c=}
\\
\langle m^2 \rangle_{\rm c} 
&= \left. \frac{d^2}{d\theta^2} \ln G(\theta) \right  |_{\theta=0}
= \frac{G^{(2)}(0)}{G(0)} - \frac{(G^{(1)}(0))^2}{(G(0))^2} 
= \langle m^2 \rangle - \langle m \rangle^2 
= \langle \delta m^2 \rangle ,
\label{eq:<m2>c=}
\end{align}
where $G^{(n)}(\theta)$ represents the $n$-th derivative of 
$G(\theta)$ and we have used $G(0)=\sum_m P(m)=1$.
In the last equality we defined 
$\delta m = m - \langle m \rangle$.
By repeating a similar manipulation, 
one can extend the relation to an arbitrary order.
The results for third- and fourth-orders are given by 
\begin{align}
\langle m^3 \rangle_{\rm c} &= \langle \delta m^3 \rangle ,
\label{eq:<m3>c=}
\\
\langle m^4 \rangle_{\rm c} 
&= \langle \delta m^4 \rangle - 3 \langle \delta m^2 \rangle^2.
\label{eq:<m4>c=}
\end{align}
Note that the first-order cumulant is equal to  the first-order 
moment, or the expectation value.
The second- and third-order cumulants are given by the central moments,
\begin{align}
\langle \delta m^n \rangle = \langle ( m - \langle m \rangle )^n \rangle. 
\end{align}
In particular, the second-order cumulant $\langle \delta m^2 \rangle$
corresponds to the {\it variance}.
This quantity is sometimes called simply fluctuation, 
because for many purposes the cumulants higher than the second-order are 
not physically significant.
The cumulants for $n\ge4$ are given by nontrivial combinations 
of central moments with $n$-th and lower orders.

All cumulants except for the first-order one are
represented by central moments and do not depend on the average 
$\langle m \rangle$.
To prove this statement,
we consider a probability distribution function
$P'(m) = P(m-m_0)$ in which the distribution is shifted by $m_0$
compared with $P(m)$.
The cumulant generating function of $P'(m)$ is calculated to be
\begin{align}
K'(\theta)
&= \ln \sum_m e^{m\theta} P'(m)
= \ln \sum_m e^{m\theta} P(m-m_0)
= \ln \sum_m e^{(m+m_0) \theta} P(m)
\nonumber \\
&= \ln \sum_m e^{m\theta} P(m) + m_0 \theta
= K(\theta) + m_0 \theta ,
\label{eq:K'=K+m_0}
\end{align}
where $K(\theta)$ is the cumulant generating function of $P(m)$.
Equation~(\ref{eq:K'=K+m_0}) shows that the difference between
$K'(\theta)$ and $K(\theta)$ is a term $m_0 \theta$.
The derivatives of $K'(\theta)$ and $K(\theta)$ higher than 
the first-order thus are equivalent.
Therefore, the cumulants higher than the first-order do not depend 
on $\langle m\rangle$, and they are represented only by central 
moments.

The expressions of moments in terms of cumulants 
are similarly obtained as follows:
\begin{align}
\langle m \rangle 
&= \left. \frac{d}{d\theta} e^{K(\theta)} \right |_{\theta=0}
= K^{(1)}(0) e^{K(0)} = \langle m \rangle_{\rm c} ,
\label{eq:<m>=K1}
\\
\langle m^2 \rangle
&= \left. \frac{d^2}{d\theta^2} e^{K(\theta)} \right |_{\theta=0}
= ( K^{(2)}(0) + (K^{(1)}(0))^2 ) e^{K(0)}
= \langle m^2 \rangle_{\rm c} + \langle m \rangle_{\rm c}^2 ,
\label{eq:<m>=K2}
\end{align}
and so forth. Up to the fourth-order, one obtains
\begin{align}
\langle m^3 \rangle 
&= \langle m^3 \rangle_{\rm c} 
+ 3 \langle m^2 \rangle_{\rm c} \langle m \rangle_{\rm c} 
+ \langle m \rangle_{\rm c}^3, \nonumber \\
\langle m^4 \rangle 
&= \langle m^4 \rangle_{\rm c} 
+ 4 \langle m^3 \rangle_{\rm c} \langle m \rangle_{\rm c} 
+ 3 \langle m^2 \rangle_{\rm c}^2
+ 6 \langle m^2 \rangle_{\rm c} \langle m \rangle_{\rm c}^2 
+ \langle m \rangle_{\rm c}^4 .
\end{align}

\subsection{Sum of two stochastic variables}
\label{sec:basic:sum}

An important property of cumulants becomes apparent when 
one considers the sum of two stochastic variables.
Let us consider two integer stochastic variables $m_1$ and $m_2$
which respectively obey probability distribution 
functions $P_1(m_1)$ and $P_2(m_2)$ which are not correlated.
Then, the probability distribution of the sum of 
two stochastic variables, $m = m_1+m_2$, is given by
\begin{align}
P(m) = \sum_{m_1,m_2} \delta_{m, m_1+m_2} P(m_1) P(m_2).
\label{eq:m=m_1+m_2}
\end{align}
(To understand Eq.~(\ref{eq:m=m_1+m_2}), one may, for example, 
imagine the probability distribution of the sum of the numbers of 
two dices.)
The moment and cumulant generating functions for $P(m)$ are
calculated to be 
\begin{align}
G(\theta) &= \sum_m e^{m\theta} P(m)
= \sum_m e^{m\theta} \sum_{m_1,m_2} \delta_{m, m_1+m_2} P_1(m_1) P_2(m_2)
\nonumber \\
&= \sum_{m_1} e^{m_1\theta} P_1(m_1) \sum_{m_2}e^{m_2\theta} P_2(m_2)
= G_1(\theta) G_2(\theta) ,
\\
K(\theta) &= \ln G(\theta)
= K_1(\theta) + K_2(\theta),
\label{eq:K=K_1+K_2}
\end{align}
where $G_i(\theta)=\sum_m e^{m\theta} P_i(m)$ and 
$K_i(\theta)=\ln G_i(\theta)$ are the moment and cumulant 
generating functions of $P_i$, respectively, for $i=1$ and $2$.
By taking $n$ derivatives of the both sides of Eq.~(\ref{eq:K=K_1+K_2}),
one finds 
\begin{align}
\langle m^n \rangle_{\rm c}
= \langle m_1^n \rangle_{\rm c} + \langle m_2^n \rangle_{\rm c}.
\label{eq:<m^n>_c=<m_1^n>_c+<m_2^n>_c}
\end{align}
This result shows that the cumulants of the probability distribution
for the sum of two independent stochastic variables 
are simply given by the sum of the cumulants.
(This is the reason why the cumulants are called in this way.)
Note that this result is obtained for two {\it independent} 
stochastic variables; when the distributions of $m_1$ and $m_2$ are 
correlated, Eq.~(\ref{eq:<m^n>_c=<m_1^n>_c+<m_2^n>_c}) no longer holds.

\subsection{Cumulants in statistical mechanics}
\label{sec:basic:stat}

In statistical mechanics, results of measurement of 
observables in a volume $V$ are fluctuating, and one can define 
their cumulants from the distribution of the results.
From Eq.~(\ref{eq:<m^n>_c=<m_1^n>_c+<m_2^n>_c}) one can argue 
an important property of the cumulants in statistical mechanics 
that the cumulants of extensive variables in grand canonical 
ensemble are {\it extensive variables}.

To see this, let us consider the number $N$ of a conserved charge
in a volume $V$ in grand canonical ensemble.
From the distribution of the result of measurements, 
one can define the cumulants $\langle N^n \rangle_{{\rm c},V}$ 
of the charge.
Next, let us consider the cumulants of the particle number in 
a twice larger volume, $\langle N^n \rangle_{{\rm c},2V}$.
This system can be separated 
into two subsystems with an equal volume $V$.
In statistical mechanics, it is usually assumed that the subsystems
are uncorrelated when the volume is sufficiently large, and the
property of the system does not depend on the shape of $V$.
Therefore, the particle number in the total system is regarded 
as the sum of the two independent particle numbers in the two subsystems.
From Eq.~(\ref{eq:<m^n>_c=<m_1^n>_c+<m_2^n>_c}), 
$\langle N^n \rangle_{{\rm c},2V}$ thus are represented as
\begin{align}
\langle N^n \rangle_{{\rm c},2V}
= 2 \langle N^n \rangle_{{\rm c},V}.
\end{align}
By similar arguments one obtains,
\begin{align}
\langle N^n \rangle_{{\rm c},\lambda V}
= \lambda \langle N^n \rangle_{{\rm c},V}
\label{eq:NV=NV'}
\end{align}
for an arbitrary number $\lambda$.
Equation~(\ref{eq:NV=NV'}) shows that the cumulants of $N$
in statistical mechanics are extensive variables.
As special cases of this property,
the average particle number $\langle N \rangle$ and the variance 
$\langle \delta N^2 \rangle$ in statistical mechanics are 
extensive variables.

From Eq.~(\ref{eq:NV=NV'}), the cumulants in volume $V$ can be
written as
\begin{align}
\langle N^n \rangle_{{\rm c},V} = \chi_n V. 
\label{eq:<N>=chiV}
\end{align}
Here, $\chi_1$ is the density of the particle, 
and $\chi_2$ is the quantity which is referred to as 
susceptibility because of the linear response relation discussed in 
Sec.~\ref{sec:equil:LRR}.
We call $\chi_n$ for $n\ge3$ as generalized susceptibilities.

Remarks on the extensive nature of the cumulants are in order.
First, the above argument is valid only when the volume of the 
system is sufficiently large.
When the spatial extent of the volume is not large enough, 
the correlation between two adjacent volumes becomes non-negligible.
This would happen when the spatial extent of the volume is comparable 
with the microscopic correlation lengths.
Next, in the above argument we have implicitly assumed grand 
canonical ensemble in addition to the equilibration.
The argument, for example, is not applicable to subvolumes in 
canonical ensemble in which the number of $N$ in the total 
system is fixed. 
In this case, the fixed total number gives rise to 
correlation between subvolumes;
because the total number is fixed, if the particle number in 
a subvolume is large the particle number in the other subvolume
tends to be suppressed.
This correlation violates the assumption of independence 
between the particle numbers in subvolumes unless the 
subvolume is small enough compared with the total volume.

\subsection{Examples of distribution functions}
\label{sec:basic:dist}

Now, we see some specific distribution functions,
which play important roles in relativistic heavy ion collisions.

\subsubsection{Binomial distribution}
\label{sec:basic:binomial}

The binomial distribution function is defined by
the number of ``successes'' of $N$ independent trials, 
each of which yields a success with probability $p$.
The binomial distribution function is given by
\begin{align}
B_{p,N} (m) = {_N C_m} p^m (1-p)^{N-m} , 
\label{eq:binomial}
\end{align}
where
\begin{align}
_N C_m = \frac{ N! }{ m! (N-m)! }
\end{align}
is the binomial coefficient.
It is easy to show that $\sum_m B_{p,N} (m) = 1$ using the binomial theorem.
The moment and cumulant generating functions of the binomial 
distribution are calculated to be
\begin{align}
G_{\rm B}(\theta) &= \sum_m e^{m\theta} B_{p,N} (m)
= \sum_m {_N C_m} (e^\theta p)^m (1-p)^{N-m}
\nonumber \\
&= ( 1 - p + e^\theta p )^N ,
\\
K_{\rm B}(\theta) &= N \ln ( 1 - p + e^\theta p ).
\label{eq:K_B}
\end{align}
From Eq.~(\ref{eq:K_B}), 
the cumulants of the binomial distribution function
are given by 
\begin{align}
\langle m^n \rangle_{\rm c} = \xi_n N
\label{eq:<m>_c=xi_n}
\end{align}
with explicit forms of $\xi_n$ up to the fourth-order 
\begin{align}
\xi_1 = p , \quad
\xi_2 = p(1-p) , \quad
\xi_3 = p(1-p)(1-2p) , \quad
\xi_4 = p(1-p)(1-6p+6p^2).
\label{eq:xi_binomial}
\end{align}
Equation~(\ref{eq:<m>_c=xi_n}) shows that 
the cumulants of the binomial distribution 
are proportional to $N$, which is a reasonable result 
from the extensive nature of cumulants.

The sum of two stochastic variables obeying independent binomial 
distribution functions with an equal probability $p$
is again distributed with a binomial one.
This can be shown by explicitly deriving the identity,
\begin{align}
B_{p,N_1+N_2}( m ) = \sum_{m_1,m_2} \delta_{m, m_1+m_2} B_{p,N_1}(m_1) B_{p,N_2}(m_2).
\end{align}
An alternative way to prove this statement is to use 
Eqs.~(\ref{eq:<m^n>_c=<m_1^n>_c+<m_2^n>_c}) and (\ref{eq:<m>_c=xi_n}).
Suppose that $m_1$ and $m_2$ obey the binomial distribution
$B_{p,N_1}(m_1)$ and $B_{p,N_2}(m_2)$, respectively.
From Eq.~(\ref{eq:<m^n>_c=<m_1^n>_c+<m_2^n>_c}), 
one finds that the cumulants of the sum $m=m_1+m_2$ are 
given by $ \langle m^n \rangle_{\rm c} = \xi_n (N_1+N_2)$, 
which are nothing but the cumulants of the binomial 
distribution $B_{p,N_1+N_2}( m )$.
Because all cumulants are those of the binomial distribution,
the distribution of $m_1+m_2$ is given by the binomial one.

\subsubsection{Poisson distribution}
\label{sec:basic:Poisson}

The Poisson distribution function is defined by 
the number of successes of independent trials, 
each of which yields a success with infinitesimal probability.
The Poisson distribution is thus obtained by taking the $p\to0$ 
limit of the binomial distribution function with fixed $\lambda=pN$.
Replacing the binomial coefficients in Eq.~(\ref{eq:binomial}) as
$_N C_m \to N^m/m!$, which is valid in this limit, 
and using the definition of Napier's number $e=\lim_{p\to0}(1+p)^{1/p}$
in Eq.~(\ref{eq:binomial}), one obtains
\begin{align}
P_\lambda (m) = \frac{\lambda^m}{m!} e^{-\lambda}.
\label{eq:Poisson}
\end{align}
The cumulant generating function of Eq.~(\ref{eq:Poisson}) is
obtained as 
\begin{align}
K_\lambda(\theta) = \lambda ( e^{\theta} -1 ).
\label{eq:K_P}
\end{align}
By taking derivatives of Eq.~(\ref{eq:K_P}), 
one finds that all the cumulants of the Poisson distribution
are the same, 
\begin{align}
\langle m^n \rangle_{\rm c} = \lambda \mbox{~~(for any $n\ge1$)}.
\label{eq:<m^n>_c=lambda}
\end{align}
This property will be used frequently in later sections.
This result also shows that the Poisson distribution is characterized
by a single parameter $\lambda$, while the binomial distribution function 
is characterized by the two parameters, $p$ and $N$.

The sum of two stochastic variables obeying two independent Poissonian 
obeys Poissonian,
\begin{align}
P_{\lambda_1+\lambda_2}( m ) = \sum_{m_1,m_2} \delta_{m, m_1+m_2} 
P_{\lambda_1}(m_1) P_{\lambda_2}(m_2), 
\end{align}
as one can explicitly show easily.
Similarly to the case of the binomial distribution, however, 
a much easier way to show this is to use 
Eqs.~(\ref{eq:<m^n>_c=<m_1^n>_c+<m_2^n>_c}) and (\ref{eq:<m^n>_c=lambda}).

The Poisson distribution is one of the most fundamental distribution 
function, as it naturally appears in various contexts.
One interesting example among them is the classical ideal gas.
One can show that the distribution of the number of a classical free 
particles in a volume in grand canonical ensemble obeys 
the Poisson distribution.
To show this, the following intuitive argument suffices.
First, consider a canonical ensemble for a volume $V_{\rm c}$ with 
a fixed particle number $N_{\rm c}$.
The grand canonical ensemble is defined by a subvolume $V$ in this 
system in the limit $r \equiv V/V_{\rm c}\to0$.
Next, consider a particle in this system.
The probability that this particle exists in the subvolume $V$ 
is given by $r$. Moreover, because we consider classical free particles, 
the probability 
that each particle is in the subvolume is uncorrelated.
The probability distribution function of the particle number $N$ 
in $V$ thus is given by the binomial distribution with 
probability $r$.
Because the grand canonical ensemble is defined by $r\to0$ limit
with fixed $N= rN_{\rm c} = VN_{\rm c}/V_{\rm c}$,
the distribution in this limit obeys Poissonian.
In Sec.~\ref{sec:equil:ideal} we will explicitly calculate the 
cumulants in the classical ideal gas and show that they satisfy
Eq.~(\ref{eq:<m^n>_c=lambda}).
When the effect of quantum statistics, Bose-Einstein or Fermi-Dirac,
shows up, this argument does not hold owing to the quantum correlation even
for ideal gas.

\subsubsection{Skellam distribution}
\label{sec:basic:Skellam}

Although the sum of stochastic variables obeying independent Poisson 
distributions again obeys Poissonian, 
the {\it difference} of two stochastic variables obeying 
independent Poisson distributions is not given by a Poisson distribution.
This is obvious from the fact that the difference can 
take a negative value, while the Poisson distribution $P_\lambda(m)$
take nonzero values only for positive $m$.
The difference,
\begin{align}
S_{\lambda_1,\lambda_2}(m) = \sum_{m_1,m_2} \delta_{m, m_1-m_2} 
P_{\lambda_1}(m_1) P_{\lambda_2}(m_2) ,
\end{align}
is called the Skellam distribution.
The generating functions of the Skellam distribution are calculated to be
\begin{align}
G(\theta) &= \sum_m e^{m\theta} \sum_{m_1,m_2} \delta_{m, m_1-m_2} 
P_{\lambda_1}(m_1) P_{\lambda_2}(m_2)
\\
&= \sum_{m_1} e^{m_1 \theta} P_{\lambda_1}(m_1) 
\sum_{m_2} e^{-m_2 \theta} P_{\lambda_2}(m_2) 
\\
&= G_{\lambda_1}(\theta) G_{\lambda_2}(-\theta) ,
\\
K(\theta) &= K_{\lambda_1}(\theta) + K_{\lambda_2}(-\theta),
\end{align}
where generating functions $G_{\lambda}(\theta)$ and 
$K_{\lambda}(\theta)$ are those of Poisson distribution.
By taking derivatives of this result with Eq.~(\ref{eq:K_P}),
one obtains
\begin{align}
\langle m^n \rangle_{\rm c} = 
\langle m_1^n \rangle_{\rm c} + (-1)^n \langle m_2^n \rangle_{\rm c}
= \lambda_1 + (-1)^n \lambda_2
\label{eq:Skellam_sum}
\end{align}
for the Skellam distribution.
This result shows that all even cumulants take a common value 
$\lambda_1+\lambda_2$, while the odd cumulants take 
$\lambda_1-\lambda_2$.
This result also shows that the Skellam distribution is 
characterized by two parameters, $\lambda_1$ and $\lambda_2$, or 
alternatively odd and even cumulants.

The explicit analytic form of $S(m)$ is given by
\begin{align}
S_{\lambda_1,\lambda_2}(m) 
= e^{-(\lambda_1+\lambda_2)} \left(\frac{\lambda_1}{\lambda_2}\right)
I_m\left( 2\sqrt{\lambda_1\lambda_2}\right),
\end{align}
where $I_m(z)$ is the modified Bessel function of the first kind.

The Skellam distribution plays an important role in 
later sections, 
because they describe the distribution of net particle number,
i.e. the difference of the numbers of 
particles and anti-particles, in the classical ideal gas.
The fluctuation of net-baryon number in hadron resonance gas, for example,
obeys the Skellam distribution.

\subsubsection{Gauss distribution}
\label{sec:basic:Gauss}

So far we have considered stochastic variables taking integer values.
An example of a distribution with a 
continuous stochastic variable is the Gauss distribution, 
which is defined by
\begin{align}
P_G(x) = \frac{1}{\sigma\sqrt{2\pi}} 
\exp\bigg[ - \frac{ ( x - x_0 )^2 }{ 2\sigma^2 } \bigg].
\end{align}
The normalization factor is required to satisfy
$\int_{-\infty}^{\infty} dx P_G(x) = 1$.
The generating functions are calculated to be
\begin{align}
G(\theta) 
&= \int dx e^{\theta x} \frac{1}{\sigma\sqrt{2\pi}} 
\exp\bigg[ - \frac{ ( x - x_0 )^2 }{ 2\sigma^2 } \bigg]
= \exp \bigg[ x_0 \theta + \frac12 \sigma^2 \theta^2 \bigg] ,
\nonumber \\
K(\theta) 
&= x_0 \theta + \frac12 \sigma^2 \theta^2.
\end{align}
We thus have
\begin{align}
\langle x \rangle &= x_0 , 
~~
\langle x^2 \rangle_{\rm c} = \sigma^2 ,
\label{eq:Gauss1,2}
\end{align}
and
\begin{align}
\langle x^n \rangle_{\rm c} &= 0 \quad \mbox{for}~ n\ge3.
\label{eq:Gauss>3}
\end{align}
The results in Eqs.~(\ref{eq:Gauss1,2}) and (\ref{eq:Gauss>3})
can, of course, also be obtained by explicitly calculating 
$\langle x \rangle = \int dx x P_G(x)$ 
and $\langle x^2 \rangle_{\rm c} = \langle \delta x^2 \rangle = 
\int dx ( x-x_0)^2 P_G(x)$, and so forth.

Equations~(\ref{eq:Gauss1,2}) and (\ref{eq:Gauss>3}) show that 
the cumulants higher than the second-order vanish for the Gauss
distribution.
In other words, nonzero higher order cumulants
characterize deviations from the Gauss distribution function.
This is the reason why the cumulants are used as 
quantities representing non-Gaussianity.

\subsection{Variance, skewness and kurtosis}
\label{sec:basic:skewkurt}

Till now, we have discussed cumulants as quantities 
characterizing distribution functions.
When one wants to describe the deviation from the Gauss distribution,
it is sometimes convenient to use 
the quantities called skewness $S$ and kurtosis $\kappa$ 
\cite{Pearson}.
These quantities are defined as 
\begin{align}
S = \frac{ \langle x^3 \rangle_{\rm c} }{ \langle x^2 \rangle_{\rm c}^{3/2}} 
= \frac{ \langle x^3 \rangle_{\rm c} }{\sigma^3} ,
\qquad 
\kappa = \frac{ \langle x^4 \rangle_{\rm c} }{ \langle x^2 \rangle_{\rm c}^2} 
= \frac{ \langle x^4 \rangle_{\rm c} }{\sigma^4} ,
\label{eq:skew-kurt}
\end{align}
where $\sigma^2$ is the variance defined by 
$\sigma^2 = \langle x^2 \rangle_{\rm c}$.
The skewness and kurtosis are interpreted as 
the third- and fourth-order cumulants of the 
renormalized stochastic variable $\tilde{x}= x/\sigma$ satisfying 
$\langle\tilde{x}^2\rangle_{\rm c}=1$, 
\begin{align}
S = \langle \tilde{x}^3 \rangle_{\rm c}, \qquad
\kappa = \langle \tilde{x}^4 \rangle_{\rm c}.
\end{align}

\begin{figure}
\begin{center}
\includegraphics*[width=8cm]{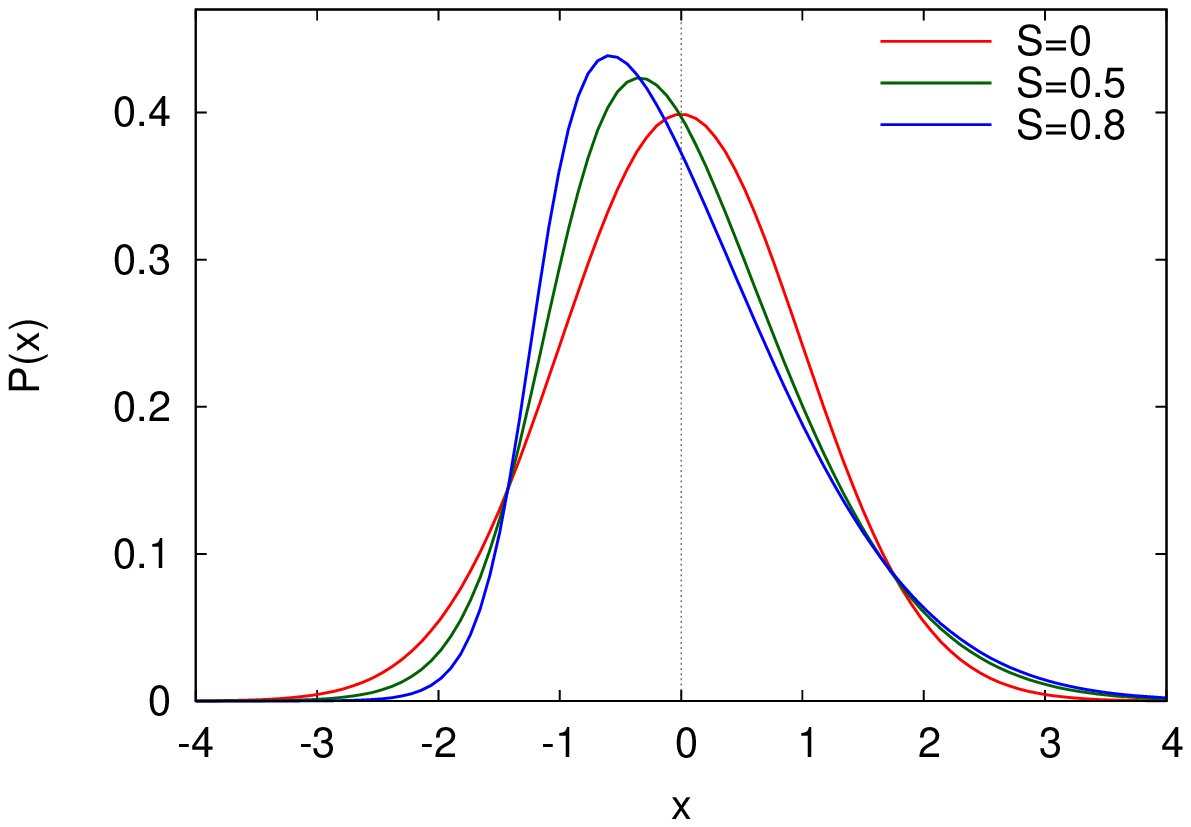}
\hspace{.5cm}
\includegraphics*[width=8cm]{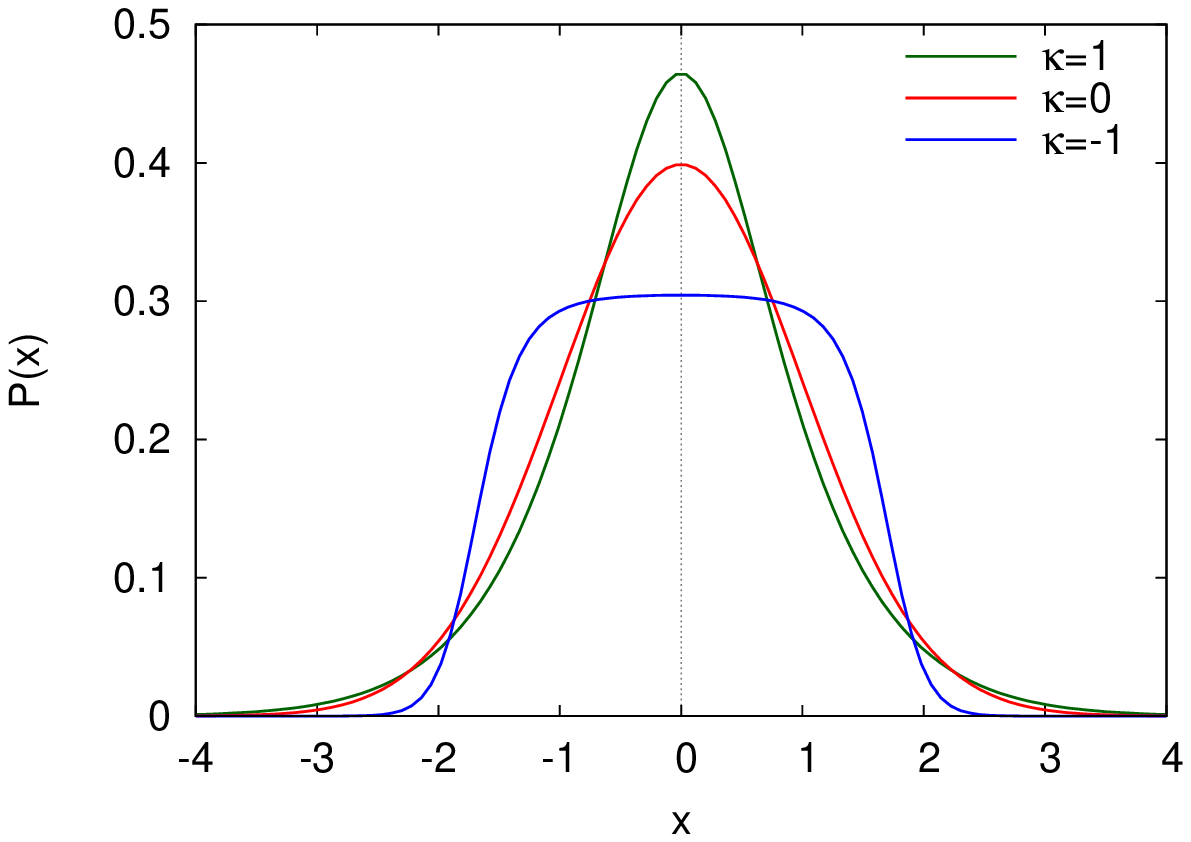}
\caption{
Typical distribution functions having nonzero skewness (left) 
and kurtosis (right).
The average and variance are set to $\langle x\rangle=0$ and 
$\sigma=1$.}
\label{fig:skew-kurt}
\end{center}
\end{figure}

In Fig.~\ref{fig:skew-kurt}, we show typical distribution functions 
having nonzero skewness and kurtosis.
All distribution functions shown in the figure satisfy 
$\langle x\rangle=0$ and $\langle x^2\rangle_{\rm c}=1$.
As shown in the left panel, the skewness represents the asymmetry of 
the distribution function.
The kurtosis, on the other hand, typically describes 
the ``sharpness'' of the distribution compared with the Gaussian one
as in the right panel.

When the non-Gaussianity of a distribution function is discussed,
the set of variables to be used, $S$ and $\kappa$, or the third and 
fourth-order cumulants, should be chosen depending on the problem.
When the distribution is expected to obey some specific one
of which the cumulants are known, 
the cumulants are much more convenient.
For example, when the distribution is expected to obey the 
Poisson one and the difference from this distribution is concerned, 
one may focus on the cumulants normalized by the average,
$\langle m^n \rangle_{\rm c}/\langle m \rangle$,
which become unity in the Poisson distribution.
The deviation from the Poisson distribution is then characterized
by the difference of the ratio from unity.
As we will see in Sec.~\ref{sec:equil}, for example, 
the fluctuation of net-baryon number in the equilibrated hadronic 
medium is well described by the Skellam distribution.
If the fluctuation carries some physics which cannot be described
by hadronic degrees of freedom in equilibrium, it may deviate
from the Skellam one.
In order to find this deviation, the best observable to be used
is the ratios of cumulants between even or odd orders 
\cite{Ejiri:2005wq}, which become unity for the Skellam distribution 
as discussed in Sec.~\ref{sec:basic:Skellam}.
Another example is the distribution of the topological charge, $Q$,  
in QCD. 
A model called the dilute instanton gas model 
for the topological sector in QCD predicts that the distribution 
of the topological charge is given by the Skellam one with 
$\langle Q \rangle=0$ \cite{GPY}.
The proximity of the ratios of the even order cumulants to 
unity thus is a useful measure to judge the validity of this model.
Recently, the measurements of the topological cumulants are
actively performed on the lattice \cite{Bonati:2013tt,
Kitano:2015fla,Borsanyi:2015cka}.

Contrary to the ratios of cumulants, 
the magnitudes of skewness and kurtosis 
in the Poisson distribution depend on the average $\lambda$ as 
\begin{align}
S = \lambda^{-1/2}, \quad
\kappa = \lambda^{-1}.
\end{align}
These quantities become arbitrary small as $\lambda$ becomes
larger.
This behavior is related to the central limit theorem in 
statistics, which states that the sum of independent 
events approaches a Gauss distribution as the 
number of events to be summed is increased.
Because the Poisson distribution can be interpreted as 
the result of the sum of independent events, it approaches
the Gauss distribution for large $\lambda$.
The $S$ and $\kappa$, the quantities characterizing 
the deviation from the Gauss distribution, become arbitrarily
small in this limit.
Similar tendency is also expected in the large volume limit 
of fluctuation observables in statistical mechanics.
Fluctuations in macroscopic systems are well 
described by Gauss distributions, and $S$ and $\kappa$ become irrelevant
in the large volume limit.
On the other hand, the higher order cumulants take nonzero 
values even in this limit.
There, however, are practical difficulties in the measurement 
of higher order cumulants in large systems.
In addition to the difficulty in the exact measurement of 
observables in large systems, there is a fundamental problem
that the statistical error of higher order cumulants grows as the 
volume increases, as discussed in the next subsection.

\subsection{Error of the cumulants}
\label{sec:basic:error}

Related to the above discussion on non-Gaussianity in large systems, 
we give some remarks on the statistical error 
associated with the measurement of higher order cumulants.

The statistical error $\Delta \hat{O}$ of an observable 
$\hat{O}$ is estimated as
\begin{align}
(\Delta \hat{O})^2 
= \frac{\langle ( \hat{O} - \langle \hat{O} \rangle )^2 \rangle}
{ N_{\rm stat}-1}
= \frac{\langle \delta \hat{O}^2 \rangle}{ N_{\rm stat}-1},
\label{eq:DeltaO}
\end{align}
where $N_{\rm stat}$ is the number of statistics, i.e. 
the number of the measurements of $\hat{O}$, and the expectation
value is taken over this statistical ensemble.

Let us consider an extensive observable $N$ in statistical mechanics,
such as particle number.
As discussed already, cumulants of $N$ are extensive
observables proportional to the volume $V$.
Now, we start from the first-order cumulant.
Usually, the density $\rho=N/V$ of this quantity is 
concerned rather than $N$.
By substituting $\rho$ into $\hat{O}$ in Eq.~(\ref{eq:DeltaO}), 
one obtains
\begin{align}
(\Delta \rho)^2 
= \frac{\langle \delta N^2 \rangle}{ V^2 N_{\rm stat} }
= \frac{ \chi_2 }{ V N_{\rm stat} },
\label{eq:Deltarho}
\end{align}
where in the last equality we used the extensive nature of 
the second-order cumulant Eq.~(\ref{eq:<N>=chiV}).
We have also assumed that $N_{\rm stat}\gg1$ and 
used an approximation $N_{\rm stat}-1 \simeq N_{\rm stat}$
in the denominator.
The result Eq.~(\ref{eq:Deltarho}) shows that the error of $\rho$ 
is proportional to $V^{-1/2}$, which is a well-known result.
With a fixed $N_{\rm stat}$, the error of $\rho$ becomes 
smaller as $V$ becomes larger.

Next, we consider the statistical error of the 
second-order cumulant of $N$.
By substituting $\delta N^2$ in $\hat{O}$, the error 
is calculated to be 
\begin{align}
(\Delta (\delta N^2) )^2 
&= \frac{\langle ( (\delta N^2) - \langle \delta N^2 \rangle )^2\rangle}{ N_{\rm stat}}
= \frac{\langle \delta N^4 \rangle - \langle \delta N^2 \rangle^2 }{ N_{\rm stat}}
= \frac{\langle N^4 \rangle_{\rm c} + 2 \langle N^2 \rangle_{\rm c}^2 }{ N_{\rm stat}}
\nonumber \\
&= \frac{ \chi_4 V + 2 ( \chi_2 V )^2 }{ N_{\rm stat}}.
\label{eq:D(dN^2)}
\end{align}
In this derivation, we have used the central moments 
as basic quantities rather than moments.
This choice suppresses the correlation between first- and
second-order moments.
The central moments are converted to cumulants in the third equality.
From Eq.~(\ref{eq:D(dN^2)}),
the error of the susceptibility $\chi_2$ is obtained by dividing 
both sides by $V^2$ as
\begin{align}
(\Delta \chi_2)^2 = \frac{ \chi_4 V^{-1} + 2 \chi_2^2 }{ N_{\rm stat}}
= \frac{ 2 \chi_2^2 }{ N_{\rm stat}} + {\cal O}(V^{-1}).
\end{align}
This result shows that the error of $\chi_2$ does not have $V$
dependence for sufficiently large $V$.
Contrary to the first-order case, the increase of $V$ does not 
reduce the statistical error of $\chi_2$.

Similarly, the error of the higher order cumulants can be 
estimated by substituting the definition of the cumulants in 
Eq.~(\ref{eq:DeltaO}).
The higher order cumulants are represented by central
moments in general as shown in sec. \ref{sec:basic:def}.
(To obtain the explicit forms, one may substitute $G^{(1)}=0$ in 
Eqs.~(\ref{eq:<m1>c=})--(\ref{eq:<m4>c=}) and so forth.)
It is therefore instructive to first see the error of the central moments.
The $n$-th order central moment is represented by the sum of 
the product of cumulants 
\begin{align}
\langle m^{n_1} \rangle_{\rm c} \langle m^{n_2} \rangle_{\rm c} \cdots 
\langle m^{n_i} \rangle_{\rm c}  ,
\end{align}
with $n=n_1+n_2+\cdots+n_i$ and $n_i\ge2$.
Moreover, one can easily show that the coefficients of all terms 
in this decomposition are positive and nonvanishing.
Now, because the cumulants are proportional to $V$,
the term which is leading in $V$ in large volume is the one containing
the product of the largest number of the cumulants.
For even $n$, it is $( \langle N^2 \rangle_{\rm c} )^{n/2}$
and one obtains the behavior of the central moments in large $V$ as
\begin{align}
\langle (\delta N)^n \rangle \sim
( \langle N^2 \rangle_{\rm c} )^{n/2} ( 1 + {\cal O}(V^{-1}) )
\sim ( \chi_2 V )^{n/2} + {\cal O}(V^{n/2-1}).
\label{eq:<dN>V}
\end{align}
For odd $n$, one obtains $\langle (\delta N)^n \rangle \sim V^{(n-1)/2}$.

Using these dependences on $V$, the statistical error of 
the central moments is given by
\begin{align}
(\Delta (\delta N^n))^2 
= \frac{ \langle \delta N^{2n} \rangle - \langle \delta N^n \rangle^2 }
{ N_{\rm stat} } 
\sim \frac{ (\chi_2 V)^n }{ N_{\rm stat} },
\label{eq:Delta<deltaN>}
\end{align}
where in the last step we have used 
Eq.~(\ref{eq:<dN>V}) and the fact that the terms proportional 
to $V^n$ never cancel out between $\langle \delta N^{2n} \rangle$
and $\langle \delta N^n \rangle^2$.
Subleading terms in $V^{-1}$ are neglected on the far right-hand side.
The error of the $n$-th order central moment thus is proportional to
$\sqrt{ V^n / N_{\rm stat} }$ for large $V$ and $N_{\rm stat}$.

Now, we come back to the statistical error of 
higher order cumulants.
As discussed already, the cumulants can be represented by
the central moments.
Substituting the cumulants in $\cal O$ into Eq.~(\ref{eq:DeltaO})
and representing them in terms of central moments, 
the error of the cumulants is given by the sum of the product 
of the central moments.
It is then concluded that the error of the cumulants
in the leading order in $V$ should be same as 
Eq.~(\ref{eq:Delta<deltaN>})
unless the highest order terms in $V$ cancel out.
The error of the generalized susceptibility thus is 
expected to behave as 
\begin{align}
\Delta \chi_n \sim \sqrt{ \frac{ \chi_2^n V^{n-2} }{ N_{\rm stat} } },
\label{eq:Deltachi}
\end{align}
for large $V$ and $N_{\rm stat}$.
Eq.~(\ref{eq:Deltachi}) shows that the error of $\chi_n$ for $n\ge3$
grows as $V$ becomes larger and
the measurement of 
non-Gaussian cumulants becomes more and more difficult 
as the spatial volume becomes larger.
This $V$ dependence is highly contrasted to the error of standard
observables Eq.~(\ref{eq:Deltarho}), which becomes smaller 
as $V$ becomes larger.
The proportionality coefficients in Eq.~(\ref{eq:Deltachi}) 
up to the fourth-order are presented in Ref.~\cite{Luo:2011tp}.

In relativistic heavy ion collisions, non-Gaussian cumulants
have been observed up to the fourth-order.
These measurements are possible because the size of the 
system observed by detectors is not large;
the particle number in each event is at most of order $10^3$.
The growth of the statistical error of higher order cumulants is 
a well known feature, and discussed in various contexts;
see for example Refs.~\cite{Morita:2013tu,Morita:2012kt,Morita:2014fda}.

\subsection{Cumulants for multiple variables}
\label{sec:basic:mult}

Next, we discuss probability distribution functions
for multiple stochastic variables and their moments and cumulants.

Let us consider a probability distribution function
$P(m_1,m_2)$ for integer stochastic variables $m_1$ and $m_2$.
The moments of this distribution are defined by
\begin{align}
\langle m_1^{n_1} m_2^{n_2} \rangle
= \sum_{m_1,m_2} m_1^{n_1} m_2^{n_2} P(m_1,m_2).
\end{align}
By defining the moment generating function as
\begin{align}
G(\theta_1,\theta_2) 
= \sum_{m_1,m_2} e^{\theta_1 m_1} e^{\theta_2 m_2} P(m_1,m_2) 
= \langle e^{\theta_1 m_1} e^{\theta_2 m_2} \rangle,
\end{align}
the moments are given by
\begin{align}
\langle m_1^{n_1} m_2^{n_2} \rangle
= \left. \frac{\partial^{n_1}}{\partial \theta_1^{n_1}} 
\frac{\partial^{n_2}}{\partial \theta_2^{n_2}} G(\theta_1,\theta_2)
\right |_{\theta_1 = \theta_2 = 0}
\end{align}
similarly to the case of a single variable.
The cumulants for $P(m_1,m_2)$ are similarly 
defined with the cumulant generating function 
$K(\theta_1,\theta_2) = \ln G(\theta_1,\theta_2)$ as
\begin{align}
\langle m_1^{n_1} m_2^{n_2} \rangle_{\rm c}
=  \left. \frac{\partial^{n_1}}{\partial \theta_1^{n_2}}
   \frac{\partial^{n_2}}{\partial \theta_2^{n_2}} 
K(\theta_1,\theta_2) \right |_{\theta_1 = \theta_2 = 0}.
\end{align}
It is easy to extend the argument in Sec.~\ref{sec:basic:def}
to relate the moments and cumulants to this case.
The relation of the cumulants with moments for $n_1=0$ or $n_2=0$ 
is equal to the previous case.
For the mixed cumulants, it is, for example, calculated to be
\begin{align}
\langle m_1 m_2 \rangle_{\rm c}
&= \left. \frac{\partial}{\partial \theta_1} 
\frac{\partial}{\partial \theta_2} \ln G(\theta_1,\theta_2)
\right |_{\theta_1 = \theta_2 = 0}
= 
\left. \frac{\partial}{\partial \theta_1} 
\bigg( \frac{\partial G(\theta_1,\theta_2)}{\partial \theta_2}
G(\theta_1,\theta_2)^{-1} \bigg) \right |_{\theta_1 = \theta_2 = 0}
\nonumber \\
&= 
\left. \bigg( G(\theta_1,\theta_2)^{-1}
\frac{\partial^2 G(\theta_1,\theta_2)}{\partial \theta_1 \partial \theta_2} 
- G(\theta_1,\theta_2)^{-2} 
\frac{\partial G(\theta_1,\theta_2)}{\partial \theta_1}
\frac{\partial G(\theta_1,\theta_2)}{\partial \theta_2} \bigg)
\right |_{\theta_1 = \theta_2 = 0}
\nonumber \\
&= 
\langle \delta m_1 \delta m_2 \rangle
\label{eq:<m1m2>}
\end{align}
and so forth.
Equation~(\ref{eq:<m1m2>}) shows that the mixed second-order 
cumulant is given by the mixed central moment, or correlation.

For a probability distribution function 
for $k$ stochastic variables $m_1, \cdots ,m_k$,
the generating functions are defined by 
\begin{align}
G(\theta_1,\cdots,\theta_k)
&= \sum_{m_1,m_2,\cdots,m_k}
\bigg( \prod_{i=1}^k e^{\theta_i m_i} \bigg) P(m_1,\cdots,m_k) ,
\\
K(\theta_1,\cdots,\theta_k) 
&= \ln G(\theta_1,\cdots,\theta_k).
\end{align}
With these generating functions, 
the moments and cumulants are defined similarly.
From these definitions, it is obvious that 
the cumulants for multi-variable distribution functions
are extensive variables in grand canonical ensembles.

\subsection{Some advanced comments}
\label{sec:basic:adv}

\subsubsection{Cumulant expansion}
\label{sec:basic:cumu-exp}

From the definition of cumulants Eq.~(\ref{eq:cumulant}),
the cumulant generating function is expanded as
\begin{align}
K(\theta) = \sum_{n=1}^\infty \frac{\theta^n}{n!} 
\langle m^n \rangle_{\rm c} .
\end{align}
By substituting $\theta=1$ in Eq.~(\ref{eq:K(theta)}), 
one obtains
\begin{align}
K(1) =
\ln \langle e^{m} \rangle = 
\ln \sum_m e^{m} P(m) = 
\sum_{n=1}^\infty \frac {\langle m^n \rangle_{\rm c}}{n!} .
\label{eq:cumulantexpansion}
\end{align}
Equation~(\ref{eq:cumulantexpansion}) is called the cumulant 
expansion, and plays effective roles in obtaining various properties 
of higher order cumulants; examples are found in 
Appendix~\ref{app:superposition} and Ref.~\cite{Kitazawa:2016awu}.

A remark here is that the cumulant expansion 
Eq.~(\ref{eq:cumulantexpansion}) has the same structure 
as the linked cluster theorem in field theory \cite{Negele}.
In fact, if one regards the ``connected part'' of correlation 
functions as the cumulant, the theorem is completely 
equivalent with the cumulant expansion.

\subsubsection{Factorial moments and factorial cumulants}
\label{sec:basic:factorial}

Up to now we have discussed moments and cumulants
as quantities characterizing probability distribution functions.
One of other sets of such quantities is
factorial moments and factorial cumulants.
The factorial moments are defined as
\begin{align}
\langle m^n \rangle_{\rm f} 
= \langle m(m-1) \cdots (m-n+1) \rangle
= \frac{d^n}{ds^n}G_{\rm f}(s)\bigg|_{s=1}
\end{align}
with the factorial moment generating function 
\begin{align}
G_{\rm f}(s) = \sum_m s^m P(m) = G(\ln s).
\end{align}
The factorial cumulants are then defined by the 
factorial cumulant generating function,
\begin{align}
K_{\rm f}(s) = \ln G_{\rm f}(s) = K(\ln s),
\end{align}
as 
\begin{align}
\langle m^n \rangle_{\rm fc} 
= \frac{d^n}{ds^n}K_{\rm f}(s).
\end{align}

In the discussion of physical property of fluctuations, 
the standard moments and cumulants tend to be
more useful than the factorial moments and cumulants.
For example, as we will see in the next section the 
cumulants of conserved charges are directly related to 
the partition function and have more apparent physical meanings
owing to the linear response relation.
For some analytic procedure and theoretical purposes, however, 
the factorial moments and cumulants make 
analyses more concise; see, for example,
Refs.~\cite{Bzdak:2012ab,Kitazawa:2013bta,
Luo:2014rea,Kitazawa:2015ira}.

To relate the moments and cumulants with factorial ones,
one may use relations between the two generating functions,
such as $K(\theta) = K_{\rm f}(e^\theta)$ or 
$K_{\rm f}(s) = K(\ln s)$:
By taking derivatives with respect to $\theta$ or $s$,
their relations are obtained similarly to the procedure 
to obtain the relations between moments and cumulants presented
in Sec.~\ref{sec:basic:def}.

\section{Bulk fluctuations in equilibrium}
\label{sec:equil}

In this section we consider the properties of thermal fluctuations,
i.e. the fluctuations in equilibrated medium.
After describing a general property of fluctuations in quantum 
statistical mechanics, we calculate the fluctuations in ideal gas.
This simple analysis allows us to understand many interesting
properties of thermal fluctuations, such as the magnitude of cumulants
in the hadronic medium \cite{Ejiri:2005wq}.
We also review the linear response relations, which connect
cumulants of conserved charges with the corresponding 
{\it susceptibilities}, i.e. magnitude of the response of 
the system against external force.
The linear response relations for higher order cumulants enable us 
to introduce physical interpretation to the behavior of cumulants 
in the QCD phase diagrams \cite{Asakawa:2009aj}.
Thermal fluctuations in the hadron resonance gas model and 
anomalous behavior of fluctuations expected to take place
near the QCD critical point are also reviewed.
We also give a brief review on the recent progress in
the study of fluctuations in lattice QCD numerical simulations.

The anomalous behaviors of fluctuation observables in 
an equilibrated medium discussed in this section
serves as motivations of experimental analyses of fluctuations.
It, however, should be remembered that 
the fluctuations discussed in this section are not 
directly observed in relativistic heavy ion collisions, 
as will be discussed in detail in Secs.~\ref{sec:e-v-e}, 
\ref{sec:diffusion} and \ref{sec:binomial}.

\subsection{Cumulants in quantum statistical mechanics}
\label{sec:equil:stat}

In this subsection, we first take a look at general properties 
of fluctuations, in particular, the cumulants of conserved charges, 
in quantum statistical mechanics.

\subsubsection{Cumulants of conserved charges}
\label{sec:equil:conserved}

We consider a system described by a Hamiltonian $H$
in a volume $V$ and assume that this system has an observable 
$\hat{N}$ which is a conserved charge.
Because $\hat{N}$ is conserved, $\hat{N}$ commutes with $H$,
\begin{align}
[ H , \hat{N} ] = 0.
\label{eq:[H,N]}
\end{align}
The grand canonical ensemble of this system with 
temperature $T$ and chemical potential $\mu$ for 
$\hat{N}$ is characterized by the density matrix
\begin{align}
\rho = \frac1Z e^{-(H-\mu \hat{N})/T},
\label{eq:rho}
\end{align}
with the grand partition function
\begin{align}
Z = {\rm tr}[ e^{-(H-\mu \hat{N})/T} ] ,
\label{eq:Z}
\end{align}
where the trace is taken over all quantum states.
The expectation value of an observable $\hat{O}$ is given by
\begin{align}
\langle \hat{O} \rangle = {\rm tr}[ \hat{O} \rho ].
\label{eq:<O>}
\end{align}

As in the previous section, one can define the moments 
and cumulants of $\hat{O}$ in quantum statistical mechanics.
The moments $\langle \hat{O}^n\rangle$ are simply given by the 
expectation values of powers of $\hat{O}$.
The cumulants are then defined from the moments and the 
relations such as Eqs.~(\ref{eq:<m1>c=}) - (\ref{eq:<m4>c=}),
which relate moments and cumulants.

We note that the moments and cumulants defined in this way are 
interpreted as those for the distribution of the particle 
number in a volume $V$.
To be more specific, imagine that you were able to count the 
particle number in $V$ in the equilibrated medium exactly at some time.
The obtained particle numbers then 
would fluctuate measurement by measurement around the average.
The moments and cumulants are those of this fluctuation.

The cumulants of the conserved charge $\hat{N}$ in quantum statistical 
mechanics are given by derivatives of the grand potential 
$\Omega = - T \ln Z$ with respect to $\mu/T$.
The first-order cumulant, i.e. the expectation value, 
$\langle N \rangle$ is given by
\begin{align}
\frac{ \partial (-\Omega/T) }{ \partial (\mu/T) }
= \frac1Z {\rm tr} [ \hat{N} e^{-(H-\mu \hat{N})/T} ] 
= \langle N \rangle.
\end{align}
The second derivative gives the second-order cumulant as
\begin{align}
\frac{ \partial^2 (-\Omega/T) }{ \partial (\mu/T)^2 }
&= \frac{ \partial}{ \partial (\mu/T) }
\left ( \frac1Z {\rm tr} [ \hat{N} e^{-(H-\mu \hat{N})/T} ] \right )
= \frac1Z {\rm tr} [ \hat{N}^2 e^{-(H-\mu \hat{N})/T} ] 
- \left( \frac1Z {\rm tr} [ \hat{N} e^{-(H-\mu \hat{N})/T} ] \right)^2 ,
\nonumber \\
&= \langle \delta \hat{N}^2 \rangle
= \langle \hat{N}^2 \rangle_{\rm c}
\end{align}
with $\delta\hat{N} = \hat{N} - \langle \hat{N} \rangle$.
Similar manipulations lead to 
\begin{align}
\langle \hat{N}^n \rangle_{\rm c}
= \frac{ \partial^n (-\Omega/T) }{ \partial (\mu/T)^n }.
\label{eq:<N^n>=Omega}
\end{align}
To show this relation, one may use the fact that 
$Z$ is the moment generating function Eq.~(\ref{eq:G(theta)})
up to normalization constant, 
\begin{align}
\langle N^n \rangle = \frac1Z {\rm tr}[ N^n e^{-(H-\mu \hat{N})/T} ]
= \frac{1}{Z}\frac{\partial^n Z}{\partial (\mu/T)^n} .
\end{align}
The normalization of Eq.~(\ref{eq:G(theta)}) affects the definition of 
the cumulant generating function in Eq.~(\ref{eq:K(theta)})
only by a constant.
Since the constant does not alter derivatives, 
derivatives of the logarithm of $Z$ give the cumulants.

As already discussed in Sec.~\ref{sec:basic:stat}, 
the cumulants of $\hat{N}$ are extensive variables.
This property can easily be shown using the fact that 
the grand potential is an extensive variable, and thus
can be written using the grand potential per unit volume, $\omega$, 
as
\begin{align}
\Omega = \omega V.
\label{eq:omegaOmega}
\end{align}
The cumulants are thus given by 
\begin{align}
\langle \hat{N}^n \rangle_{\rm c}
= \frac{ \partial^n (-\omega/T) V }{ \partial (\mu/T)^n } 
\equiv \chi_n V,
\label{eq:<N^n>=omega}
\end{align}
where on the far right-hand side 
we introduced the cumulant per unit volume,
\begin{align}
\chi_n = \frac{\partial^n (-\omega/T)}{\partial (\mu/T)^n}.
\label{eq:chi_n}
\end{align}
The quantities $\chi_n$ defined here is called susceptibilities,
as the reason will be explained in Sec.~\ref{sec:equil:LRR}.

The extensive property of cumulants Eq.~(\ref{eq:<N^n>=omega})
gives a constraint on the correlation function of particle 
number density.
The particle number $\hat{N}$ in a volume $V$
is related to the particle density $n(\bm{x})$ as 
\begin{align}
\hat{N} = \int_V d\bm{x} n(\bm{x}).
\end{align}
The extensive property Eq.~(\ref{eq:<N^n>=omega})
then implies that 
\begin{align}
\langle n(\bm{x}_1) n(\bm{x}_2) \cdots n(\bm{x}_i) \rangle_{\rm c}
= \chi_i \delta(\bm{x}_1-\bm{x}_2) \delta(\bm{x}_2-\bm{x}_3) 
\cdots \delta(\bm{x}_{i-1}-\bm{x}_i) ,
\label{eq:<nn...n>}
\end{align}
because this is the only choice that is consistent with 
Eq.~(\ref{eq:<N^n>=omega}) for any choice of volume $V$.
Equation~(\ref{eq:<nn...n>}) shows that the particle densities 
at different positions in coordinate space have no correlations.
This result is consistent with Eq.~(\ref{eq:omegaOmega}) 
and the discussion in Sec.~\ref{sec:basic:stat}.
For the validity of Eq.~(\ref{eq:omegaOmega}), 
the volume $V$ has to be large enough.
When the volume is not large enough so that the microscopic 
correlation length is not negligible, 
Eq.~(\ref{eq:omegaOmega}) is no longer valid.
In such a case, Eqs.~(\ref{eq:<N^n>=omega}) or
(\ref{eq:<nn...n>}) is not satisfied, either.

\subsubsection{Cumulants of non-conserved quantities}

Here, it is worth emphasizing that the above discussion
is applicable only for conserved charges, because it 
makes full use of the commutation relation Eq.~(\ref{eq:[H,N]}).
To see this, let us consider the cumulants of a non-conserved
quantity $\hat{N}'$, which does not commute with $H$.
Even for this case,
one can define a {\it would-be} grand partition function,
\begin{align}
Z' = {\rm tr} [ e^{-(H-\mu' \hat{N}')/T} ] ,
\label{eq:Z'}
\end{align}
where $\mu'$ would be interpreted as something like
the chemical potential for $\hat{N}'$.
With this definition, however, Eq.~(\ref{eq:Z'}) is {\it not} 
the moment generating function of $\hat{N}'$.
In fact, derivatives of $Z'$ with respect to $\mu'/T$ are
calculated to be
\begin{align}
\frac{\partial}{\partial(\mu'/T)} Z'
&= {\rm tr} [ \hat{N}' e^{-(H-\mu' \hat{N}')/T} ] = Z' \langle \hat{N}' \rangle,
\label{eq:dZ'}
\\
\frac{\partial^2}{\partial(\mu'/T)^2} Z'
&= \int_0^{1/T} d\tau {\rm tr} [ \hat{N}' e^{-(H-\mu' \hat{N}')\tau} \hat{N}'
e^{-(H-\mu' \hat{N}')(1/T-\tau)} ] \ne Z' \langle \hat{N}'^2 \rangle,
\label{eq:dZ'2}
\end{align}
where in Eqs.~(\ref{eq:dZ'}) and (\ref{eq:dZ'2}) 
we used the cyclic property of trace and a relation 
\begin{align}
\frac{d}{dt} e^{M(t)} = \int_0^1 ds e^{sM} \frac{dM}{dt} e^{(1-s)M}
\end{align}
for a linear operator $M$.
Equation~(\ref{eq:dZ'2}) shows that the second derivative of $Z'$
does not give the second moment.
Similarly, higher derivatives of $Z'$ do not give the moments
$\langle \hat{N}'^n \rangle_{\rm c}$.
As shown in Eq.~(\ref{eq:dZ'}) only the first derivative 
still gives the expectation value $\langle \hat{N}' \rangle$.
Because $Z'$ is no longer proportional to the moment generating 
function, $\Omega' = - T \ln Z'$ is not the cumulant generating
function any more, either.

Because the cumulants of conserved charges are determined from 
the partition function or the grand potential by taking derivatives,
the cumulants of the conserved charge are defined unambiguously
once the grand potential is given in a theory.
This, however, is not true for non-conserved quantities.

When the construction of grand potential is difficult,
one may use an alternative way to define the cumulants.
When the operator $\hat{O}$ for an observable is explicitly known
in a theory, in principle one can calculate the moments of $\hat{O}$ by
calculating the expectation value 
$\langle \hat{O}^n \rangle$.
The cumulants are then also determined using the relations in
Sec.~\ref{sec:basic:def}.
This method is applicable to both conserved and 
non-conserved quantities.
To use this method, however, one has to have 
the explicit form of the operator, as well as their powers,
in the theory.
The conserved charges are related to 
the symmetry of the theory via Noether's theorem 
and their operators can be usually defined as the Noether currents.
For non-conserved quantities, however, 
the corresponding operator is sometimes unclear.
For example, the operator for the ``total pion number''
in QCD is not known.
This problem makes the concept of cumulants of 
non-conserved quantities ambiguous.
In lattice QCD Monte Carlo simulations, for example, 
one can calculate the cumulants of net-electric charge,
which is a conserved charge in QCD, while the cumulants
of total pion number, which is not conserved in QCD,
cannot be determined.
It is also worthwhile to note that the powers of an operator,
$\hat{O}^n$, can be nontrivial in field theory because of 
ultraviolet divergence, even when the 
operator $\hat{O}$ is known.
In this case an appropriate regularization is required 
to define them.
In this sense, the definition of the cumulant using the 
grand potential Eq.~(\ref{eq:<N^n>=Omega}) is convenient 
because it only requires the grand potential.
Analyses of the cumulants of conserved charges in Lattice QCD, 
for example, make use of this definition \cite{Ding:2015ona}.

\subsubsection{Linear response relation}
\label{sec:equil:LRR}

An important property of the cumulants of conserved charges 
in thermal medium is the linear response relation.
From Eq.~(\ref{eq:<N^n>=Omega}) one obtains
\begin{align}
\chi_2 =
\frac{\langle \hat{N}^2 \rangle_{\rm c} }V
= \frac{ \partial }{ \partial (\mu/T) } \frac{\langle \hat{N} \rangle }V .
\label{eq:LRR}
\end{align}
The right-hand side of Eq.~(\ref{eq:LRR}) 
represents the magnitude of the variation of density 
$\langle \hat{N} \rangle/V$ induced by the change of the corresponding 
external force, $\mu$.
In this sense, $\chi_2$ is called susceptibility.
The relation Eq.~(\ref{eq:LRR}) shows that the susceptibility
is {\it equivalent} to the fluctuation of $\hat{N}$ 
per unit volume.
One can generally derive similar relations between 
a susceptibility and fluctuation of conserved charges, 
which are referred to as the linear response relations.

\begin{figure}
\begin{center}
\includegraphics*[width=6cm]{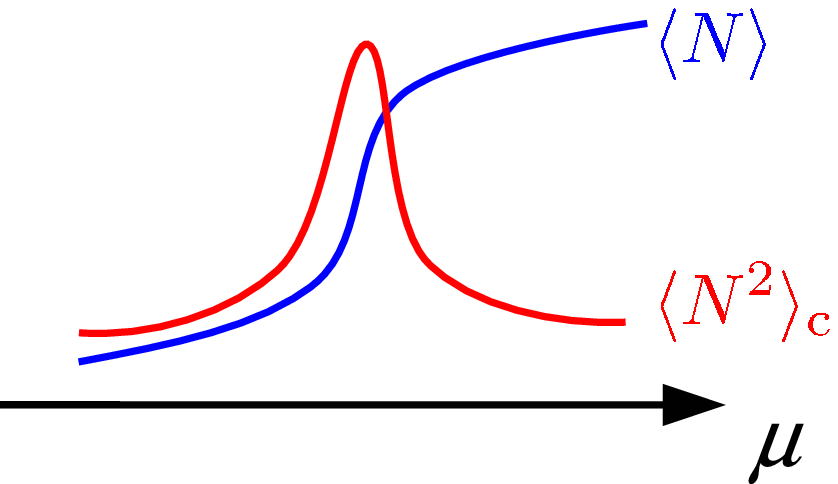}
\hspace{1.5cm}
\includegraphics*[width=6cm]{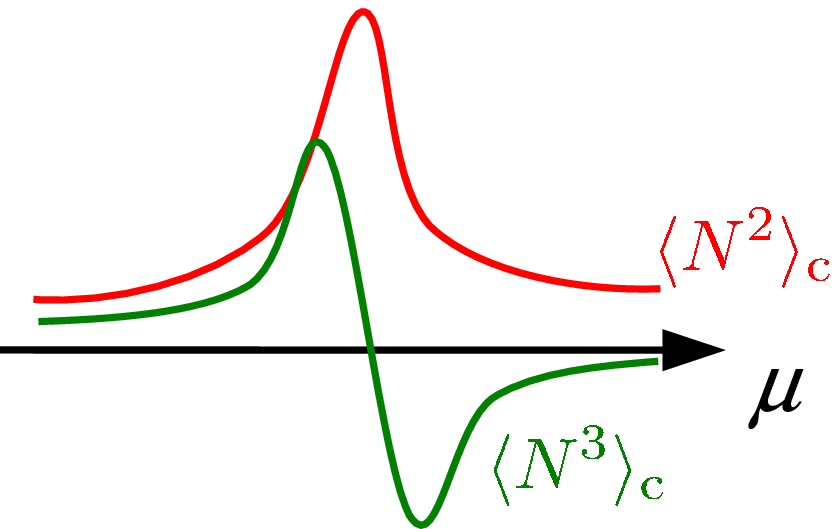}
\caption{
Dependences of various cumulants on $\mu$.}
\label{fig:mu-dep}
\end{center}
\end{figure}

The linear response relation allows us to obtain 
a geometrical interpretation for the behavior of 
the second order cumulant.
Suppose that we know $\langle\hat{N}\rangle$ as a function of 
$\mu$ for a given $T$.
Then, Eq.~(\ref{eq:LRR}) tells us that 
$\langle \hat{N}^2 \rangle_{\rm c}$ is enhanced for $\mu$ at which 
$\langle \hat{N} \rangle$ has a steep rise;
see the left panel of Fig.~\ref{fig:mu-dep}.
On the other hand, if $\langle \hat{N}^2 \rangle_{\rm c}$ has 
a peak structure at some $\mu$, it means that $\langle\hat{N}\rangle$ 
has a steep rise around this $\mu$.
As we will see in Sec.~\ref{sec:equil:CP}, 
this simple argument is quite useful in interpreting the behavior of
fluctuations near the QCD critical point.

The linear response relation can be extended to higher order cumulants.
From Eq.~(\ref{eq:<N^n>=Omega}) one also obtains the relations
for higher orders,
\begin{align}
\chi_{n+1} 
= \frac{ \langle \hat{N}^{n+1} \rangle_{\rm c} }V
= \frac{ \partial }{ \partial (\mu/T) }
\frac{ \langle \hat{N}^n \rangle_{\rm c} }V
= \frac{ \partial \chi_n }{ \partial (\mu/T) }.
\label{eq:LRRn}
\end{align}
This relation shows that the $(n+1)$-th order cumulant plays 
a role of the susceptibility of $n$-th order one.
In this sense it is reasonable to call Eq.~(\ref{eq:chi_n}) as 
(generalized) susceptibility.
Moreover, because the behavior of the $(n+1)$-th order cumulant is 
proportional to the $\mu$ derivative of the $n$-th order one,
similarly to the second-order case this relation introduces 
an geometric interpretation for the behaviors of the higher order cumulants.
For example, as shown in the right panel in Fig.~\ref{fig:mu-dep},
when the second-order cumulant $\langle \hat{N}^2 \rangle_{\rm c}$ 
has a peak structure as a function of $\mu$ the third-order one
$\langle N^3 \rangle_{\rm c}$ changes the sign there \cite{Asakawa:2009aj}.
Note that $\langle \hat{N}^2 \rangle_{\rm c}$ is positive definite 
but the cumulants higher than second-order can take positive and 
negative values.
Because of the positivity of $\langle \hat{N}^2 \rangle_{\rm c}$,
$\langle\hat{N}\rangle$ as a function of $\mu$ 
is concluded to be a monotonically-increasing function, but 
the second and higher order cumulants are not.

We finally note that 
the relation like Eq.~(\ref{eq:LRR}) is not valid 
for non-conserved quantities, 
because $\mu'$ derivatives in Eq.~(\ref{eq:dZ'2}) do not 
give the moment.

\subsubsection{Cumulants of energy}
\label{sec:Equil:E}

Similarly to conserved charges, it is possible to relate
the cumulants of energy, which is also a conserved charge, with 
the partition function $Z$ \cite{Asakawa:2009aj}.
To obtain the relations,
one takes $1/T$ derivative of $Z$ with $\mu/T$ fixed.
This can be done by introducing two independent variables
\begin{align}
\beta = 1/T, ~~ \alpha = \mu/T,
\end{align}
and take $\beta$ derivative of $Z$.
Because $\beta$ derivative is given by 
\begin{align}
\frac{\partial}{\partial \beta} 
= \frac{\partial (1/T)}{\partial \beta} \frac{\partial}{\partial (1/T)}
+ \frac{\partial \mu}{\partial \beta} \frac{\partial}{\partial \mu}
= \frac{\partial}{\partial (1/T)} - T\mu
\frac{\partial}{\partial \mu} ,
\end{align}
the derivative is calculated to be
\begin{align}
\frac{\partial Z}{\partial (-1/T) } \bigg|_{\mu/T}
&= - \frac{\partial}{\partial \beta} Z 
= \left( - \frac{\partial}{\partial (1/T)} + T\mu
\frac{\partial}{\partial \mu} \right) Z 
= {\rm tr}[ H e^{-(H-\mu \hat{N})/T} ] 
\nonumber \\
&= Z \langle H \rangle.
\end{align}
Similarly, one can show 
$\displaystyle {\frac{\partial^n Z}{\partial (-\beta)^n} 
= Z \langle H^n \rangle }$.
The cumulants of energy are then obtained as
\begin{align}
\langle H^n \rangle_{\rm c} 
= \frac{\partial^n \ln Z}{\partial (-1/T)^n} \bigg|_{\mu/T} 
= \left( - \frac{\partial}{\partial (1/T)} + T\mu
\frac{\partial}{\partial \mu} \right)^n \frac{-\Omega}T.
\end{align}
It is also possible to construct the mixed cumulants between 
energy and the conserved charge as 
\begin{align}
\langle \hat{N}^{n_1} H^{n_2} \rangle_{\rm c} 
= \frac{\partial^{n_1}}{\partial (\mu/T)^{n_1}} \langle H^{n_2} \rangle_{\rm c} .
\end{align}

The linear response relation for this case is given by
\begin{align}
\frac{\partial \langle H \rangle }{\partial T} \bigg|_{\mu/T} 
= \frac{\langle H^2 \rangle_{\rm c}}{T^2}.
\label{eq:<H^2>}
\end{align}
The left-hand side of this equation is the increase of the energy
per unit $T$, i.e. the specific heat with fixed $\mu/T$. 
Equation~(\ref{eq:<H^2>}) thus shows that the fluctuation 
of energy divided by the square of the temperature
is equal to specific heat.

\subsection{Ideal gas}
\label{sec:equil:ideal}

Next, let us consider the cumulants of a conserved charge in ideal gas.
We start from ideal gas composed of a single species of particles.
Because the particles do not interact with each other, 
the particle number $N$ is 
automatically a conserved charge.
The grand potential of the ideal gas per unit volume $\omega$ is given by
\cite{Yagi}.
\begin{align}
-\frac{\omega}T = g \int \frac{d^3 p}{(2\pi)^3} 
\ln ( 1 \mp e^{-(E(p)-\mu)/T} )^{\mp 1},
\label{eq:omega-free}
\end{align}
where $g$ represents the degeneracy such as spin degrees of freedom.
The minus and plus signs on the right-hand side corresponds to 
bosons and fermions, respectively.
$E(p)$ is the energy dispersion of the particle;
for non-relativistic and relativistic cases with mass $m$, 
$E(p)=p^2/2m$ and $E(p)=\sqrt{m^2+p^2}$, respectively.

Cumulants of $N$ per unit volume is given by taking $\mu$ 
derivatives of Eq.~(\ref{eq:omega-free}).
For example, the first-order cumulant gives the density $\rho$ as
\begin{align}
\rho = \frac{\langle N \rangle }V
= \frac{\partial (-\omega/T)}{\partial (\mu/T)}
= g \int \frac{d^3 p}{(2\pi)^3} \frac1{e^{(E(p)-\mu)/T} \mp 1},
\label{eq:ideal1}
\end{align}
which is the well-known Bose-Einstein and Fermi-Dirac distribution
functions.
The second-order cumulant is similarly calculated as 
\begin{align}
\chi_2 = \frac{\langle N^2 \rangle_{\rm c} }V
= \frac{\partial^2 (-\omega/T)}{\partial (\mu/T)^2}
= g \int \frac{d^3 p}{(2\pi)^3} 
\frac{e^{(E(p)-\mu)/T}}{(e^{(E(p)-\mu)/T} \mp 1)^2}.
\label{eq:ideal2}
\end{align}

Next, let us consider the dilute limit $\rho/T^3\ll1$.
From Eq.~(\ref{eq:ideal1}), this limit is realized 
when $e^{(E(p)-\mu)/T} \gg 1$, or equivalently 
\begin{align}
E(p)-\mu \gg T,
\label{eq:(E-mu)gg1}
\end{align}
for all $p$.
In this case, the integrand in Eq.~(\ref{eq:omega-free}) 
is approximated as 
$\ln ( 1 \mp e^{-(E(p)-\mu)/T} )^{\mp1} \simeq e^{-(E(p)-\mu)/T}$
and Eq.~(\ref{eq:omega-free}) reduces to the 
grand potential of the free classical (Boltzmann) gas,
\begin{align}
-\frac{\omega}T = g \int \frac{d^3 p}{(2\pi)^3} e^{-(E(p)-\mu)/T}
= g e^{\mu/T} \int \frac{d^3 p}{(2\pi)^3} e^{-E(p)/T} .
\label{eq:omega-Boltzmann}
\end{align}
For the relativistic case with $E(p)=\sqrt{m^2+p^2}$, 
Eq.~(\ref{eq:(E-mu)gg1}) is satisfied for 
$m-\mu \gg T$.
In this case, the integral in Eq.~(\ref{eq:omega-Boltzmann}) is rewritten 
using the modified Bessel function of the second kind $K_n(x)$
\cite{Kapusta} as
\begin{align}
\omega = -g \frac{m^2 T^2}{2\pi^2} e^{\mu/T} K_2\left ( \frac{m}T \right ) .
\end{align}

With Eq.~(\ref{eq:omega-Boltzmann}), 
higher order susceptibilities are easily calculated to be
\begin{align}
\frac{\langle N^n \rangle_{\rm c}}V
= \frac{\partial^n (-\omega/T)}{\partial (\mu/T)^n}
= -\frac{\omega}T, 
\label{eq:omega-Boltzmann2}
\end{align}
for all $n\ge1$.
This result shows that all cumulants are identical.
From this result and the discussion in Sec.~\ref{sec:basic:Poisson},
the distribution of the particle number
for the free Boltzmann gas is given by the Poisson distribution.
An intuitive explanation of this result is given 
in Sec.~\ref{sec:basic:Poisson}.

In relativistic quantum field theory, all charged particles are
always accompanied by their antiparticles carrying the opposite charge.
For a particle with chemical potential $\mu$, 
the chemical potential of its antiparticle is $-\mu$.
The grand potential is thus given by 
\begin{align}
-\frac{\omega}T = g \int \frac{d^3 p}{(2\pi)^3} 
\left[
\ln ( 1 \mp e^{-(E(p)-\mu)/T} )^{\mp 1} + 
\ln ( 1 \mp e^{-(E(p)+\mu)/T} )^{\mp 1} \right].
\label{eq:omega-free-rela}
\end{align}
For massless particles with $E(p)=p$, the integral in 
Eq.~(\ref{eq:omega-free-rela}) can be carried out analytically.
For fermions, the result is 
\begin{align}
\omega = -g \left( 
\frac{7\pi^2}{360} T^4 + \frac1{12} T^2\mu^2 + \frac1{24\pi^2}\mu^4
\right).
\label{eq:omega-massless}
\end{align}

\subsubsection{Particles with non-unit charge}

Next, we consider an ideal gas in which the charge carried by 
the particles is not unity in some unit.
Such a case is realized, for example, when two particles each
of which carry a unit charge form a molecule.
If all particles are confined into molecules and the residual interaction
between the molecules is weak enough, the system can be regarded as 
free gas of doubly charged particles.

Let us consider free gas of particles with charge $r$ 
that is a rational number.
Because of the definition of chemical potential,
the chemical potential of the particle is $r\mu$.
If the system is dilute enough so that it can be regarded as 
free Boltzmann gas, the grand potential is given by 
\begin{align}
-\frac{\omega}T = g \int \frac{d^3 p}{(2\pi)^3} e^{-(E(p)-r\mu)/T}
= g e^{r\mu/T} \int \frac{d^3 p}{(2\pi)^3} e^{-E(p)/T}.
\label{eq:omega-r}
\end{align}
The expectation value of the total charge $Q$ is obtained 
by $\mu/T$ derivative of Eq.~(\ref{eq:omega-r}) as
\begin{align}
\frac{\langle Q \rangle}V
= \frac{\partial (-\omega/T)}{\partial (\mu/T)}
= r g e^{r\mu/T} \int \frac{d^3 p}{(2\pi)^3} e^{-E(p)/T}
= r \rho
\label{eq:<Q>r}
\end{align}
with $\rho=\langle N \rangle/V$ being the particle density.
Similarly, the cumulants of the total charge $Q$ 
are obtained by taking $\mu/T$ derivatives of 
Eq.~(\ref{eq:omega-r}) as 
\begin{align}
\frac{\langle Q^n \rangle_{\rm c}}V
= \frac{\partial^n (-\omega/T)}{\partial (\mu/T)^n}
= r^n \rho.
\label{eq:<Q^n>}
\end{align}
We thus obtain the relation between cumulants
\begin{align}
\frac{\langle Q^{n_1} \rangle_{\rm c}}{\langle Q^{n_2} \rangle_{\rm c}}
= r^{n_1-n_2}.
\end{align}

This result shows that the magnitude of cumulants
is drastically changed when the charge carried by the effective
degrees of freedom in the system changes.
This property of cumulants leads to possibility to
diagnose the quasi-particle property of the system with cumulants.

In relativistic heavy ion collisions, 
the use of this property of cumulants as diagnostic tools for the 
deconfinement transition, at which quarks carrying fractional
charges are liberated, was first suggested in 
Refs.~\cite{Asakawa:2000wh,Jeon:2000wg} for the second order cumulant.
The idea is then extended to higher order ones in Ref.~\cite{Ejiri:2005wq}.
We will discuss these studies in more detail later.

\subsubsection{Mixture of differently charged particles and net particle number}
\label{sec:equil:mult}

Next, let us consider the ideal classical gas composed of 
several particle species with different charges.
To be specific, we consider a system composed of 
particles with charges $r_1$, $r_2,~ \cdots,~r_n$.
Using the chemical potential $\mu$ of the charge,
the grand potential per unit volume is given by
\begin{align}
-\frac{\omega}T = \sum_i g_i \int \frac{d^3 p}{(2\pi)^3} e^{-(E(p)-r_i\mu)/T}. 
\label{eq:omega-r2}
\end{align}
By taking $\mu/T$ derivatives of Eq.~(\ref{eq:omega-r2}),
cumulants of the charge, $Q$, are obtained as 
\begin{align}
\langle Q^n \rangle_{\rm c}
= \sum_i r_i^n \rho_i , 
\label{eq:<Q^n>-12}
\end{align}
with $\rho_i=\langle N_i \rangle/V$ being the density of 
particles labeled by $i$.
Note that uncharged particles
do not contribute to $\langle Q^n \rangle_{\rm c}$.

In QCD, conserved charges are given by 
the net-particle number, i.e. the difference between
particle and antiparticle numbers.
From Eq.~(\ref{eq:<Q^n>-12}), the cumulants of the net-particle 
number $Q$ are given by 
\begin{align}
\langle Q^n \rangle_{\rm c}
= \langle N \rangle + (-1)^n \langle \bar{N} \rangle ,
\label{eq:<Q^n>-net}
\end{align}
where $\langle N \rangle$ and $\langle \bar{N} \rangle$ are 
the numbers of particles and antiparticles, respectively.
This result shows that the distribution of the net-particle number 
is given by the Skellam one.

\subsubsection{Shot noise}
\label{sec:equil:shot}

At this point, it is worthwhile to comment on 
an example of fluctuations in completely different 
physical systems.
We consider the shot noise, i.e. 
the fluctuation of electric currents in electric circuits 
with a potential barrier.
It is known that the magnitude of the shot noise is proportional 
to the charge of the elementary excitation of the system.
As we show in this subsection, this property comes from 
the Poissonian nature of the electric current, and 
shares the same mathematics as we have discussed above.

The shot noise is the fluctuation of the electric current 
at a potential barrier that 
an electron can pass through with a small probability per unit time.
(In the original study of Schottky, fluctuation in a 
vacuum tube was investigated \cite{Schottky}.)
In typical experiments, correlation between electrons is well 
suppressed so that the probability of electrons to pass through 
the potential barrier can be regarded as independent with one another.
Then, the number of electrons which pass through 
the barrier in some time interval is given by the Poisson distribution
to a good approximation.
The time evolution of the number of electrons which go 
through the barrier is well 
described by the Poisson {\it process} \cite{Gardiner}.
In typical experiments, the magnitude of the fluctuation of 
electric current is measured from the power spectrum in 
frequency space assuming white noise.
Here, however, for simplicity we focus on the 
amount of the charge in a time interval and do not consider
time evolution.

Because the number of electrons $N$ which pass through the 
potential barrier in a time interval $\Delta t$ is given by 
the Poisson distribution, the cumulants of $N$ satisfy
$\langle N^n \rangle_{\rm c} = \langle N \rangle$ for $n\ge1$.
The amount of charge $Q$ which passes through the barrier 
in $\Delta t$ is related to $N$ as 
$Q = e N$ with the charge of the electron, $e$.
The second-order cumulant of $Q$ is then given by
\begin{align}
\frac{ \langle Q^2 \rangle_{\rm c} }{ \langle Q \rangle }
= \frac{ e^2 \langle N^2 \rangle_{\rm c} }{ e \langle N \rangle }
= e ,
\label{eq:shotnoise}
\end{align}
where in the second equality we have used 
the Poissonian nature of $N$.
Equation~(\ref{eq:shotnoise}) suggests that the ratio 
$\langle Q^2 \rangle_{\rm c} / \langle Q \rangle$ can be used
to measure the charge of the electron \cite{Schottky}.

In superconducting materials, electrons
are ``confined'' into Cooper pairs, which are doubly charged,
when the material undergoes the phase transition to superconductor.
The charge $e$ in Eq.~(\ref{eq:shotnoise}) then should be replaced 
with the one of the Cooper pairs, $2e$, 
and the ratio $\langle Q^2 \rangle_{\rm c} / \langle Q \rangle$ 
should become twice larger than that in the normal material.
In fact, such a behavior is observed experimentally \cite{Jehl}.
The shot noise is also successfully applied to investigate 
the fractional quantum Hall effect, in which the shot noise behaves
as if the elementary charge became fractional \cite{FQHE}.
While the shot noise is usually measured up to the second-order, 
in some systems higher order cumulants are observed
\cite{Gustavssona}.

Finally, we remark that the physics considered here 
is completely different from thermal fluctuations;
the former is the fluctuations associated with nonequilibrium 
diffusion processes,
while the latter is the fluctuations in equilibrated media.
Nevertheless, there is a common feature, proportionality between
the fluctuation and the elementary charge, owing to
the common underlying mathematics.

\subsection{Fluctuations in QCD}
\label{sec:equil:QCD}

Now we turn to the thermal fluctuations in QCD.
After a review on fluctuations in the hadron resonance gas 
model, which well describes thermodynamics of QCD at 
low temperature and density, we consider the 
behaviors of fluctuations in the deconfined medium and 
near the QCD critical point.
Recently, the cumulants of conserved charges have been 
actively studied in lattice QCD numerical simulations 
\cite{Ejiri:2005wq,Gavai:2010zn,Schmidt:2010xm,
Mukherjee:2011td,Borsanyi:2011sw,Nagata:2012pc,Bazavov:2012jq,
Bazavov:2012vg,Bazavov:2013dta,Nakamura:2013ska,Borsanyi:2013hza,
Bellwied:2013cta,Bazavov:2013uja,Borsanyi:2014ewa,
Bazavov:2014yba,Bazavov:2014xya,Gupta:2014qka,Nakamura:2015jra,
Bazavov:2015zja};
see for reviews Refs.~\cite{Ding:2015ona,Borsanyi:2015axp}.
The established knowledge in this field is also summarized 
in the following.

\subsubsection{Hadron resonance gas (HRG) model}
\label{sec:equil:HRG}

We first consider the medium at low temperature and density.
It is well known that the medium at low $T$
and chemical potentials is well described by hadronic
degrees of freedom.
When the condition Eq.~(\ref{eq:(E-mu)gg1}) is well 
satisfied for all hadrons and the
interactions between hadrons can be neglected,
the contributions of individual hadrons on thermodynamics
can be regarded as those in free Boltzmann gas.
In this case, the grand potential is given by
\begin{align}
-\frac{\omega_{\rm HRG}}T 
= \sum_i g_i \int \frac{d^3 p}{(2\pi)^3} 
e^{-(E(p)-q_{\rm B}^{(i)}\mu_{\rm B}-q_{\rm Q}^{(i)} \mu_{\rm Q}-q_{\rm S}^{(i)} \mu_{\rm S})/T},
\label{eq:omega-HRG}
\end{align}
where $i$ runs over all species of hadrons.
Here, we have introduced three chemical potentials, 
$\mu_{\rm B}$, $\mu_{\rm Q}$ and $\mu_{\rm S}$, which are baryon,
electric charge and strange chemical potentials, respectively.
$g_i$ is the degeneracy of the hadron labeled by $i$, and 
$q_c^{(i)}$ with $c={\rm B}$, ${\rm Q}$ and ${\rm S}$ represents 
the baryon, electric charge and strange numbers carried by the hadron $i$,
respectively.
The grand potential Eq.~(\ref{eq:omega-HRG})
containing all known hadrons \cite{PDG} 
as free particles is called the hadron resonance gas (HRG) model 
\cite{BraunMunzinger:2003zd}.
From the chemical freezeout temperature and chemical potential
determined in relativistic heavy ion collisions \cite{Cleymans:1998fq},
it is known that the condition Eq.~(\ref{eq:(E-mu)gg1}) is well
satisfied for the hot medium after chemical freezeout 
for all hadrons except for pions with the mass $m_\pi\simeq140$~MeV.
It is known that the HRG model well reproduces the 
thermodynamic quantities calculated on the lattice QCD 
below the pseudo-critical temperature $T\lesssim T_c \simeq 150$ MeV 
for vanishing chemical potentials \cite{Ding:2015ona}.
Although the HRG model contains only free particles, 
there is an argument that by incorporating resonance states
the effects of interaction between hadrons is effectively taken
into account \cite{Landau1,Landau2}.

Now let us calculate the cumulants of conserved charges,
net-baryon, net-electric charge and net-strange numbers
in the HRG model based on Eq.~(\ref{eq:omega-HRG}).
We first consider the net-baryon number cumulants, which are 
obtained by taking $\mu_{\rm B}$ derivatives of Eq.~(\ref{eq:omega-HRG}).
The baryon number carried by baryons and anti-baryons are
$q_{\rm B}^{(i)}=+1$ and $q_{\rm B}^{(i)}=-1$, respectively,
while the mesonic degrees of freedom do not carry the
baryon number and
do not contribute to the fluctuation of net-baryon number 
in the HRG model.
The net-baryon number cumulants thus are calculated to be
\begin{align}
\langle N_{\rm B,net}^n \rangle_{\rm c}
&= \frac{\partial^n (-\omega_{\rm HRG}/T )}{\partial (\mu_{\rm B}/T)^n}
= - \frac{\omega_{\rm B}}T - (-1)^n \frac{\omega_{\bar{\rm B}}}T 
\nonumber \\
&= \langle N_{\rm B} \rangle + (-1)^n \langle N_{\rm \bar{B}} \rangle ,
\label{eq:<N_Bnet^n>c}
\end{align}
where $\omega_{\rm B}$ and $\omega_{\bar{\rm B}}$ denote the 
contributions of all baryons and anti-baryons to the grand potential
$\omega_{\rm HRG}$, respectively, and $N_{\rm B}$ and $N_{\rm \bar{B}}$
are the baryon and anti-baryon numbers, respectively.
The result Eq.~(\ref{eq:<N_Bnet^n>c}) shows that the fluctuation of 
the net-baryon number in the HRG model is given by the Skellam 
distribution and the ratios of cumulants between even or odd orders 
are unity \cite{Ejiri:2005wq},
\begin{align}
\frac{\langle N_{\rm B,net}^{n+2m} \rangle_{\rm c}}
{ \langle N_{\rm B,net}^n \rangle_{\rm c}} =1 ,
\label{eq:<NB>/<NB>=1}
\end{align} 
for integer $n$ and $m$.

Next, we consider the net-electric charge.
Contrary to the net-baryon number, there are hadronic resonances 
having $q_{\rm Q}^{(i)}=\pm2$, such as $\Delta^{++}(1232)$, 
in addition to $q_{\rm Q}^{(i)}=0$ and $\pm1$ states.
Due to these resonances, the fluctuation of net-electric 
charge is not given by the Skellam distribution.
The cumulants of net-electric charge is given by 
\begin{align}
\langle N_{\rm Q,net}^n \rangle_{\rm c}
= \langle N_{q_{\rm Q}=1} \rangle 
+ (-1)^n \langle N_{q_{\rm Q}=-1} \rangle
+ 2^{n-1} \left\{ \langle N_{q_{\rm Q}=2} \rangle 
+ (-1)^n \langle N_{q_{\rm Q}=-2} \rangle \right\},
\end{align}
with $\langle N_{q_{\rm Q}=m} \rangle$ being the density of
hadrons having the electric charge $q_{\rm Q}=m$.
Owing to the $q_{\rm Q}=\pm2$ states, higher order cumulants
tend to become larger than the Skellam one.
The same argument applies to the net-strange number, while the
contribution of $q_{\rm S}=\pm2$ states is a little more suppressed 
because of their heavy masses; the lightest of such baryons is
$\Xi^0$ and $\bar{\Xi}^0$ with the mass 1315 MeV.
For net-electric charge fluctuations, 
Bose-Einstein statistics of pions also gives rise to deviations
from the Skellam distribution, while the effect of 
Fermi-Dirac statistics of baryons on Eq.~(\ref{eq:<NB>/<NB>=1})
is well suppressed in the hadronic medium relevant to 
relativistic heavy ion collisions.

Next we consider the mixed cumulants in the HRG model.
As an example, the correlation of net-baryon and net-strange
numbers is given by
\begin{align}
\langle N_{\rm B,net} N_{\rm S,net} \rangle_{\rm c}
= \frac{\partial^2 (\omega_{\rm HRG}/T )}
{\partial (\mu_{\rm B}/T)\partial (\mu_{\rm S}/T)}
= \sum_i q_{\rm B}^{(i)} q_{\rm S}^{(i)} \langle N_i \rangle .
\label{eq:<NBNS>}
\end{align}
Equation~(\ref{eq:<NBNS>}) shows that the contributions 
to $\langle N_{\rm B,net} N_{\rm S,net} \rangle_{\rm c}$
come from hadrons with $q_{\rm B}^{(i)}\ne0$ and $q_{\rm S}^{(i)}\ne0$.
The lightest hadron having nonzero $q_{\rm B}$ and $q_{\rm S}$
is $\Lambda(1115)$ and $\bar{\Lambda}(1115)$ with the mass $m_\Lambda=1115$ MeV.
Because the mass of the nucleon and the mass of the lightest
strange meson, $K^\pm$, are smaller than $m_\Lambda$,
the density of $\Lambda$ is suppressed compared with 
hadrons having only either $q_{\rm B}^{(i)}\ne0$ or $q_{\rm S}^{(i)}\ne0$.
Therefore, the magnitude of 
$\langle N_{\rm B,net} N_{\rm S,net} \rangle_{\rm c}$
should be suppressed 
\begin{align}
\langle N_{\rm B,net} N_{\rm S,net} \rangle_{\rm c}
\ll \langle N_{\rm B,net}\rangle \langle N_{\rm S,net} \rangle_{\rm c} ,
\end{align}
in the HRG model with small $\mu_{\rm B}$ and $\mu_{\rm S}$
\cite{Koch:2005vg}.
Various mixed cumulants in the HRG model can be understood 
in a similar manner.

If the fluctuation observables in relativistic heavy ion collisions 
are well described by the hadronic degrees of freedom in equilibrium, 
the cumulants of conserved charges should be consistent with 
those in the HRG model.
Conversely, if the fluctuations show deviation from those in the HRG
model, they serve as experimental signals of non-hadronic and/or
non-thermal physics.
In this sense, the cumulants in the HRG model are usually 
compared with the experimental results as the ``baseline''
\cite{Karsch:2010ck}.
In particular, because the ratios between cumulants take a simple 
form in the HRG model, they are useful observables to investigate
these nontrivial physics.
In Fig.~\ref{fig:STARfluc}, the ratio of net-proton
number cumulants, $\kappa\sigma^2=\langle N_{\rm p,net}^4\rangle_{\rm c} /
\langle N_{\rm p,net}^2\rangle_{\rm c}$ and $S\sigma/({\rm Skellam})
=\langle N_{\rm p,net}^3\rangle_{\rm c} / \langle N_{\rm p,net}\rangle$
are plotted\footnote{In QCD, net-proton number is not a
  conserved charge, and the definition of its cumulants is ambiguous.
  In the HRG model, on the other hand, the net-proton number is a 
  conserved charge because protons and anti-protons in the HRG model 
  are free particles, and its cumulants are well defined. 
  For the same reason as for net-baryon number, this model gives
  $\langle N_{\rm p,net}^4\rangle_{\rm c} / \langle N_{\rm p,net}^2\rangle_{\rm c} =
  \langle N_{\rm p,net}^3\rangle_{\rm c} / \langle N_{\rm p,net}\rangle = 1$.
  However, they, of course, are not the net-baryon number in QCD
  \cite{Kitazawa:2011wh,Kitazawa:2012at}.
  Problems with the use of net-proton number as a proxy of net-baryon number
  will be discussed in Secs.~\ref{sec:e-v-e} and \ref{sec:binomial}.
}.
The figure shows that the ratios of the cumulants show
statistically-significant deviations from the HRG baseline.
This experimental result clearly shows that the ratio carries 
information on physics which cannot be described by the HRG model, 
such as the onset of the deconfined phase, existence of QCD 
critical point, or some non-equilibrium phenomena.

\subsubsection{Onset of deconfinement transition}
\label{sec:equil:deco}

Next, we focus on the deconfined medium.
At extremely high temperature, the quarks can be regarded as 
approximately free particles owing to asymptotic freedom.
The cumulants of conserved charges thus are given by 
those of ideal quark gas in this case.
If Fermi-Dirac statistics were negligible there,
the cumulants of net-quark number would become the Skellam ones,
\begin{align}
\langle N_{\rm q,net}^n \rangle_{\rm c}
= \langle N_{\rm q} \rangle + (-1)^n \langle N_{\rm \bar{q}} \rangle ,
\label{eq:Nq}
\end{align}
with $N_{\rm q}$ ($N_{\rm \bar{q}}$) being the quark (anti-quark) number.
By recalling that all quarks and anti-quarks carry $\pm1/3$ baryon 
number,
Eq.~(\ref{eq:Nq}) is converted to net-baryon number cumulant as 
\begin{align}
\langle N_{\rm B,net}^n \rangle_{\rm c}
= \frac1{3^n} \left[ \langle N_{\rm q} \rangle 
+ (-1)^n \langle N_{\rm \bar{q}} \rangle \right],
\label{eq:NBNq}
\end{align}
where we have used $N_{\rm q,net}=3N_{\rm B,net}$.
From this result, the ratio of the net-baryon number is given by 
\begin{align}
\frac{ \langle N_{\rm B,net}^{n+2} \rangle_{\rm c} }
{ \langle N_{\rm B,net}^n \rangle_{\rm c} } = \frac19,
\label{eq:1/9}
\end{align}
i.e. compared with Eq.~(\ref{eq:<NB>/<NB>=1}) the ratio 
is about one order suppressed when the deconfined medium
is realized \cite{Ejiri:2005wq}.
This change of the ratio is reminiscent of the shot 
noise in the fractional quantum Hall effect discussed in 
Sec.~\ref{sec:equil:shot}.

The above argument, however, is modified by 
the Fermi-Dirac statistics of quarks, because the
masses of quark quasi-particles $m_{\rm q}$ would not satisfy
$m_{\rm q}/T \gg 1$ in the deconfined phase.
For massless quarks, the net-quark number 
cumulants can be calculated with Eq.~(\ref{eq:omega-massless}).
By taking $\mu/T$ derivatives of Eq.~(\ref{eq:omega-massless}),
we have
\begin{align}
\frac{\langle N_{\rm q,net}^3 \rangle_{\rm c}}
{\langle N_{\rm q,net} \rangle_{\rm c}} 
= \frac6{\pi^2} \left( 1 + \frac1{\pi^2} \frac{\mu^2}{T^2} \right)^{-1},
\quad
\frac{\langle N_{\rm q,net}^4 \rangle_{\rm c}}
{\langle N_{\rm q,net}^2 \rangle_{\rm c}} 
= \frac6{\pi^2} \left( 1 + \frac3{\pi^2} \frac{\mu^2}{T^2} \right)^{-1},
\end{align}
while all cumulants $\langle N_{\rm q,net}^n \rangle_{\rm c}$
for $n\ge5$ vanish.
This result shows that for massless quarks the ratios of 
cumulants are suppressed compared with the Skellam value, 
Eq.~(\ref{eq:1/9}).
The inclusion of Fermi-Dirac statistics thus does not alter the
suppression of higher order cumulants in the deconfined medium.

The electric charge $q_f$ carried by quarks are 
$\pm1/3$ and $\pm2/3$ depending on the flavor $f$.
Assuming the Boltzmann statistics we have 
\begin{align}
\langle N_{\rm Q,net}^n \rangle_{\rm c}
= \sum_f q_f^2 \big( \langle N_f \rangle 
+ (-1)^n \langle N_{\bar{f}} \rangle \big) .
\end{align}
Because of quarks with $q_f=\pm2/3$, especially
up quarks, the suppression of higher order cumulants compared
with the HRG values is much milder compared with the net-baryon 
number in Eq.~(\ref{eq:NBNq}).
The baryon number cumulants thus are better observable 
to see the deconfinement phase transition.

\begin{figure}
\begin{center}
\includegraphics*[width=8cm]{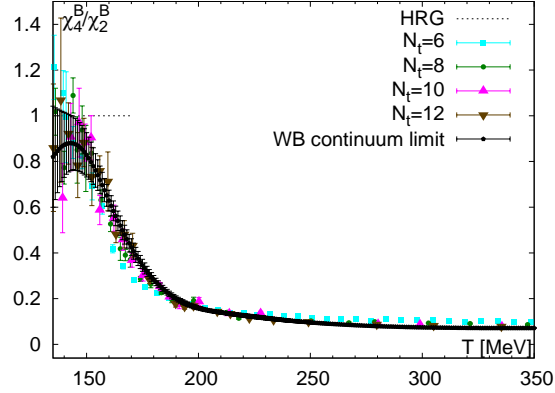}
\caption{
Ratio of the fourth- and second-order baryon number 
susceptibilities $\chi_4^{\rm B}/\chi_2^{\rm B}=
\langle N_{\rm B,net}^4 \rangle_{\rm c}/\langle N_{\rm B,net}^2 \rangle_{\rm c}$ 
calculated on the lattice \cite{Borsanyi:2013hza}.
Points labeled by ``WB continuum limit'' are the value of 
$\chi_4^{\rm B}/\chi_2^{\rm B}$ after taking the continuum
(small lattice spacing) limit.
}
\label{fig:c4/c2}
\end{center}
\end{figure}

The ratios of cumulants have been actively analyzed in lattice QCD
numerical simulations.
In Fig.~\ref{fig:c4/c2} we show an example of the recent analysis 
on the ratio of net-baryon number cumulants 
$\langle N_{\rm B,net}^4 \rangle_{\rm c}/\langle N_{\rm B,net}^2 \rangle_{\rm c}$
\cite{Borsanyi:2013hza}.
The figure shows that the ratio is consistent with the 
HRG value Eq.~(\ref{eq:<NB>/<NB>=1}) below the pseudo-critical 
temperature $T_c\simeq150-160$~MeV, while the ratio suddenly 
drops above $T_c$ and approaches the free Fermi gas value at high $T$.

\begin{figure}
\begin{center}
\includegraphics*[width=8cm]{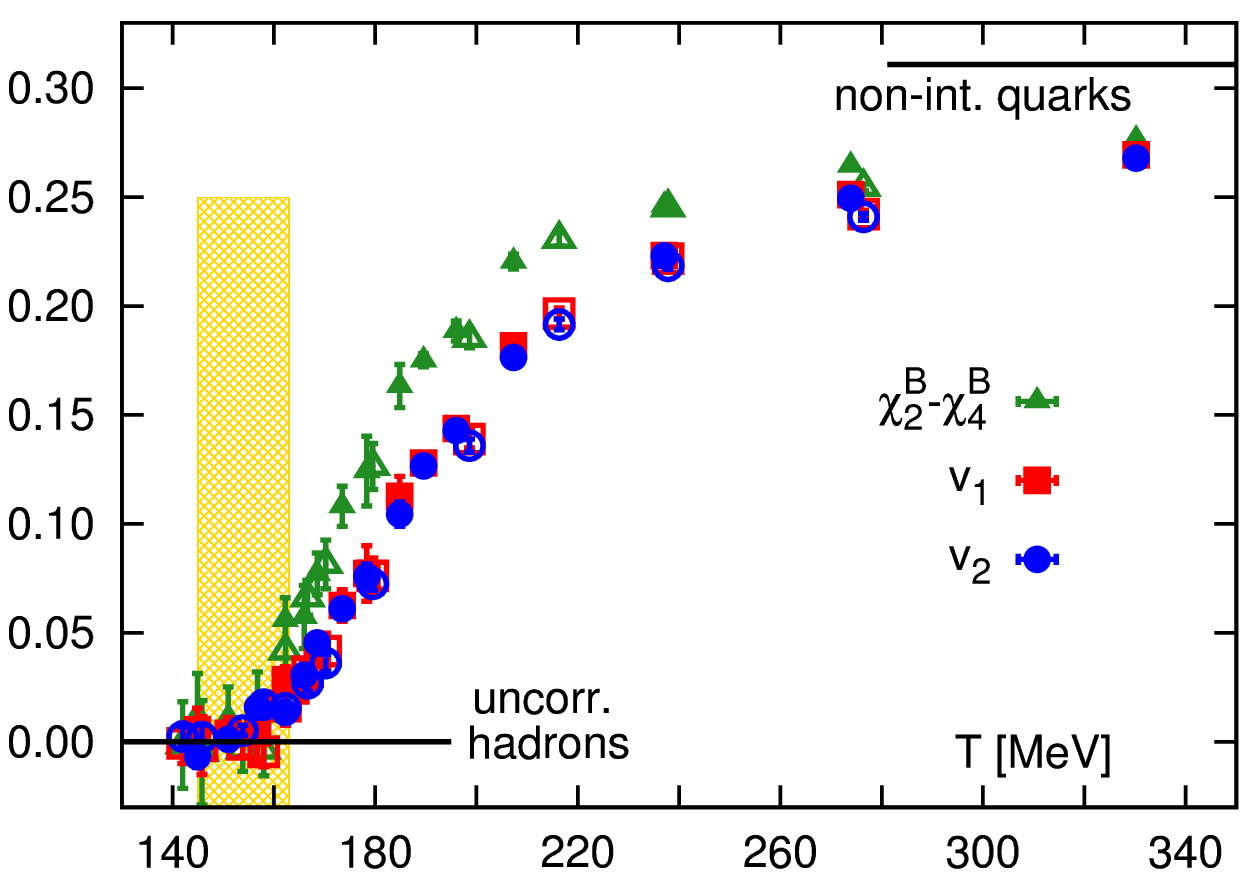}
\caption{
Combination of the net-baryon and net-strange number cumulants,
$\langle N_{\rm B,net}^4 \rangle_{\rm c}-\langle N_{\rm B,net}^2 \rangle_{\rm c}$, 
$v_1$ and $v_2$ which are chosen so that they vanish in the HRG model
\cite{Bazavov:2013dta}.
The horizontal line labeled ``uncorr. hadrons'' indicates 
the values in the HRG model.
}
\label{fig:v1v2}
\end{center}
\end{figure}

Recently, more sophisticated ways to investigate the medium 
properties near $T_c$ using the higher order and mixed cumulants 
have been studied in the lattice community.
In Ref.~\cite{Bazavov:2013dta}, various (mixed) cumulants are 
investigated in combinations which vanish in the HRG model.
For example, from Eq.~(\ref{eq:<NB>/<NB>=1}) one finds that 
$\langle N_{\rm B,net}^4 \rangle_{\rm c}
-\langle N_{\rm B,net}^2 \rangle_{\rm c}=0$ in the HRG model.
Similar combinations are found also by considering mixed cumulants
such as Eq.~(\ref{eq:<NBNS>}).
In Fig.~\ref{fig:v1v2}, three combinations of 
net-baryon and net-strange number cumulants 
which vanish in the HRG model obtained on the lattice are plotted 
\cite{Bazavov:2013dta}.
The figure shows that these combinations are indeed consistent
with zero for $T\lesssim T_c$, but suddenly becomes nonzero 
above $T_c$.
This result suggests that the medium is well described by 
the hadronic degrees of freedom up to near $T_c$, but the 
non-hadronic physics shows up around $T_c$.
Similar ideas are applied to investigate the 
flavor hierarchy in the breakdown of the HRG model
\cite{Bazavov:2013dta,Bellwied:2013cta,Bazavov:2014yba}.
These studies suggest that fluctuations are
useful observables to understand quasi-particle properties
in the medium although they are static quantities.

Finally, we note that the cumulants at extremely high $T$ and 
$\mu_{\rm B}$ can be analyzed perturbatively 
\cite{Mogliacci:2013mca,Haque:2013qta,Haque:2013sja}.
The comparisons between these perturbative analyses with 
lattice results have been done \cite{Bazavov:2013dta}.

\subsubsection{QCD critical point}
\label{sec:equil:CP}

\begin{figure}
\begin{center}
\includegraphics*[width=8cm]{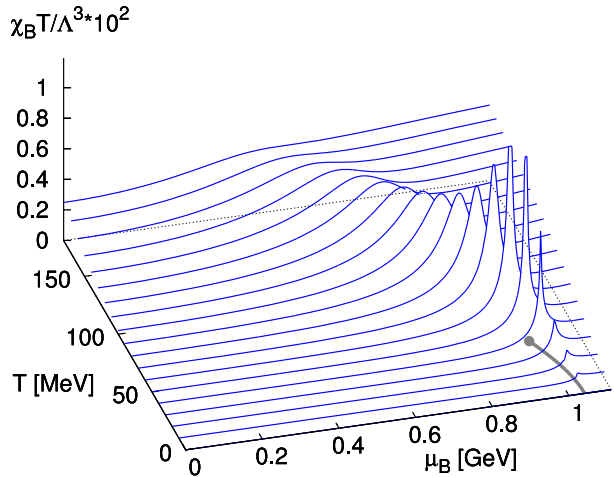}
\caption{
Plot of the baryon number susceptibility 
$\chi_{\rm B}=\langle N_{\rm B,net}^2 \rangle/V$
as a function of $T$ and $\mu_{\rm B}$ obtained in a 
chiral effective model \cite{Asakawa:2009aj}.}
\label{fig:chi3d}
\end{center}
\end{figure}

Around the boundary between the confined and deconfined phases,
anomalous behaviors of fluctuation observables associated with 
the phase transition are expected to occur.
In particular, the QCD phase diagram in the $T$--$\mu_{\rm B}$ plane 
is expected to have the QCD critical point(s), which is the endpoint 
of the first-order phase transition line 
\cite{Asakawa:1989bq,Kitazawa:2002bc,Stephanov:2007fk}.
(An example of the phase diagram is shown in the bottom surface
of Fig.~\ref{fig:chi3d}.)
Because the phase transition at this point is 
of second-order, various fluctuation observables 
diverge there.
It is known that this point belongs the same universality class
as in the 3d $Z_2$ Ising model, which is the same universality class as 
the critical point in the phase diagram of water belongs to,
and the critical exponents are determined 
by the universality arguments\footnote{
It is known \cite{Son:2004iv,Minami:2011un} that the dynamical 
universality class of this point belongs to that of the model H in the 
classification of Hohenberg and Halperin \cite{Hohenberg:1977ym}.
}.
The anomalous behaviors associated with the critical point 
should become experimental observables to find this point 
\cite{Stephanov:1998dy,Stephanov:1999zu,Hatta:2003wn}.

An order parameter to characterize the QCD critical point
is the chiral condensate $\sigma=\langle \bar\psi\psi \rangle$.
Associated with the softening of the effective potential, the 
correlation length and fluctuation of the $\sigma$ field diverge
at the critical point.
Owing to this divergence, fluctuations of fields which couple to 
$\sigma$ also diverges.
Because the net-baryon number has a coupling with the
$\sigma$ field \cite{Hatta:2002sj,Fujii:2003bz,Son:2004iv,Fujii:2004jt}
with nonzero current quark mass and $\mu_{\rm B}$, 
the net-baryon number fluctuation diverges at the critical point.
From studies on the baryon number susceptibility 
\cite{Kunihiro:1991qu} based on chiral effective models 
\cite{Hatta:2002sj,Sasaki:2006ws,Sasaki:2006ww,Fukushima:2008wg,
Schaefer:2009ui,Fu:2009wy,
Asakawa:2009aj,Skokov:2010uh,Ichihara:2015kba},
it is known that the baryon number susceptibility has a 
ridge structure along the crossover line \cite{Hatta:2002sj} 
as illustrated in Fig.~\ref{fig:chi3d}.

Near the QCD critical point, higher order cumulants of 
conserved charges also behave anomalously.
In Ref.~\cite{Stephanov:2008qz}, the higher order cumulants
are calculated as a function of the correlation length of 
$\sigma$ field up to the fourth-order.
It is pointed out that the higher order cumulants are 
more sensitive to the correlation length.
Another interesting property of higher order cumulants is 
that they change the sign near the critical point 
\cite{Asakawa:2009aj,Friman:2011pf,Stephanov:2011pb}.
As discussed in Sec.~\ref{sec:equil:LRR}, the net-baryon number 
cumulants are given by the $\mu_{\rm B}$ derivative of the cumulant 
lower by one order as 
\begin{align}
\langle N_{\rm B,net}^{n+1} \rangle_{\rm c}
= \frac{ \partial \langle N_{\rm B,net}^n \rangle_{\rm c} }
{\partial(\mu_{\rm B}/T)} .
\end{align}
From this relation and the ridge structure of the baryon number
susceptibility $\chi_2^{\rm B} 
= \langle N_{\rm B,net}^2 \rangle_{\rm c}/V$
along the phase boundary as shown in 
Fig.~\ref{fig:chi3d}, it is immediately concluded that 
$\chi_3^{\rm B} = \langle N_{\rm B,net}^3 \rangle_{\rm c}/V$
changes the sign at the phase boundary near the QCD
critical point \cite{Asakawa:2009aj}.
The same conclusion is also obtained for 
mixed cumulants between net-baryon and energy $E$ 
such as $\langle N_{\rm B,net}^2 E \rangle_{\rm c}/V$ 
\cite{Asakawa:2009aj}.
By taking one more $\mu_{\rm B}$ derivative, 
one can also conclude similarly that 
$\chi_4^{\rm B} = \langle N_{\rm B,net}^4 \rangle_{\rm c}/V$
becomes negative along the phase boundary near the QCD critical point.
This behavior can also be confirmed by the universality
argument of the 3d $Z_2$ universality class \cite{Stephanov:2011pb}.

For $\mu_{\rm B}=0$, from the mapping of the scaling parameters in
the QCD phase diagram, similar argument is applicable to 
$T$ derivatives of even order cumulants, which leads to 
$\chi_{2(n+1)}^{\rm B} \sim \partial \chi_{2n}^{\rm B} / \partial T$
\cite{Skokov:2010uh}. 
From this relation it is pointed out that $\chi_6^{\rm B}$
becomes negative near the phase boundary for small $\mu_{\rm B}$
\cite{Friman:2011pf}.

Besides the baryon number cumulants,
those of electric charge and strangeness also 
behave anomalously near the critical point.
The anomalous behaviors in these quantities, however, are weaker 
than those of the baryon number cumulants.
The electric charge susceptibility 
$\chi_2^{\rm Q}=\langle N_{\rm Q,net}^2 \rangle_{\rm c}/V$, 
for example, is obtained by $\mu_{\rm Q}$
derivatives of $\omega$.
This derivative is rewritten in terms of the baryon and 
isospin chemical potentials, $\mu_{\rm B}$ and $\mu_{\rm I}$, as
$\partial/\partial \mu_{\rm Q} 
= ( \partial/\partial \mu_{\rm B} + \partial/\partial \mu_{\rm I} )/2$
\cite{Asakawa:2009aj}.
The electric susceptibility thus is given by 
\begin{align}
\chi_2^{\rm Q} 
= \frac{\partial^2 (-\omega/T)}{\partial (\mu_{\rm Q}/T)^2 }
= \frac14 \left( \frac{\partial}{\partial (\mu_{\rm B}/T) } 
+ \frac{\partial}{\partial (\mu_{\rm I}/T) } \right)^2 \frac{-\omega}T
= \frac14 ( \chi_2^{\rm B} + \chi_2^{\rm I} ),
\label{eq:chi_2^Q}
\end{align}
where $\chi_2^{\rm I}= (\partial^2 (-\omega/T))/(\partial (\mu_{\rm I}/T)^2)$
is the isospin susceptibility.
In the last equality in Eq.~(\ref{eq:chi_2^Q}) 
we have neglected the cross term
$(\partial^2 \omega)/(\partial \mu_{\rm B} \partial \mu_{\rm I})$
which vanishes in the isospin symmetric medium.
The isospin susceptibility does not have
a divergence at the critical point in an isospin symmetric medium
\cite{Hatta:2003wn}.
Equation~(\ref{eq:chi_2^Q}) shows that the effect of 
the divergence in $\chi_2^{\rm B}$ is relatively suppressed 
in $\chi_2^{\rm Q}$.

In order to compare the cumulants of conserved charges obtained 
in effective models or lattice QCD numerical simulations 
with experimental data, the ratios of cumulants are 
calculated on the chemical freezeout line \cite{Cleymans:1998fq}
in the literature \cite{Gupta:2011wh,Athanasiou:2010kw,
BraunMunzinger:2011dn,BraunMunzinger:2011ta,Morita:2012kt,
Morita:2014fda,Fukushima:2014lfa,Bluhm:2014wha,Albright:2015uua}.
Recently, the use of fluctuation observables for the determination
of the freezeout temperature (of fluctuations) is also 
proposed \cite{Karsch:2012wm,Ding:2015ona,Bazavov:2015zja}.
These analyses would be used as a qualitative guide to 
understand the experimental results on the ratios of
cumulants.
As will be discussed in the next sections, however, the 
event-by-event analyses in relativistic heavy ion collisions 
measure the fluctuations in the final state, which are 
not the thermal fluctuation at some early stage in the time 
evolution of the hot medium.
Because of this difference, their direct comparison with theories
may lead to wrong conclusions; 
while the event-by-event fluctuations would carry information 
on early thermodynamics, this information must be extracted 
by correcting various contributions 
associated with the experiment.
This is the subject which will be addressed in the following
sections.
We also note that the theoretical studies on the 
net-baryon number cumulants are sometimes compared with the 
experimental results on net-proton number cumulants, which are 
not the same quantities as the former; this problem will be discussed 
in Sec.~\ref{sec:binomial}.

\section{Event-by-event fluctuations}
\label{sec:e-v-e}

In Sec.~\ref{sec:equil}, we have seen that the cumulants 
of conserved charges in an equilibrated medium behave 
characteristically reflecting the onset of deconfinement
or the existence of the critical point.
The goal of the measurement of fluctuations in relativistic
heavy ion collisions is to find these behaviors in fluctuation
observables.
In these experiments, fluctuations are observed by
event-by-event analyses.
In this method, the numbers of particles of identified species 
are observed in some coverage of a detector in each event.
The distribution of the numbers for individual events is 
called event-by-event fluctuation.
It is believed that fluctuations observed in this way 
carry information on the thermal fluctuation in early
stage in the time evolution of the hot medium.

In this and succeeding two sections, we discuss
event-by-event analysis.
In particular, we take a closer look at the relation between 
event-by-event and thermal fluctuations.
In this section, we first give an
overview of event-by-event analysis, 
and discuss why, when and how event-by-event fluctuations 
can be compared with thermal ones\footnote{
The experimental setting of event-by-event analysis 
and associated problems is summarized nicely 
in a review Ref.~\cite{Koch:2008ia}.}.
Two specific problems are discussed in later sections separately.
In Sec.~\ref{sec:diffusion}, we address the effect of 
the diffusion of fluctuations in later stages.
It will be discussed that the rapidity window dependences
of fluctuation observables can be used to understand this effect.
We then consider the problem of efficiency correction in 
event-by-event analysis in Sec.~\ref{sec:binomial}. 
The difference between net-baryon and net-proton number 
cumulants will also be discussed there.

\subsection{Event-by-event analysis}
\label{sec:e-v-e:eve}

\begin{figure}
\begin{center}
\includegraphics*[width=8cm]{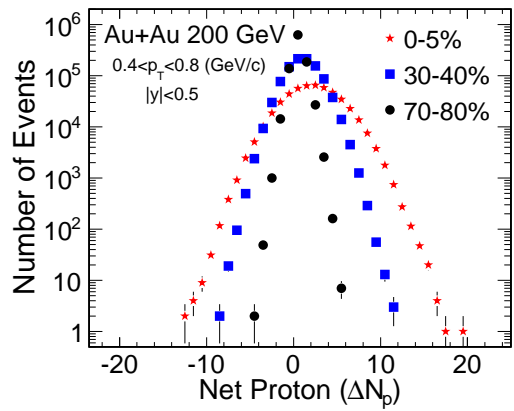}
\caption{
Event-by-event histogram of net-proton number cumulants
measured by STAR collaboration at RHIC \cite{STAR-fluc}.
}
\label{fig:hist}
\end{center}
\end{figure}

In event-by-event analysis of fluctuation observables 
in relativistic heavy ion collisions,
some observables, such as the numbers of specific particles,
are observed in some coverage of a detector in each event.
The numbers then take different values for each event.
As an example, in Fig.~\ref{fig:hist} we show the event-by-event 
histogram of net-proton number observed by STAR collaboration
\cite{STAR-fluc}.
Regarding the histogram as the probability distribution function,
one can construct the cumulants of the particle number 
from this event-by-event distribution\footnote{
In actual experimental analyses, the procedure to obtain the 
cumulants is more complicated. 
For details, see Refs.~\cite{Luo:2011tp,Luo:2014rea}.
}.

\subsubsection{Conserved charges in heavy ion collisions}

Among various fluctuation observables, 
those of conserved charges are believed to have suitable 
properties to investigate early thermodynamics.
First, as we have discussed in Sec.~\ref{sec:equil:stat}, 
the cumulants of conserved charges in thermal medium 
are well defined and can be obtained unambiguously 
in a given theory.
Moreover, through the linear response relations 
the cumulants of conserved charges are directly related 
to the property of the medium.
Second, as we will discuss in Sec.~\ref{sec:diffusion} in detail, 
the time evolution of fluctuations is typically slow for conserved
charges.
The slow variation enables us to investigate the 
medium property in the early stage from event-by-event
analysis, although the measurement is performed for
the final state \cite{Asakawa:2000wh,Jeon:2000wg}.

In QCD, the net-flavor numbers are conserved charges
besides energy, momentum and angular momentum.
In heavy ion collisions, the numbers
of net-baryon, net-electric charge and net-strangeness,
which are given by the linear combinations of net-flavor numbers,
are frequently used instead of the net-flavor numbers.
Among these three charges, the net-electric charge is 
most directly observable in heavy ion collisions, because the 
detectors for heavy ion collisions can observe almost 
all charged particles entering the detector with particle identification.
Figure~\ref{fig:ALICEfluc} is an example of the experimental 
result of net-electric charge fluctuation \cite{Adamczyk:2013dal}.
Higher order cumulants of net-electric charge have been also 
measured by STAR collaboration recently \cite{Adamczyk:2014fia}.

The measurement of net-baryon number is more difficult,
because the typical detectors cannot 
identify neutral baryons, in particular neutrons which account 
for about half of the total baryons.
Because of this problem, net-proton number cumulants are 
measured in experiments and used as a proxy of net-baryon number;
an example is shown in Fig.~\ref{fig:STARfluc} \cite{Adamczyk:2013dal}.
It, however, should be remembered that the net-proton number
is not a conserved charge and different from net-baryon number
\cite{Kitazawa:2011wh,Kitazawa:2012at}.
In fact, it is shown in Refs.~\cite{Kitazawa:2011wh,Kitazawa:2012at}
that the net-proton and net-baryon number cumulants can take 
significantly different values and thus the substitution of 
the former for the latter cannot be justified.
In this study it is also shown that the net-baryon number cumulants 
can be determined experimentally without measuring of neutrons.
We will discuss these issues in Sec.~\ref{sec:binomial}.

The experimental measurement of net-strange number has a 
difficulty at a more fundamental level.
The strange charges in heavy ion collisions are dominantly
carried by kaons.
Among them, charged kaons $K^+$ and $K^-$ carrying 
strange numbers $s=\pm1$, respectively, can be measured by detectors.
The strange number is also carried by the neutral kaons, $K^0$ 
and $\bar{K}^0$ having strange numbers $s=\pm1$, respectively.
The decays of these neutral kaons undergo the weak interaction,  
in which $K^0_{\rm L}$ and $K^0_{\rm S}$ are eigenstates
but not eigenstates of strange number.
Because of this mixing, the net-strangeness
carried by $K^0_{\rm L}$ and $K^0_{\rm S}$ can not be observable,
even if the detectors measure these particles.
Fluctuations of net kaon number, the difference of
the numbers of $K^-$ and $K^+$, 
are often considered experimentally as a proxy of net strangeness.
One, however, should keep in mind that the substitution of the 
net-kaon number for net-strangeness contains the same problem 
as that of net-proton number for net-baryon number.

\subsubsection{Coverage to count the particle number}
\label{sec:e-v-e:rapidity}

In event-by-event analysis, the particle number arriving
at some range of the detector is counted in each event.
In order to compare event-by-event fluctuations with 
thermal ones, it is desirable to choose this range so that 
it corresponds to a spatial volume of the hot medium.
The detectors in heavy ion collisions, however, 
can only measure the momentum of particles in the final state.

The momentum range in the final state can be related to a spatial 
volume of the hot primordial medium in coordinate space in some special
cases such as the Bjorken scaling flow \cite{Yagi}.
In the Bjorken picture, which is well justified for large 
$\sqrt{s_{\rm NN}}$, the hot medium has boost invariance along 
the longitudinal direction.
To describe such systems, it is convenient to introduce 
the rapidity $y$ and the coordinate-space rapidity $Y$,
\begin{align}
y = \frac12 \ln \frac{1+\beta}{1-\beta} = \tanh^{-1} \beta,
\quad
Y = \frac12 \ln \frac{t+z}{t-z},
\end{align}
respectively.
Here, $\beta$ is the velocity along the longitudinal direction; 
for particles it is given by $\beta=p_z/E$ with the longitudinal 
momentum $p_z$ and energy $E$.
$t$ and $z$ represent the time and longitudinal coordinate, respectively.
Rapidities $y$ and $Y$ obey an additional law under the Lorentz boost;
with the boost of a system with velocity $\beta_0$, 
rapidities are transformed as 
\begin{align}
y' = y + \beta_0, \quad Y'= Y + \beta_0.
\end{align}

Because of the boost invariance, 
in the Bjorken picture the rapidity $y$ of a 
{\it fluid element} is equal to its coordinate-space rapidity $Y$.
Therefore, up to the thermal motion of individual particles
the rapidity of particles at $y$ is identical to
the coordinate-space one, $Y$.
Using this correspondence, one can count the particle number
in an (approximate) coordinate space interval from the 
momentum distribution in the final state.
In typical event-by-event analyses, 
the range to count the particle number is chosen to a (pseudo-)rapidity 
interval $\Delta y$, while the transverse momentum $p_T$ and the 
azimuthal angle are integrated out.
This range then can approximately be identified with the spatial 
volume in the corresponding
coordinate-space rapidity $\Delta Y = \Delta y$

\begin{figure}
\begin{center}
\includegraphics*[width=8cm]{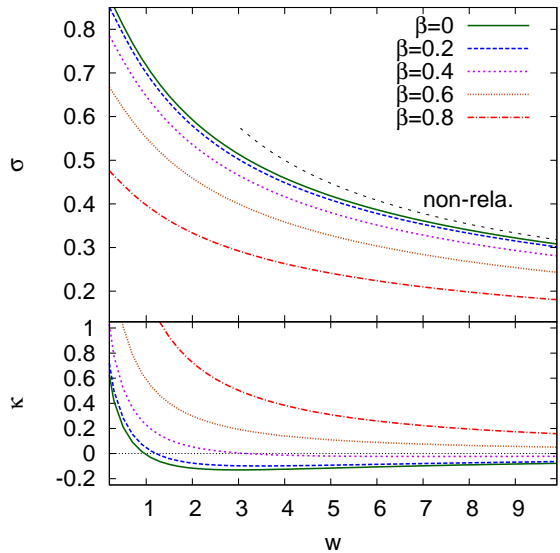}
\caption{
Width $\sigma$ and kurtosis $\kappa$ of the thermal distribution 
in rapidity space calculated in a blast wave model with 
several values of radial velocity $\beta$ \cite{Asakawa:QM15,Ohnishi}.
The horizontal axis shows $w=m/T_{\rm kin}$ with the mass of particle $m$ and 
the freezeout temperature $T_{\rm kin}$.}
\label{fig:ywidth}
\end{center}
\end{figure}

It, however, should be remembered that the correspondence between 
$y$ and $Y$ is valid only for the fluid element, and  it does not
hold for the rapidities of individual particles in the volume element,
because individual particles have nonzero velocity against the fluid 
element due to the thermal motion.
Because of the thermal motion, the correspondence between
the particle distribution in $Y$ space and in $y$ space is blurred.
The use of $y$ as a proxy of $Y$ is valid only up to the 
resolution brought by this thermal motion.
The rapidity window $\Delta y$ to measure event-by-event particle
distribution has to be taken sufficiently large 
compared with the resolution width brought by
the thermal motion in rapidity space 
$\Delta y_{\rm thermal}$ in order to suppress the effect of the
blurring; otherwise, the information of fluctuations
in a volume in $\Delta Y$ is lost.
In the upper panel of Fig.~\ref{fig:ywidth}, we plot the width 
of the thermal motion $\sigma=\Delta y_{\rm thermal}$ 
calculated in a simple blast wave model \cite{Asakawa:QM15,Ohnishi}.
The parameter $w=m/T_{\rm kin}$ is the ratio between the particle mass $m$ and 
the kinetic-freezeout temperature $T_{\rm kin}$,
while $\beta$ is the radial velocity.
For central collisions at the top-RHIC and LHC energies, the blast wave
fit to the transverse momentum spectra gives
$\beta\simeq0.6$ and $T_{\rm kin}\simeq100$ MeV \cite{Abelev:2013vea}.
Assuming that particles are emitted from the medium at the 
kinetic-freezeout surface with the thermal distribution and 
using these values of $\beta$ and $T_{\rm kin}$,
one obtains that the thermal width is 
$\Delta y_{\rm thermal}\simeq0.5$ for 
pions ($w=m/T_{\rm kin}\simeq1.5$) and $\Delta y_{\rm thermal}\simeq0.25$ 
for nucleons ($w\simeq9$); see Fig.~\ref{fig:ywidth}.
On the other hand, 
the maximum rapidity window of the STAR detector is $\Delta y=1.0$.
For the analysis of net-electric charge fluctuation, this width
is comparable with $\Delta y_{\rm thermal}$, and thus 
may not be large enough to suppress the effect of 
thermal motion \cite{Asakawa:QM15,Ohnishi}.
More quantitative estimate of this effect on cumulants 
\cite{Asakawa:QM15,Ohnishi,Kitazawa:2015ira} will be discussed
in Sec.~\ref{sec:diffusion}.
In the lower panel of Fig.~\ref{fig:ywidth}, the kurtosis of the 
thermal distribution in rapidity space is also plotted. 
The panel suggests that the non-Gaussianity of the distribution 
is not large for the kinetic freezeout parameters except for 
the case of pions. 
In fact, it is shown from an explicit calculation of the cumulants
that the effects of the non-Gaussianity are well suppressed 
in the cumulants \cite{Asakawa:QM15,Ohnishi}.

We emphasize that the above argument on the approximate corresponsence
between $y$ and $Y$ is based on 
the Bjorken picture for medium expansion. 
When the Bjorken picture breaks down for low energy collisions,
further subtle discussion is needed for the relation
between $y$ and $Y$, and the interpretation of event-by-event 
fluctuations in a rapidity interval $\Delta y$.
This would particularly be the case for the BES energy, 
$\sqrt{s_{\rm NN}} \simeq 7.7 - 20$~GeV, at RHIC.
In the interpretation of fluctuation observables in this energy
range (for example, Fig.~\ref{fig:STARfluc}), therefore, 
careful arguments on these issues, which should be carried 
out with the aid of dynamical models and information of various
observables, is required.

The other remark concerning the choice of $\Delta y$ is that 
the observed event-by-event fluctuations are those in the final state 
in heavy ion collisions, 
and not those in some early stage, such as at chemical freezeout time, 
in the time evolution.
Even if clear signals in fluctuation observables are 
well developed reflecting thermodynamics in some early time, 
they are modified during the time evolution in later stages
before the detection.
Here, it is worth emphasizing that event-by-event fluctuations 
continue to change until {\it kinetic} freezeout.
This is contrasted with {\it average} particle abundances, which are 
almost frozen at {\it chemical} freezeout time.
The modification of event-by-event fluctuations after
chemical freezeout is easily understood from the fact that the particle 
number in a coordinate-space rapidity interval $\Delta Y$ 
in a collision event is modified due to the 
motion of individual particles.
When one compares event-by-event fluctuations with thermal 
fluctuations in early stage, this effect has to be also taken 
into account.
In Sec.~\ref{sec:diffusion}, we argue that the effects of 
the time evolution after chemical freezeout and the thermal width 
$\Delta y_{\rm thermal}$ are simultaneously described 
as a diffusion process.
It will also be discussed that the magnitude of these effects can 
be examined experimentally from the $\Delta y$ dependences 
of the cumulants.

Finally, we note that in general the coverages in the $p_T$ and
azimuthal direction of detectors are not perfect,
although the coverages are desirable to be taken as large 
as possible to measure the event-by-event fluctuations in 
a volume in coordinate-space.
The imperfect coverages also modify fluctuation observables.
In fact, in a recent analysis of net-proton number cumulants by STAR 
Collaboration the dependence on the $p_T$ coverage
is reported \cite{Luo:2015ewa}; see the right panel in 
Fig.~\ref{fig:STARfluc}.
It should also be remembered that the azimuthal coverage 
of the PHENIX detector is significantly limited 
compared with the STAR detector.
The effect of finite acceptance is in part treated as an 
efficiency effect as discussed in Sec.~\ref{sec:binomial}.
The effects of the $p_T$ cut on event-by-event analysis 
are recently investigated, for example, in 
Refs.~\cite{Alba:2014eba,Morita:2014nra,Karsch:2015zna}.

\subsubsection{Canceling spatial volume dependences}

The cumulants of thermal fluctuations are proportional to 
the spatial volume of the system as discussed in Sec.~\ref{sec:equil:stat}.
In order to give a physical meaning to the magnitude of 
cumulants observed by the event-by-event analyses, therefore,
one has to know the spatial volume of the hot medium besides
the value of the cumulants.
Moreover, because the hot medium created in relativistic heavy 
ion collisions is an expanding system, the spatial volume is 
changing with time evolution.

In order to eliminate the spatial volume dependence, 
in relativistic heavy ion collisions cumulants are 
usually discussed in terms of the ratios between extensive 
observables.
Here, the extensive observables have to be taken as 
conserved quantities or their cumulants.
By taking the ratio of these quantities observed in a
rapidity window $\Delta y$, 
the effect of the spatial volume and its time evolution 
can be canceled out under the Bjorken picture as follows.
As discussed previously, the rapidity window $\Delta y$
approximately corresponds to the one of 
coordinate-space rapidity $\Delta Y = \Delta y$.
Although the spatial volume of the medium in $\Delta Y$ 
changes during the time evolution in a nontrivial way,
the amount of a conserved quantity in $\Delta Y$ is conserved
if we neglect the effect of diffusion, i.e. exchange 
of the quantity with adjacent volume elements
\cite{Asakawa:2000wh,Jeon:2000wg}.
The ratio of conserved quantities in $\Delta Y$, 
therefore, does not depend on the time evolution of spatial volume 
when the effect of diffusion is negligible.

Before the measurement of higher order cumulants has started,
the magnitude of the second-order cumulant was discussed in terms of 
the ratio with entropy density \cite{Asakawa:2000wh,Jeon:2000wg},
which is a conserved quantity in non-dissipative hydrodynamic expansion.
The use of a quantity called the $D$-measure,
\begin{align}
D = 4 \frac{ \langle N_{\rm Q}^2 \rangle_{\rm c}}
{ \langle N_{\rm tot} \rangle},
\label{eq:Dmeasure}
\end{align}
as an experimental observable with the net-electric charge $N_{\rm Q}$ 
and the number of total charge $N_{\rm tot}$,
i.e. the sum of the positively and negatively charged particle numbers,
was suggested
in Ref.~\cite{Jeon:2000wg}.
Here, $N_{\rm tot}$ is used as a proxy of the entropy density.
It is estimated that the D-measure takes $D=3\sim4$ in the 
HRG model while the value of $D$ is about twice or more smaller if 
the deconfined medium is created \cite{Jeon:2000wg}.
Figure~\ref{fig:ALICEfluc} shows the D-measure measured 
by ALICE collaboration \cite{ALICE}.
The experimental result shows the suppression of 
this quantity compared with the hadronic value especially
for large $\Delta\eta$.
This result suggests the survival of the thermal fluctuation 
created in the deconfined medium.
The origin of the $\Delta\eta$ dependence in the figure 
will be discussed in Sec.~\ref{sec:diffusion}.
Later, the use of the ratio between cumulants of conserved 
charges was proposed in Ref.~\cite{Ejiri:2005wq}.
This choice removes the ambiguity in the relation between 
the entropy and $N_{\rm tot}$ in the definition of the $D$-measure.
The top and bottom panels in the right figure of 
Fig.~\ref{fig:STARfluc}, for example, show the ratio of 
net-proton number cumulants,
\begin{align}
\kappa \sigma^2 = 
\frac{ \langle N_p^4 \rangle_{\rm c} }{ \langle N_p^2 \rangle_{\rm c} },
\quad
\frac{S\sigma}{\rm Skellam} =
\frac{ \langle N_p^3 \rangle_{\rm c} }{ \langle N_p \rangle_{\rm c} },
\end{align}
respectively, where the skewness $S$ and kurtosis $\kappa$ are 
defined in Sec.~\ref{sec:basic:skewkurt}.
As discussed in Sec.~\ref{sec:equil}, these ratios become
exactly unity in the HRG model.
These ratios thus are suitable in exploring the existence of 
physics which cannot be described by the HRG model.
The experimental results in Fig.~\ref{fig:STARfluc} show that
these ratios show statistically-significant deviation
from unity.
This result thus cannot be described solely by the thermal 
fluctuation in the hadronic medium.
Non-hadronic or non-thermal physics come into play in these 
observables, which should be understood in future studies.

\subsection{Global charge conservation}

Above, we discussed that larger rapidity window $\Delta y$
has to be taken to suppress the burring effects due to 
the thermal motion of individual particles.
When $\Delta y$ becomes larger, however, another problem due to 
the finiteness of the system, or global charge conservation,
shows up.

The hot medium created by relativistic heavy ion collisions
is a finite-size system.
If one measures a conserved charge in 
the total system, it is always fixed to the sum of 
those in the colliding nuclei and does not fluctuate.
This fact is called the global charge conservation.
Conserved charges can have event-by-event fluctuations 
when measurement is performed for subsystems.
Even if measurement is performed for a subsystem, 
when the subsystem is not small enough compared 
with the total system the global charge conservation affects
event-by-event fluctuations.

The effect of the global charge conservation on fluctuation 
observables in an equilibrated medium is investigated 
in Refs.~\cite{Jeon:2000wg,Bleicher:2000ek,Begun:2004gs,
Begun:2006uu,Bzdak:2012an}.
These analyses suggest that the effect of global charge 
conservation tends to reduce the magnitude of fluctuation 
observables.
In Ref.~\cite{Sakaida:2014pya}, the effect of global 
charge conservation is studied by incorporating 
non-equilibrium effects in the time evolution 
of the hadronic medium.
In this study it is argued that the effect of the global charge 
conservation tends to be suppressed owing to non-equilibrium 
effects, which stem from the finiteness of the diffusion speed,
compared with those in Refs.~\cite{Jeon:2000wg,Bleicher:2000ek}.
In particular, it is pointed out that the experimental result
on net-electric charge fluctuation at ALICE \cite{ALICE} 
is not affected by this effect \cite{Sakaida:2014pya}.

As the collision energy $\sqrt{s_{_{\rm NN}}}$ is lowered,
the length of the hot medium along the rapidity 
direction becomes smaller, and the effect of the global
charge conservation becomes more prominent with fixed 
rapidity window $\Delta y$.
It is not a priori clear whether there exists a range of $\Delta y$
in which both the effects of thermal blurring and global charge
conservation are well suppressed in low energy collisions as realized
in the BES program, or not.
The breakdown of the Bjorken picture would also modify the 
justification of the use of $\Delta y$ in place of $\Delta Y$.
An answer to this question would be obtained by experimental 
study of the $\Delta y$ dependences of various cumulants
\cite{Kitazawa:2013bta,Sakaida:2014pya,Kitazawa:2015ira} and
by theoretical study of the transportation of conserved
charges without assuming the Bjorken picture.

\subsection{Non-thermal event-by-event fluctuations}

Collision events in heavy ion collisions have 
various event-by-event fluctuations besides the thermal ones
discussed so far.
For example, the energy per unit rapidity and shape of the initial 
state just after the collision are fluctuating 
\cite{Cao:2015cba,Niemi:2015qia}, even with 
fixed collision energy and after a centrality selection.
These fluctuations give rise to additional event-by-event 
fluctuations besides the thermal ones.
Production of jets in high energy collisions will be another 
source of event-by-event fluctuations.
When experimentally observed event-by-event fluctuations are 
directly compared with thermal ones, it is implicitly assumed 
that these non-thermal fluctuations are well suppressed 
compared with thermal fluctuations.
This assumption, however, has to be examined carefully.
When the contribution of non-thermal fluctuations is not small,
they have to be eliminated in order to isolate
the thermal fluctuations.

An example of non-thermal event-by-event fluctuations 
is that of the energy density in the initial state.
Even with fixed impact parameter, the energy density of the 
produced medium after the collision is fluctuating event by event.
Under the Bjorken picture, the event-by-event fluctuation of 
energy density per unit rapidity is translated to that of
spatial volume with a fixed temperature.
Because cumulants of thermal fluctuations are proportional to 
the spatial volume, the energy fluctuation in the initial state 
directly affects the magnitude of event-by-event fluctuations 
\cite{Skokov:2012ds,Alba:2015iva}.
Similarly, the finite size of the centrality bins also gives
rise to event-by-event fluctuations of the volume.
In experimental analyses, in order to remove this fluctuation 
as much as possible analyses of cumulants are performed 
with narrow centrality bins, and then summed up 
\cite{Luo:2011tp}.
On the theoretical side, the effect of the volume fluctuation 
on cumulants can be estimated by superposing the events 
with different spatial volumes.
The event-by-event cumulants are then represented by the cumulants 
of thermal and volume fluctuations; see Appendix~\ref{app:superposition}.
Another possibility to take account of the volume fluctuation
is to employ so-called strongly-intensive quantities 
\cite{Gorenstein:2011vq,Begun:2012rf,Sangaline:2015bma},
which are combinations of cumulants in which the spatial volume
fluctuations are canceled out.
Although strongly-intensive quantities are usually defined
for non-conserved quantities \cite{Gorenstein:2011vq,Begun:2012rf}, 
if they were defined solely with conserved quantities 
their physical meaning would become more apparent.

In high energy collisions, production of jets would also 
disturb fluctuation observables in a rapidity window $\Delta y$.
The total conserved charges carried by jets are typically small,
because the charges carried by primary partons (gluon or quark)
except energy, momentum, and angular momentum, are
negligibly small.
As a first approximation, therefore, the effects of jets on the 
conserved-charge fluctuations would be well suppressed.
They, however, give rise to fluctuation of energy density 
along the rapidity direction, and would modify the cumulants 
in a rapidity window.
To the best of the authors' knowledge, these effects 
have not been discussed in the literature.

A qualitative observation on these non-thermal event-by-event 
fluctuations is that they tend to enhance the magnitude of 
fluctuations.
In particular, for the second-order cumulant both thermal and 
non-thermal fluctuations take positive values.
Assuming that they are uncorrelated, 
the total cumulant is simply given by their sum.
In this sense, the suppression of the second-order cumulant of 
net-electric charge observed by ALICE collaboration 
\cite{ALICE} in Fig.~\ref{fig:ALICEfluc} 
is quite interesting; 
because all non-thermal fluctuations tend to enhance 
event-by-event fluctuations, 
the suppression is most probably attributed
to the thermal fluctuation \cite{Kitazawa:2014nja}.

\subsection{Other problems}

Besides the above problems, 
there are various issues which should be considered
in the interpretation of the event-by-event fluctuations.
\begin{enumerate}

\item
{\bf Fluctuations in the pre-equilibrium stage:}
For $\sqrt{s_{\rm NN}}$ larger than the top RHIC energy, 
the pre-equilibrium state is dominated by gluons.
Because gluons do not carry conserved charges except
energy, momentum and angular momentum, with the absence of quarks
the fluctuations of conserved charges in the pre-equilibrium 
system should be small.
Only after the pair creation of quarks, which carry conserved charges,
fluctuations of conserved charges start to increase.
If the equilibration of fluctuations is not established
during the time evolution, the cumulants in the final 
state would be suppressed as a remnant of the small 
fluctuation in the pre-equilibrium stage.
In low energy collisions, on the other hand, the 
magnitude of fluctuations in the pre-equilibrium system 
would be determined by the fluctuation of baryon stopping.

\item 
{\bf Limitation of detector's ability:}
The finite efficiency and acceptance of detectors 
modify the result of event-by-event analysis 
\cite{Kitazawa:2011wh,Kitazawa:2012at,Bzdak:2012ab}.
These effects will be addressed in Sec.~\ref{sec:binomial}.
The effect of the particle misidentifications by detectors 
on event-by-event analysis has to be investigated separately
from efficiency problems \cite{Anticic:2013htn,OAK}.

\item
{\bf Final state hadronic interactions:}
The decays of resonance states modify event-by-event
fluctuation in a rapidity window \cite{Kitazawa:2012at,Nahrgang:2014fza}.
To the first approximation, the effect of the resonance decays can be 
treated as a part of diffusion.
In Ref.~\cite{OAK}, the effect of secondary (knockout) protons 
is investigated.
The effect of deuteron formation is studied 
in Ref.~\cite{Feckova:2015qza}.

\end{enumerate}

\section{Time evolution of fluctuations in diffusive processes}
\label{sec:diffusion}

In Sec.~\ref{sec:equil} we have seen that the cumulants in 
equilibrated QCD medium show characteristic behaviors
when the medium undergoes phase transitions.
The goal of the measurement of fluctuations in relativistic
heavy ion collisions is to find these behaviors in the cumulants
measured by event-by-event analysis.
As we have already discussed in Sec.~\ref{sec:e-v-e}, however, 
fluctuations measured by event-by-event analysis 
are not the same as the thermal fluctuations.
In particular, it is to be remembered that the system created by 
the collisions is a dynamical system, although all analyses
discussed in Sec.~\ref{sec:equil} assumes equilibration.
Besides the caveats listed in Sec.~\ref{sec:e-v-e},
here we emphasize that 
in the comparison of thermal fluctuations with 
event-by-event one, the following two assumptions
are implicitly made:
\begin{enumerate}
\item
The medium establishes a (near-){\it equilibration of fluctuations}
in the early stage. 
\item
The signals developed in the early stage survive until the final state;
although the fluctuations tend to be shifted toward the
equilibrated values by final state interactions in the hadronic medium
and thermal blurring, 
this effect is assumed to be well suppressed.
\end{enumerate}
The purpose of this section is to consider these issues.
To this end, we first introduce a simple stochastic model
to illustrate a diffusive process of fluctuations.
We elucidate
the concept of {\it equilibration of fluctuations of conserved charges}.
It will be shown that this concept can be different 
from local equilibration; the time scale to establish the former
can be significantly longer than that for the latter.

A characteristic feature of the time evolution of conserved-charge 
fluctuations compared with non-conserved quantities is that 
the time evolution of the former is typically slow 
because their evolution is governed by hydrodynamic equations.
In particular, it can become arbitrary slow as the spatial volume 
to define the fluctuation becomes larger.
This property is highly contrasted with the one of non-conserved
quantities, whose typical time scales are typically short and 
insensitive to the spatial volume.

In this section, we consider the stochastic diffusion equation 
(SDE), which is a stochastic version of the diffusion equation
with a Langevin-type stochastic term.
This equation, which is a part of theory of hydrodynamic fluctuations 
\cite{Landau2,Kapusta:2011gt}, is suitable to describe the evolution of
conserved-charge fluctuations 
in diffusion processes for second order.
By describing the non-equilibrium time evolution of 
conserved-charge fluctuations in diffusive systems in this model,
we illustrate key ingredients associated with the time
evolution of fluctuations.
We also discuss that this formalism cannot describe 
the nonzero non-Gaussian fluctuations in equilibrium in a 
straightforward manner, and thus is not suitable for the description 
of the non-Gaussianity in relativistic heavy ion collisions.
A model to describe the diffusion of non-Gaussianity is introduced
in Sec.~\ref{sec:diffusion:DME}.

\subsection{Langevin equation for Brownian motion}
\label{sec:diffusion:Langevin}

Before starting the discussion of diffusive processes, 
however, for pedagogical purposes we first briefly take a look at
the Langevin 
equation for a single Brownian particle, which is a simple stochastic 
equation.
Readers who are familiar with the Langevin equation can skip this 
subsection.

Let us consider a heavy particle floating in a fluid.
We call this particle the Brownian particle in the following.
For simplicity we consider the velocity $v$ of this particle only
along a one-dimensional direction.
When this particle moves with a nonzero velocity it
receives a drag force proportional to $v$ from the fluid.
The equation of motion of this particle is then given by
\begin{align}
  m \frac{dv}{dt} = -\gamma v,
  \label{eq:gamma}
\end{align}
where $\gamma$ is the drag coefficient.
The solution of this equation with an initial condition
$v=v_0$ at $t=0$ is easily obtained as
\begin{align}
v(t)=v_0 e^{-\gamma' t}, 
\label{eq:v=v_0e}
\end{align}
with $\gamma'=\gamma/m$.
This solution shows that the velocity of the particle becomes 
arbitrary slow as $t$ becomes large, and vanishes 
in equilibrium defined by $t\to\infty$.

On the other hand, statistical mechanics tells
us that particles in an equilibrated medium have thermal
motion; from the equi-partition principle in classical statistical
mechanics the expectation value of the square of the velocity is 
given by $\langle v^2 \rangle_{\rm eq} = T/m$ in equilibrium, where 
$\langle \cdot \rangle_{\rm eq}$ denotes the thermal average.
When one is concerned with the velocity of order $v \simeq \sqrt{T/m}$,
therefore, the solution Eq.~(\ref{eq:v=v_0e}) is not satisfactory.

The thermal motion of Brownian particles comes from
the interaction with atoms composing the fluid \cite{Einstein}.
Because the thermal motion of individual atoms 
is random and not controlled by macroscopic quantities, 
even if we start from initial conditions with the same macroscopic
observables the motion of the particles would fluctuate to result 
in different time evolutions of Brownian particles
around the solution Eq.~(\ref{eq:v=v_0e}).
To incorporate such stochastic effects in Eq.~(\ref{eq:gamma}),
one may promote Eq.~(\ref{eq:gamma}) to a Langevin equation by adding
a stochastic term $\xi(t)$ as
\begin{align}
  \frac{dv}{dt} = -\gamma' v + \xi(t).
  \label{eq:gamma_xi}
\end{align}
Because the average motion of Brownian particles should be
well described by Eq.~(\ref{eq:gamma}) even with $\xi(t)$,
the effect of $\xi(t)$ should vanish on average.
We thus require 
\begin{align}
\langle \xi(t) \rangle=0,
\label{eq:<xi>=0}
\end{align}
where the expectation value is taken for different time evolutions.

Next, the correlation of the stochastic terms described 
by the two-point function $\langle \xi(t_1) \xi(t_2) \rangle$ can 
take nonzero values.
The correlation, however, would vanish when $t_1$ and $t_2$ are 
well separated, because the stochastic terms which come from microscopic
interactions should be uncorrelated for macroscopic time separation.
When the typical time scale for the variation of $v$ is sufficiently 
long compared with the one for $\xi(t)$,
the correlation function should
be well approximated by the delta function,
\begin{align}
\langle \xi(t_1) \xi(t_2) \rangle = A \delta( t_1 - t_2 ),
\label{eq:<xixi>=delta}
\end{align}
with an unknown coefficient $A$, which will be determined later.
The stochastic term obeying Eq.~(\ref{eq:<xixi>=delta}) is 
called the white noise, because the Fourier transform of 
Eq.~(\ref{eq:<xixi>=delta}) is constant as a function of 
frequency.

The solution of Eq.~(\ref{eq:gamma_xi}) 
with an initial condition $v(0)=v_0$ is given by
\begin{align}
v(t) = v_0 e^{-\gamma' t} + \int_0^t dt' e^{-\gamma'(t-t')} \xi(t').
\label{eq:v(t)}
\end{align}
The average of $v(t)$ is obtained
by taking the expectation values of both sides as
\begin{align}
\langle v(t) \rangle = \langle v_0 \rangle e^{-\gamma' t}, 
\label{eq:<v(t)>}
\end{align}
where the second term in Eq.~(\ref{eq:v(t)}) vanishes
by Eq.~(\ref{eq:<xi>=0}).
In Eq.~(\ref{eq:<v(t)>}) we assumed that
the statistical average is taken over both the stochastic effect
and the fluctuation of the initial condition, and replaced
$v_0$ by its average.
The result Eq.~(\ref{eq:<v(t)>}) is equivalent with Eq.~(\ref{eq:v=v_0e});
the introduction of the stochastic term does not modify
the average of the velocity.

The characteristic of Eq.~(\ref{eq:gamma_xi}) is that its
solution can fluctuate around the average Eq.~(\ref{eq:<v(t)>}).
To see the magnitude of the fluctuation, 
we take the average of the square of $v(t)$, 
\begin{align}
\langle (v(t))^2 \rangle
= \langle v_0^2 \rangle e^{-2\gamma' t} 
+ 2 e^{-\gamma' t} \int_0^t dt' e^{-\gamma'(t-t')} \langle v_0 \xi(t') \rangle
+ \int_0^t dt_1 dt_2 e^{-\gamma'(t-t_1)} e^{-\gamma'(t-t_2)}
\langle \xi(t_1) \xi(t_2) \rangle.
\end{align}
Since the stochastic term originates from microscopic effects,
$\xi(t)$ would be uncorrelated with the fluctuation of 
the initial condition $v_0$.
Under this assumption, $\langle v_0 \xi(t') \rangle$ 
in the second term on the right-hand side vanishes.
Substituting Eq.~(\ref{eq:<xixi>=delta}) into the last term,
and subtracting the average Eq.~(\ref{eq:<v(t)>}),
one obtains
\begin{align}
\langle (\delta v(t))^2 \rangle
= \langle ( v(t) - \langle v(t) \rangle )^2 \rangle
= \langle \delta v_0^2 \rangle e^{-2\gamma' t} 
+ \frac{A}{2\gamma'} ( 1 - e^{-2\gamma' t} ),
\label{eq:<dv^2>=A}
\end{align}
where $\langle \delta v_0^2 \rangle
= \langle v_0^2 \rangle - \langle v_0 \rangle^2$
is the fluctuation in the initial condition.
This result shows that the fluctuation of $v(t)$ relaxes
from the initial value to the equilibrated one, $A/2\gamma'$,
with the relaxation time $1/2\gamma'$.

The fluctuation in equilibrium is given by 
the equi-partition principle as $\langle v^2 \rangle_{\rm eq} = T/m$.
Because this condition should be satisfied in the $t\to\infty$
limit in Eq.~(\ref{eq:<dv^2>=A}), one obtains 
$A/2\gamma' = \langle v^2 \rangle_{\rm eq} = T/m$, or 
\begin{align}
  \langle \xi(t_1) \xi(t_2) \rangle = 2 \gamma' \langle v^2 \rangle_{\rm eq}
  \delta(t_1-t_2)
= \frac{ 2 \gamma' T}m \delta(t_1-t_2).
\label{eq:FDR1}
\end{align}
This relation is known as the fluctuation dissipation relation
(of first-kind), which relates the magnitude of the microscopic 
random force with macroscopic observables \cite{Einstein}.

Several comments are in order.
First, the stochastic process described by the Langevin equation 
Eq.~(\ref{eq:gamma_xi}) with a white noise is a Markov process. 
In fact, Eq.~(\ref{eq:gamma_xi}) can be converted to an equivalent 
Fokker-Planck equation, which is a partial differential equation
for a distribution function, which is of the first-order in time derivative
\cite{Gardiner}.
This means that the stochastic process is a Markov process.
Second, having obtained the fluctuation dissipation relation
for Gaussian fluctuation, Eq.~(\ref{eq:FDR1}), it may look possible 
to extend this relation to higher order cumulants, in such a way that 
the non-Gaussian cumulants $\langle v^n \rangle_{\rm c,eq}$ would
be related to higher order correlation 
$\langle \xi(t_1) \xi(t_2) \cdots \xi(t_n) \rangle_{\rm c}$.
This idea, however, results in failure.
It is shown that the stochastic term should be of 
Gaussian and all correlations higher than the second-order vanish 
\begin{align}
\langle \xi(t_1) \xi(t_2) \cdots \xi(t_n) \rangle_{\rm c}=0
\quad \mbox{ ($n\ge3$)},
\label{eq:<xi...xi>}
\end{align}
for Markov processes if the stochastic variable $v(t)$ varies 
continuously \cite{Gardiner}.
With Eq.~(\ref{eq:<xi...xi>}), it is easy to check that all higher 
order cumulants of $v$ vanish in the $t\to\infty$ limit.
For the third-order case, for example, 
$\langle v(t)^3 \rangle_{\rm c}$
is calculated to be 
\begin{align}
\langle (v(t))^3 \rangle
=& \langle v_0^3 \rangle e^{-3\gamma' t} 
+ 3 \int_0^t dt' \langle v_0^2 \xi(t') \rangle e^{-2\gamma' t} e^{-\gamma'(t-t')} 
\nonumber \\
&
+ 3 \int_0^t dt_1 dt_2 \langle v_0 \xi(t_1)\xi(t_2) \rangle 
e^{-\gamma' t} e^{-\gamma'(t-t_1)} e^{-\gamma'(t-t_2)} 
\nonumber \\
&+ \int_0^t dt_1 dt_2 dt_3 \langle \xi(t_1)\xi(t_2)\xi(t_3) \rangle 
e^{-\gamma'(t-t_1)} e^{-\gamma'(t-t_2)} e^{-\gamma'(t-t_3)} .
\end{align}
In the large $t$ limit the first three terms depending on the initial 
condition vanish owing to the $e^{-\gamma't}$ factor, while the last 
term also vanishes because of Eq.~(\ref{eq:<xi...xi>}).
This result shows that the Langevin equation Eq.~(\ref{eq:gamma_xi}) 
is not suitable for the description of the relaxation of non-Gaussian
fluctuations toward nonzero equilibrated values.

\subsection{Stochastic diffusion equation}
\label{sec:diffusion:SDE}

\subsubsection{Formalism}

Now, we have come to the main subject of this section, 
the stochastic process in diffusive systems.
To treat this problem, we introduce the stochastic diffusion equation (SDE).
This formalism, which can be regarded as a counterpart of the 
theory of hydrodynamic fluctuations \cite{Landau2,Kapusta:2011gt},
serves as a useful tool to describe the time evolution of 
fluctuations around the solution of the diffusion equation.
Although we limit our attention to one dimensional cases, 
the generalization to multi-dimensional ones is straightforward.

The time evolution of a conserved charge $n(x,t)$ is given by
the equation of continuity,
\begin{align}
\frac{\partial}{\partial t} n(x,t)
= - \frac{\partial}{\partial x} j(x,t),
\label{eq:dn=dj}
\end{align}
with the current $j(x,t)$.
It is phenomenologically known that $j(x,t)$ in various systems
well obeys the constitutive equation called the Fick's law,
\begin{align}
j = - D \frac{\partial}{\partial x} n(x,t),
\label{eq:Fick}
\end{align}
where $D$ is the diffusion constant.
By combining Eqs.~(\ref{eq:dn=dj}) and (\ref{eq:Fick}),
one obtains the diffusion equation 
\begin{align}
\frac{\partial}{\partial t} n(x,t)
= D \frac{\partial^2}{\partial x^2} n(x,t).
\label{eq:diffusion}
\end{align}
The solution of Eq.~(\ref{eq:diffusion}) is easily obtained
in Fourier space.
In the $t\to\infty$ limit, $n(x,t)$ approaches a uniform
form $n(x,t)=n_0$ without fluctuations.

In thermal systems, on the other hand, $n(x,t)$ should be 
fluctuating.
In fact, the integral of $n(x,t)$ in some spatial extent with 
length $L$,
\begin{align}
Q_L(t) = \int_0^L dx n(x,t),
\label{eq:Q_L}
\end{align}
is the number of the conserved charge in $L$, and 
as we have seen in Sec.~\ref{sec:equil} this quantity 
is fluctuating in equilibrium.
This means that $n(x,t)$ is also fluctuating in equilibrium.
When one considers the time evolution of $n(x,t)$ with a 
resolution that this thermal fluctuation is not negligible, 
the property of Eq.~(\ref{eq:diffusion}) that $n(x,t)$ becomes
static in the $t\to\infty$ is not satisfactory.

To describe the approach of fluctuations toward the thermal one 
in diffusive systems, one may modify 
Eq.~(\ref{eq:diffusion}) by introducing a stochastic term
similarly to the procedure in Sec.~\ref{sec:diffusion:Langevin}.
The stochastic term should be introduced in the constitutive
equation Eq.~(\ref{eq:Fick}), because this is a
phenomenological relation:
Although the current $j(x,t)$ in a macroscopic scale 
should obey Eq.~(\ref{eq:Fick}),
microscopically $j(x,t)$ should be fluctuating 
around Eq.~(\ref{eq:Fick}), because of the random thermal
motions of the microscopic constituents of the fluid.
We thus modify Eq.~(\ref{eq:Fick}) as 
\begin{align}
j(x,t) = - D \frac{\partial}{\partial x} n(x,t) + \xi(x,t),
\label{eq:Fick_xi}
\end{align}
where $\xi(x,t)$ represents the stochastic effect depending on 
$x$ and $t$.
The conservation law Eq.~(\ref{eq:dn=dj}), on the other hand, is 
an exact relation and should not be altered.
Substituting Eq.~(\ref{eq:Fick_xi}) into Eq.~(\ref{eq:dn=dj}),
one obtains
\begin{align}
\frac{\partial}{\partial t} n(x,t)
= D \frac{\partial^2}{\partial x^2} n(x,t) 
- \frac{\partial}{\partial x} \xi(x,t).
\label{eq:SDE}
\end{align}

Because the stochastic term should not modify the average motion
of $n(x,t)$, the average of the stochastic term should vanish,
\begin{align}
\langle \xi(x,t)\rangle = 0 ,
\label{eq:<xi>}
\end{align}
where the meaning of the statistical average is understood similarly
to the one in the previous subsection.
The two-point correlation $\langle \xi(x_1,t_1)\xi(x_2,t_2) \rangle$,
on the other hand, can take nonzero values.
We require that the correlation of the stochastic term is temporarily
and spatially local,
i.e.
\begin{align}
\langle \xi(x_1,t_1)  \xi(x_2,t_2)\rangle = A \delta(x_1-x_2) \delta(t_1-t_2).
\label{eq:<xixi>}
\end{align}
This requirement will be justified if the length and temporal scales 
of the motion of $n(x,t)$ described by Eq.~(\ref{eq:SDE}) is 
sufficiently large compared with the microscopic scales 
responsible for $\xi(x,t)$.
The overall coefficients $A$ will be determined later.
Equation~(\ref{eq:SDE}) together with Eqs.~(\ref{eq:<xi>}) and 
(\ref{eq:<xixi>}) is referred to as the SDE.

Equation~(\ref{eq:SDE}) is solved in Fourier space similarly to 
Eq.~(\ref{eq:gamma_xi}) in the previous subsection.
By defining the Fourier transform of $n(x,t)$ as
\begin{align}
n(k,t) = \int dx e^{-ikx} n(x,t),
\end{align}
we obtain
\begin{align}
\frac{\partial}{\partial t} n(k,t) = -Dk^2 n(k,t) + ik\xi(k,t).
\label{eq:dndt}
\end{align}
The solution of Eq.~(\ref{eq:dndt}) with the initial 
condition $n(k,0)=n_0(k)$ is given by
\begin{align}
n(k,t) = n_0(k) e^{-Dk^2t} + \int_0^t dt' e^{-Dk^2(t-t')} ik\xi(k).
\label{eq:n(k,t)}
\end{align}
The Fourier transform of the stochastic term satisfies
\begin{align}
\langle \xi(k,t) \rangle &= \int dx e^{-ikx} \langle \xi(x,t) \rangle = 0,
\label{eq:<xi_k>}
\\
\langle \xi(k_1,t_1) \xi(k_2,t_2) \rangle 
&= \int dx_1 dx_2 e^{-ik_1x_1-ik_2x_2} \langle \xi(x_1,t_1) \xi(x_2,t_2) \rangle
\nonumber \\
&= \int dx_1 dx_2 e^{-ik_1x_1-ik_2x_2} A \delta(x_1-x_2) \delta(t_1-t_2)
\nonumber \\
&= 2 \pi A \delta(k_1+k_2) \delta(t_1-t_2).
\label{eq:<xixi_k>}
\end{align}

By taking the expectation value of Eq.~(\ref{eq:n(k,t)}) and using 
Eq.~(\ref{eq:<xi_k>}), one obtains
\begin{align}
  \langle n(k,t) \rangle = \langle n_0(k) \rangle e^{-Dk^2 t} ,
  \label{eq:<n(k,t)>}
\end{align}
which is equivalent with the solution of the diffusion equation.
Next, by taking the average of the product of Eq.~(\ref{eq:n(k,t)}) 
with Eq.~(\ref{eq:<xixi_k>}), one obtains
\begin{align}
\langle n(k_1,t) n(k_2,t) \rangle 
= \langle n_0(k_1) n_0(k_2) \rangle e^{-D(k_1^2+k_2^2) t}
+ \frac{ \pi A }D \delta(k_1+k_2) ( 1 - e^{-2Dk_1^2 t}) ,
\end{align}
where it has been assumed that the stochastic term is uncorrelated with 
the initial condition $n_0(k)$, i.e. 
$\langle n_0(k_1) \xi(k_2,t)\rangle=0$.
This result with Eq.~(\ref{eq:<n(k,t)>}) gives
\begin{align}
\langle \delta n(k_1,t) \delta n(k_2,t) \rangle 
= \langle \delta n_0(k_1) \delta n_0(k_2) \rangle e^{-D(k_1^2+k_2^2) t}
+ \frac{ \pi A }D \delta(k_1+k_2) ( 1 - e^{-2Dk_1^2 t}) .
\label{eq:<dnkdnk>}
\end{align}

Next, let us consider the large $t$ limit of Eq.~(\ref{eq:<dnkdnk>}).
In this limit, the fluctuation of $n(x,t)$ should approach the 
equilibrium one.
The first term in Eq.~(\ref{eq:<dnkdnk>}) is suppressed in this limit
owing to the exponential factor;
although the zero mode fluctuation 
$\langle \delta n_0(0) \delta n_0(0) \rangle$ 
in the initial condition is not damped 
in this limit, this term has a negligible contribution 
unless the initial condition has a long range correlation.
One then obtains
\begin{align}
\lim_{t\to\infty} \langle \delta n(k_1,t) \delta n(k_2,t) \rangle 
= \frac{ \pi A }D \delta(k_1+k_2).
\end{align}
The correlation function in spatial coordinate in this limit is given by 
\begin{align}
\lim_{t\to\infty} \langle \delta n(x_1,t) \delta n(x_2,t) \rangle 
&= \lim_{t\to\infty} \int \frac{dk_1}{2\pi} \frac{dk_2}{2\pi}
e^{ik_1 x_1} e^{ik_2 x_2} \langle \delta n(k_1,t) \delta n(k_2,t) \rangle 
\nonumber \\
&= \frac{A}{2D} \delta(x_1-x_2).
\label{eq:lim<nn>}
\end{align}
This form of the correlation function is consistent
with Eq.~(\ref{eq:<nn...n>}), which is the correlation
function in an equilibrated medium.
The coefficient of this term is thus identified as the susceptibility,
\begin{align}
\frac{A}{2D} = \chi_2 .
\label{eq:ADchi2}
\end{align}
With Eqs.~(\ref{eq:lim<nn>}) and (\ref{eq:ADchi2}), 
the fluctuation of $Q_L$ in Eq.~(\ref{eq:Q_L}) is calculated to be
\begin{align}
\lim_{t\to\infty}\langle (\delta Q_L(t))^2 \rangle 
= \lim_{t\to\infty} 
\int_0^L dx_1 dx_2 \langle \delta n(x_1,t) \delta n(x_2,t) \rangle
= \chi_2 L.
\label{eq:<Q^2>ADL}
\end{align}
This result shows that the fluctuation of $Q_L$ in equilibrium 
is proportional to $L$, which is the result obtained 
in Sec.~\ref{sec:equil:conserved}.
Substituting Eq.~(\ref{eq:ADchi2}) into Eq.~(\ref{eq:<xixi>}), one obtains
\begin{align}
\langle \xi(x_1,t_1) \xi(x_2,t_2) \rangle 
= 2D\chi_2 \delta(x_1-x_2) \delta(t_1-t_2).
\label{eq:FDRdiffusion}
\end{align}
In Eq.~(\ref{eq:FDRdiffusion}), the property of the stochastic term 
is given through the macroscopic observables $D$ and $\chi_2$.
This is the fluctuation dissipation relation 
in the SDE.

\subsubsection{Time evolution in SDE}

Next, we investigate the time evolution of fluctuation
in the SDE.
To simplify the problem, in this article we limit our attention 
to the solution of the SDE for an initial condition satisfying
\begin{align}
\langle \delta n(x_1,0) \delta n(x_2,0) \rangle 
= \sigma_0 \delta(x_1-x_2).
\label{eq:sigma_0}
\end{align}
In this initial condition we assume that the fluctuation is 
local similarly to the equilibrium case Eq.~(\ref{eq:<nn...n>}),
but the proportionality coefficient $\sigma_0$ takes a 
different value from the susceptibility $\chi_2$.
With Eq.~(\ref{eq:sigma_0}), the fluctuation of $Q_L$ is an 
extensive variable, $\langle (\delta Q_L)^2 \rangle = \sigma_0 L$
in the initial condition.
By substituting Eq.~(\ref{eq:sigma_0}) into Eq.~(\ref{eq:<dnkdnk>}),
the solution in Fourier space for this initial condition is given by 
\begin{align}
\langle \delta n(k_1,t) \delta n(k_2,t) \rangle 
= \pi\left( \sigma_0 e^{-2Dk_1^2 t} + \chi_2 ( 1 - e^{-2Dk_1^2 t}) \right)
\delta(k_1+k_2) .
\end{align}
The fluctuation of $Q_L$ is obtained by performing the 
inverse Fourier transformation as
\begin{align}
\langle (\delta Q_L)^2 \rangle 
= L( \sigma_0 F_2(X) + \chi_2 ( 1-F_2(X)) ),
\label{eq:<Q_L^2>=F}
\end{align}
where we have introduced a dimensionless variable 
$X=\sqrt{2Dt}/L$ and 
\begin{align}
F_2(X) &= \int dz [ I_X(z/L)]^2 ,
\label{eq:F_2}
\\
I_X(\zeta) &= \int_{-1/2}^{1/2} d\xi \int \frac{dp}{2\pi}
e^{-X^2 p^2/2} e^{ip(\xi+\zeta)}
= \frac12 \left(
{\rm erf} \left ( \frac{ \zeta+1/2 }{\sqrt2 X} \right )
- {\rm erf} \left ( \frac{ \zeta-1/2 }{\sqrt2 X} \right ) \right),
\label{eq:I_X}
\end{align}
with the error function ${\rm erf}(x)= (2/\sqrt{\pi})\int_0^x dt e^{-t^2}$;
see Ref.~\cite{Kitazawa:2015ira} for manipulations.
The same result is obtained in Ref.~\cite{Shuryak:2000pd} using
the Fokker-Planck equation.
We remark that the solution depends on $t$ and $L$ only through $X$.
Here, $\sqrt{2Dt}$ is the diffusion distance of particles in 
the system described by Eq.~(\ref{eq:diffusion}).

\begin{figure}
\begin{center}
\includegraphics*[width=7cm]{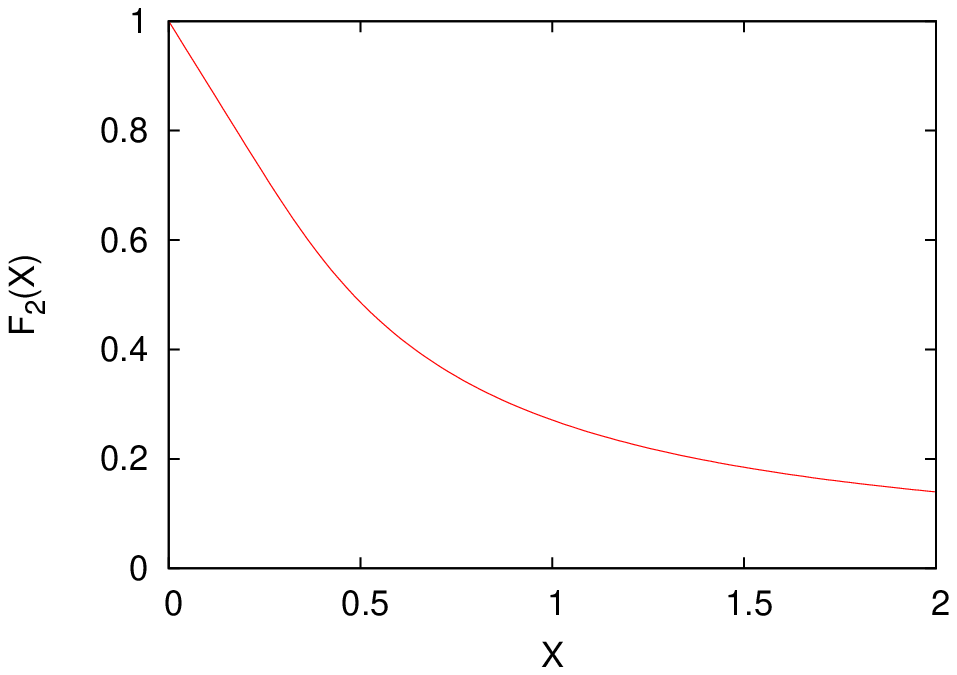}
\includegraphics*[width=7cm]{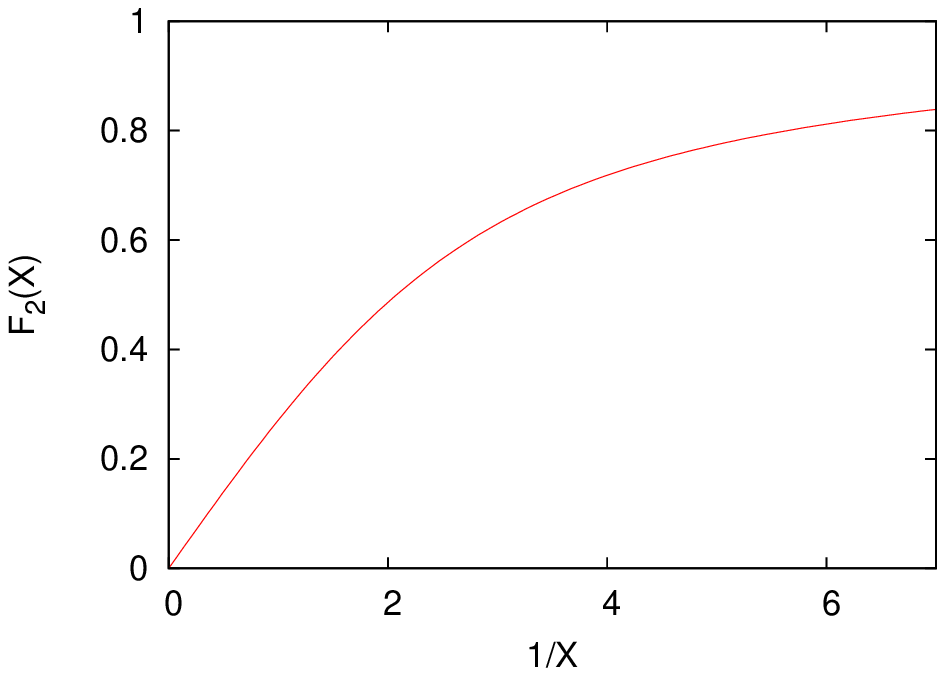}
\caption{
Function $F_2(X)$ in Eq.~(\ref{eq:F_2}).
}
\label{fig:F2}
\end{center}
\end{figure}

In Fig.~\ref{fig:F2}, we plot $F_2(X)$ as functions of $X$ and $1/X$. 
The left panel can be interpreted as the $\sqrt{t}$ dependence of 
$\langle (\delta Q_L)^2 \rangle$ with a fixed $L$.
Substituting this behavior of $F_2(X)$ into  Eq.~(\ref{eq:<Q_L^2>=F}),
one sees that $\langle (\delta Q_L)^2 \rangle$ approaches the 
equilibrated value $\chi_2$ from the initial value $\sigma_0$ 
as $t$ becomes larger.
Because Eq.~(\ref{eq:<Q_L^2>=F}) depends on $t$ only through $X$,
the variation is slower for larger $L$.
From the right panel of Fig.~\ref{fig:F2}, 
one can read off the $L$ dependence of $\langle (\delta Q_L)^2 \rangle$ 
for a given $t$.
The panel shows that the value of 
$\langle (\delta Q_L)^2 \rangle$ with fixed $t$ approaches the 
equilibrated (initial) value as $L$ becomes smaller (larger).

The right panel of Fig.~\ref{fig:F2} with Eq.~(\ref{eq:<Q_L^2>=F}) 
also shows that $\langle (\delta Q_L)^2 \rangle$ is not 
proportional to $L$ for finite $t>0$.
This behavior is contrasted to the extensive nature of 
fluctuations in equilibrium.
The $L$ dependence of $\langle (\delta Q_L)^2 \rangle/L$ comes from 
the nonzero correlation $\langle n(x_1,t)n(x_2,t)\rangle$ at $x_1\ne x_2$,
which vanishes in the large $t$ limit.
Intuitively, the ``memory'' of the initial condition gives rise to 
this correlation.
Only when the particles composing the system completely forget 
the initial condition in the $t\to\infty$ limit, 
the equilibrium property Eq.~(\ref{eq:lim<nn>}) is realized.

An important comment here is that the local equilibration and 
the equilibration of conserved-charge fluctuations are 
different concepts.
In particular, the latter is not a necessary condition of 
the former.
To see this, we note that the SDE is reasonably obtained if 
one assumes the local equilibration of the medium.
In fact, the local equilibration well justifies the use of 
the modified Fick's law Eq.~(\ref{eq:Fick_xi}) with a diffusion 
constant.
Moreover, the value of susceptibility in equilibrium is used
in SDE through the fluctuation dissipation relation, although
the SDE is an equation to describe non-equilibrium density 
profile $n(x,t)$.
A reasonable justification for the use of the equilibrium values 
would be to assume that the microscopic time scales establishing the 
local equilibration is sufficiently shorter than the typical 
one of $n(x,t)$.
The difference between the time scales comes from the conserving
nature of the charge. 
Because the charge is conserved, the variation of the local density $n(x,t)$ 
is achieved only through diffusion, i.e. transfer of charges.
It, however, can become arbitrary slow as the length 
scale becomes longer, as shown in the $L$ dependence of 
Eq.~(\ref{eq:<Q_L^2>=F}).
For non-conserving quantities, on the other hand, the variation of 
the local density is typically insensitive to the length scale.

\subsubsection{Discussions}

Next, we comment on the description of non-Gaussian cumulants
of $Q_L$ or $n(x,t)$ in the SDE.
As in the previous section, it is easy to show that 
the fluctuation of $Q_L$ becomes of Gaussian in the 
$t\to\infty$ limit; $\langle Q_L^n \rangle_{\rm c}=0$ for $n\ge3$
in this limit.
When one wants to describe the time evolution of 
non-Gaussian cumulants toward the {\it nonzero} equilibrated value,
the SDE Eq.~(\ref{eq:SDE}) with Eq.~(\ref{eq:FDRdiffusion})
is not suitable.
One may try to introduce higher order correlation of $\xi(x,t)$
so that $Q_L$ in the $t\to\infty$ limit becomes non-Gaussian.
However, similarly to the argument in Sec.~\ref{sec:diffusion:Langevin}
it is rigorously shown that the stochastic term in the SDE 
must be of Gaussian and higher order correlations vanish 
\cite{Gardiner}
\begin{align}
\langle \xi(x_1,t_1) \cdots \xi(x_n,t_n) \rangle_{\rm c} = 0
\mbox{ ~~ for $n\ge3$. }
\label{eq:<xi...xi>=0}
\end{align}
It is possible to introduce nonzero higher order correlation
of $\xi(x,t)$
by brute force ignoring the theorem.
This trial, however, results in failure with unphysical long-range 
correlations in higher order correlations of $\xi(x,t)$
\cite{Ono:master}.
Another way to make the distribution of $Q_L$ in equilibrium 
non-Gaussian in the SDE is to make the susceptibility 
$n(x,t)$ dependent.
With $n(x,t)$ dependent $\chi_2$, $Q_L$ in equilibrium becomes 
non-Gaussian even with the Gaussian stochastic term.
In this case, however, the SDE becomes nonlinear and analytic 
treatment of the equation becomes difficult.
In Sec.~\ref{sec:diffusion:DME}, we consider another 
model for diffusion processes which can describe
nonzero non-Gaussian fluctuations in equilibrium. 

The fluctuation dissipation relation for the SDE discussed 
in this subsection can be generalized to hydrodynamic equations
in a similar manner \cite{Landau2}.
Besides Eq.~(\ref{eq:dn=dj}), 
the hydrodynamic equations are constructed from 
the conservation law of energy-momentum tensor
\begin{align}
\partial_\mu T_{\mu\nu}=0,
\label{eq:dT=0}
\end{align}
and the constitutive equations for the elements of $T_{\mu\nu}$.
In a usual dissipative hydrodynamic equations, 
the constitutive equations are deterministic.
With a motivation similar to that with which we have introduced the SDE, 
it is possible to promote these equations to those
with stochastic terms.
Assuming the locality of the stochastic terms, their 
magnitudes are determined by macroscopic observables
including viscosity via the fluctuation dissipation relation
\cite{Landau2}.
This procedure is extended to relativistic systems in 
Ref.~\cite{Kapusta:2011gt}, in which an application of 
the stochastic equations to relativistic heavy ion collisions 
is addressed.
The stochastic equations constructed in this way are
called theory of hydrodynamic fluctuations or stochastic
hydrodynamics.
The SDE is regarded as a counterpart of this formalism.

\subsection{Diffusion of fluctuations in heavy ion collisions}
\label{sec:diffusion:hadronic}

The purpose of event-by-event analysis in relativistic heavy ion
collisions is to
measure anomalous behaviors in thermal fluctuations 
which occur in the early stage of the time evolution.
As discussed in Sec.~\ref{sec:e-v-e} already,
however, the experimental detectors can only measure 
the fluctuations in the final state.
Even if anomalous thermal fluctuations are well developed in the 
early stage, the fluctuations are modified in the hadronic
medium owing to diffusion.
For conserved charges, this time evolution is caused by
diffusion process
\cite{Asakawa:2000wh,Jeon:2000wg,Shuryak:2000pd},
and would be well described by the SDE.

Under the Bjorken expansion, the diffusion of conserved charges 
takes place in coordinate-space rapidity space, and 
rapidity window $\Delta Y$ to count the particle
number serves as the length $L$ in Eq.~(\ref{eq:Q_L}).
It is thus expected that the magnitude of conserved-charge 
fluctuations is more QGP like as $\Delta Y$ becomes wider,
while the magnitude becomes more hadronic for narrower
$\Delta Y$.
Assuming the correspondence between $\Delta Y$ and the 
(momentum-space) rapidity interval $\Delta \eta$,
such a behavior is in fact observed by ALICE collaboration
in net-electric charge fluctuation as shown in Fig.~\ref{fig:ALICEfluc}
\cite{ALICE}.
In the figure, the right vertical axis represents the $D$-measure
Eq.~(\ref{eq:Dmeasure}), a quantity proportional to 
net-electric charge fluctuation 
$\langle N_{\rm Q}^2 \rangle_{\rm c}/\Delta\eta$.
The figure shows that as $\Delta \eta$ becomes larger 
the magnitude of $\langle N_{\rm Q}^2 \rangle_{\rm c}/\Delta\eta$ 
is more suppressed.
If the magnitude of $\langle N_{\rm Q}^2 \rangle_{\rm c}$ is
small in the early stage in the time evolution,
this $\Delta \eta$ dependence is reasonably understood 
as a result of the diffusion process discussed in the previous 
subsection.

In Sec.~\ref{sec:e-v-e:rapidity}, we discussed that the experimental 
measurements are performed in a (pseudo-)rapidity window, $\Delta \eta$.
On the other hand, thermal fluctuations are defined in a 
coordinate-space rapidity window $\Delta Y$, and the diffusion 
process also takes place in coordinate space.
As discussed in Sec.~\ref{sec:e-v-e:rapidity}, 
there is only an approximate correspondence between the
rapidities $Y$ and $\eta$ even if the Bjorken picture
holds for the flow of the medium;
the measurement in rapidity $\eta$ is 
accompanied with a blurring arising from the thermal motion of 
individual particles at kinetic freezeout.
However, the effect of the blurring due to the rapidity conversion 
can be regarded as if it were
a part of the diffusion effects, 
because the distribution of the thermal motion in $y$ space is almost 
Gaussian as already discussed in Sec.~\ref{sec:e-v-e:rapidity}.
Therefore, the above interpretation on the $\Delta \eta$ dependence of 
Fig.~\ref{fig:ALICEfluc} hardly changes even after including 
the effect of blurring, although in this case 
the value of diffusion length $\sqrt{2Dt}$ has to be 
understood as the one including the effects of blurring 
\cite{Asakawa:QM15,Ohnishi}.

The time evolution of fluctuations of conserved charges 
has also been investigated in molecular dynamical models in 
Refs.~\cite{Bleicher:2000ek,Haussler:2007un}.
Extension of the SDE to include the memory effect
by introducing higher order time derivative(s) to 
the diffusion equation 
is discussed in Refs.~\cite{Aziz:2004qu,Young:2014pka}.

The $D$-measure has also been measured at RHIC
\cite{Adcox:2002mm,Adams:2003st,Abelev:2008jg}.
Contrary to the ALICE result, 
these results are consistent with the value in the 
hadronic medium ($\sqrt{s_{_{\rm NN}}}$ dependence of the $D$-measure 
is nicely summarized in Fig.~4 in Ref.~\cite{ALICE}).
The maximum $\Delta \eta$ in these experiments, $\Delta \eta=1.0$,
determined by the structure of the detectors, however, is smaller 
than the one of the ALICE, $\Delta \eta=1.6$.
As is evident in Fig.~\ref{fig:ALICEfluc}, the narrower 
$\Delta \eta$ makes $\langle N_{\rm Q}^2 \rangle_{\rm c}$ 
more hadronic.
This difference in $\Delta \eta$ in these experiments would be
one of the origins of the contradiction.

When the hot medium passes through near the QCD critical point, 
besides the diffusion of conserved charges, the dynamical time 
evolution of sigma field $\sigma=\langle \bar\psi \psi \rangle$, 
which is the order parameter of the chiral phase transition, has to be
considered simultaneously.
Although the fluctuation of $\sigma$ diverges at the critical
point in equilibrium, in dynamical system the approach to the 
equilibrium value is limited due to the critical slowing down.
The growth of the fluctuation of $\sigma$ is investigated 
in terms of the correlation length of $\sigma$ field 
in Refs.~\cite{Berdnikov:1999ph,Nonaka:2004pg}.
These studies suggest that the growth of the correlation
length is limited to $\xi\simeq2$ fm
even if the system passes exactly on the critical point.
The time evolution of the third- and fourth-order cumulants of $\sigma$
near the critical point is discussed in Ref.~\cite{Mukherjee:2015swa}.
To describe the time evolution of conserved-charge fluctuations
near the critical point, the coupling of $\sigma$ with the 
conserved charge \cite{Fujii:2003bz,Fujii:2004jt} should play
a crucial role.
Attempts to model the time evolution of fluctuations incorporating 
both $\sigma$ and conserved-charge fields in a stochastic 
formalism are made in Refs.~\cite{Stephanov:2009ra,Nahrgang:2011mg,
Nahrgang:2011vn,Herold:2014zoa}.
When the hot medium undergoes a first-order phase transition,
the fluctuation would be enhanced owing to domain formation
\cite{Herold:2014zoa} and spinodal instabilities 
\cite{Sasaki:2007db,Steinheimer:2012gc,Steinheimer:2013xxa}.
Understanding these highly dynamical processes, especially the 
growth of fluctuations and their effects on experimental signals, 
is an interesting future subject.

\subsection{$\Delta\eta$ dependence of higher order cumulants}
\label{sec:diffusion:DME}

In Sec.~\ref{sec:diffusion:SDE}, we have seen that 
the SDE is suitable to describe
the diffusion process of Gaussian fluctuations.
As discussed there, however, in this model it is difficult to 
describe nonzero non-Gaussian fluctuations in equilibrium.
Accordingly, this formalism is not suitable to describe the 
approach of non-Gaussian cumulants toward nonzero equilibrated value,
which would happen in heavy ion collisions.

\begin{figure}
\begin{center}
\includegraphics*[width=8cm]{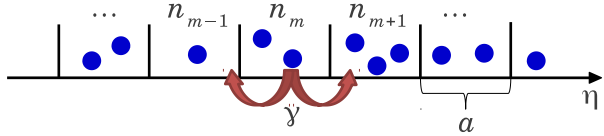}
\caption{
System described by the diffusion master 
equation Eq.~(\ref{eq:DME}).}
\label{fig:DME}
\end{center}
\end{figure}

For a description of the time evolution of 
non-Gaussian cumulants in diffusive systems,
a model called the diffusion master equation (DME) \cite{Gardiner} 
is employed in Refs.~\cite{OAK,Sakaida:2014pya,Kitazawa:2015ira}.
A feature of these studies is that the discrete nature
of particle number is explicitly treated.
In the DME for one dimensional problems, the coordinate is 
divided into discrete cells with an equal length $a$
(see, Fig.~\ref{fig:DME}).
We denote the number of particles in each cell, 
labeled by an integer $m$, as $n_m$.
We then introduce the probability distribution function 
$P(\bm{n},t)$ that each cell contains $n_m$ particles at time $t$
with $\bm{n}=(\cdots,n_{m-1},n_{m},n_{m+1},\cdots )$.
It is also assumed that each particle moves to 
adjacent cells with a probability $\gamma$ per 
unit time, as a result of microscopic 
interactions and random motion.
The probability $P(\bm{n},t)$ then obeys 
the differential equation 
\begin{align}
\partial_\tau P(\bm{n},\tau)
=& \gamma(t) \sum_m [
( n_{m} + 1 )  
\{ P(\bm{n}+\bm{e}_{m}-\bm{e}_{m+1},\tau)
+ P(\bm{n} +\bm{e}_{m}-\bm{e}_{m-1},\tau) \} 
\nonumber \\
& -2 n_m P(\bm{n},\tau) ],
\label{eq:DME}
\end{align}
where $\bm{e}_{m}$ is the vector that all components are
zero except for the $m$th-one, which takes unity.
Equation (\ref{eq:DME}) is referred to as the
DME \cite{Gardiner}.

The time evolution of the cumulants and correlation functions
of particle number described by the DME, Eq.~(\ref{eq:DME}), can be
solved analytically \cite{Kitazawa:2013bta}.
To obtain the solution  for arbitrary initial conditions,
it is convenient to use the formula of superposition of 
probability distribution functions given in 
Appendix~\ref{app:superposition} \cite{Kitazawa:2013bta,Kitazawa:2015ira}.
One then takes the continuum limit, $a\to0$, of this 
solution.
It is shown that the time evolution of average density
$\langle n(x,t)\rangle$ after taking the continuum limit is
consistent with the one obtained with the diffusion equation
Eq.~(\ref{eq:diffusion}) with the diffusion constant
$D(t)=\gamma(t) a^2$.
Moreover, the time evolution of the second-order cumulant
with the DME is also consistent with that with the SDE.

In the DME, motion of the individual particles composing the system 
is given by random walk without correlations with one another.
The time evolution of the particle distribution thus is given
by the superposition of these uncorrelated particles.
In this sense, it is reasonable that the solution of the DME 
is consistent with those in the diffusion equation and the SDE.
In the $t\to\infty$ limit, each particle exists any position
with an equal probability irrespective of the initial condition,
and they are uncorrelated with each other.
The particle number $Q_L$ in an interval $L$, therefore, is
simply given by the Poisson distribution in this limit 
when $L$ is sufficiently smaller than the length of the system
(see, Sec.~\ref{sec:basic:Poisson}).
The cumulants of $Q_L$ in this limit thus are given by 
\begin{align}
  \langle Q_L^n \rangle_{\rm c} = \langle Q_L \rangle = \rho L
  \label{eq:Q_L^n}
\end{align}
with the average density $\rho$.

The time evolution of the second-order cumulant in the DME
agrees with that in the SDE with $\chi_2=\rho$.
On the other hand, higher order cumulants in the DME take nonzero 
values in equilibrium as shown in Eq.~(\ref{eq:Q_L^n})
contrary to the case of the SDE.
Therefore, the DME is regarded as an extension of the SDE
to describe the approach of the higher order cumulants
toward nonzero equilibrium values.

The DME can be extended to the system with multi-particle species.
This allows us to define the difference of two particle numbers.
Because the distributions of two particle numbers in an interval $L$ 
become Poissonian in equilibrium,
the difference of the particle numbers in the interval is given 
by the the Skellam distribution in equilibrium.
This property is suitable for the description of diffusion
process of net-baryon number in the hadronic medium in heavy ion 
collisions, because its fluctuations in the HRG
model are given by the Skellam distribution as discussed in 
Sec.~\ref{sec:equil:HRG}.

\begin{figure}
\begin{center}
\includegraphics*[width=7cm]{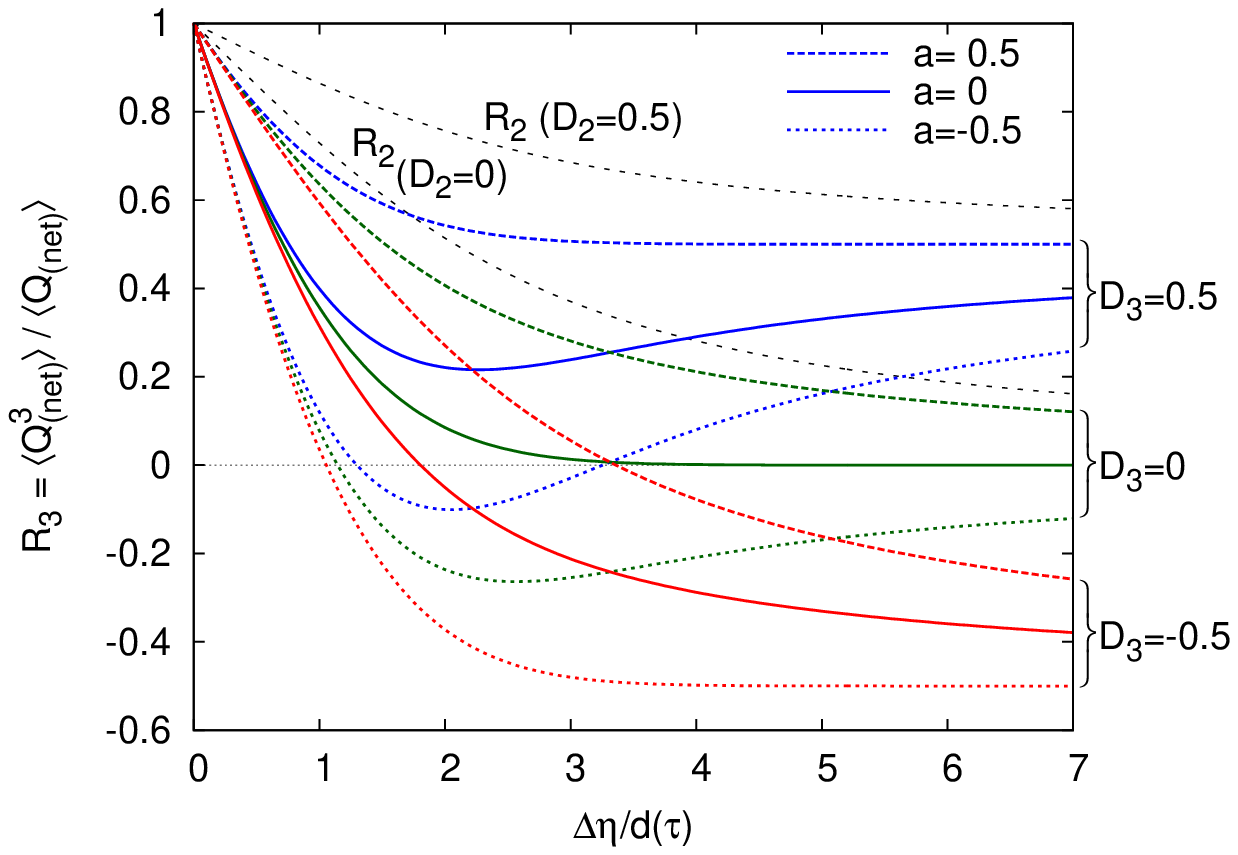}
\hspace{.5cm}
\includegraphics*[width=7cm]{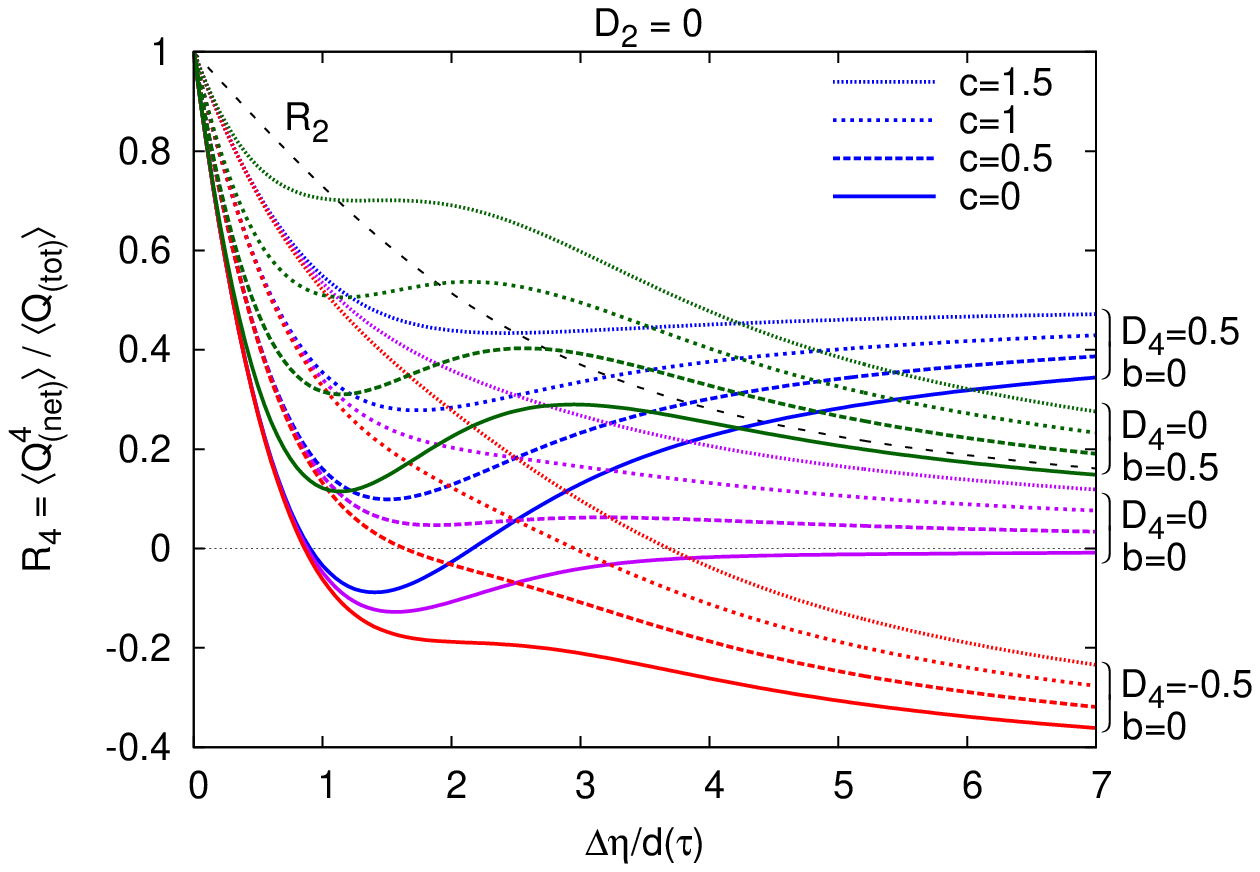}
\caption{
Example of the solution of Eq.~(\ref{eq:DME}) \cite{Kitazawa:2015ira}.
Behaviors of the third- and fourth-order cumulants 
as functions of rapidity window ($\Delta\eta$)
for various initial conditions.}
\label{fig:dydep}
\end{center}
\end{figure}

\begin{figure}
\begin{center}
\includegraphics*[width=7cm]{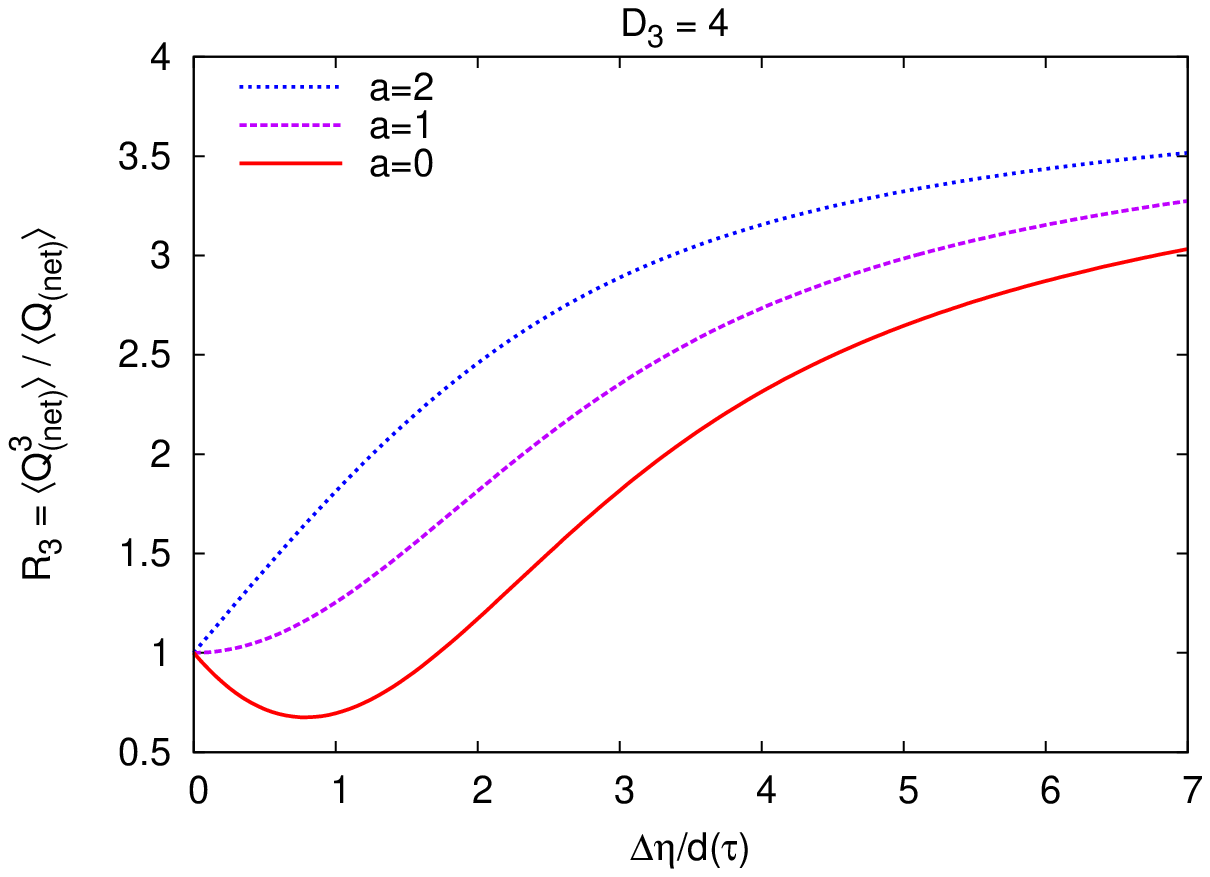}
\hspace{.5cm}
\includegraphics*[width=7cm]{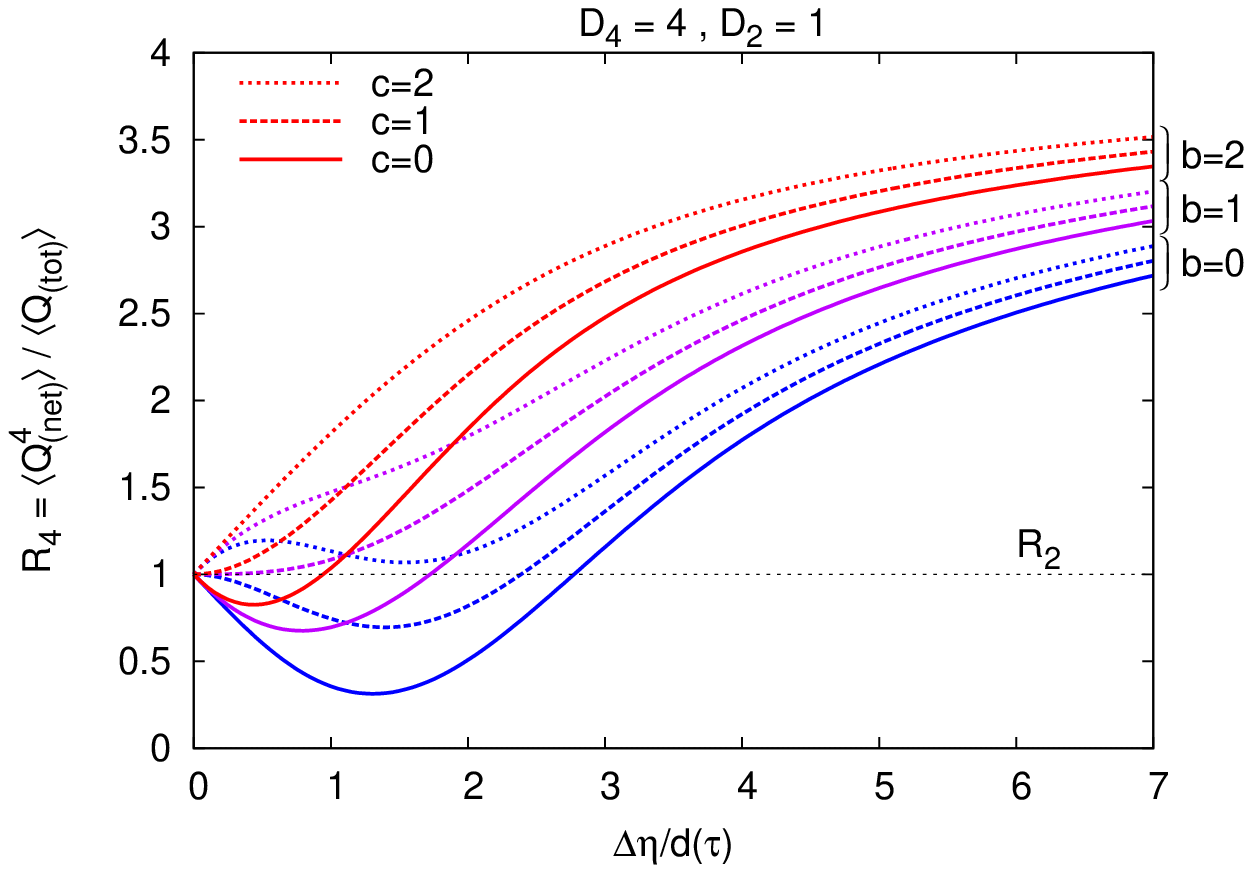}
\caption{
Same as Fig.~\ref{fig:dydep}, but with initial 
condition with larger higher order cumulants.
}
\label{fig:dydep4}
\end{center}
\end{figure}

Now we consider the time evolution in the DME.
To compare the result with fluctuations in heavy ion collisions,
we call the spatial coordinate as rapidity $\eta$, 
and the difference of the particle numbers as net-particle number.
We also denote the net-particle number in 
a rapidity interval $\Delta\eta$ as $Q_{\rm (net)}$.
Similarly to the case of the solution of SDE in Sec.~\ref{sec:diffusion:SDE}, 
the time evolution of the cumulants $\langle Q_{\rm (net)}^n \rangle_{\rm c}$ 
depends on $\Delta\eta$ and the proper time $\tau$ only through 
a combination $X=d(\tau)/\Delta\eta$,
where $d(\tau)$ is the diffusion distance in rapidity space,
\begin{align}
d(\tau) = \sqrt{ 2 \int_{\tau_{\rm initial}}^\tau d\tau D(\tau)}.
\end{align}
As discussed in Sec.~\ref{sec:diffusion:hadronic}, one has 
to distinguish the coordinate- and momentum-space rapidities.
Because the effect of thermal blurring accompanied with the conversion
of these rapidities can be regarded as a part of diffusive process
as discussed in Sec.~\ref{sec:diffusion:hadronic},
in this subsection we do not distinguish the rapidities.
We, however, have to reinterpret the meaning of $d(\tau)$
to include the effect of the blurring.

In Figs.~\ref{fig:dydep} and \ref{fig:dydep4}, we show 
some examples of the $\Delta\eta$ dependences of third- (left) 
and fourth- (right) order cumulants obtained in the DME
\cite{Kitazawa:2015ira}.
In these figures, the cumulants are normalized by their 
equilibrated values.
These quantities thus become unity at $\Delta\eta=0$.
The initial conditions are chosen so as to 
satisfy the locality condition Eq.~(\ref{eq:<nn...n>}),
with the coefficients in front of the delta-function, 
$D_2$, $D_3$, $D_4$, $a$, $b$ and $c$, are treated as free 
parameters.
Here, $D_n$ is the $n$th-order susceptibility of the net charge 
in the initial condition.
Figure~\ref{fig:dydep} shows the results for initial conditions 
with small susceptibilities $D_3$ and $D_4$, while 
in Fig.~\ref{fig:dydep4} $D_3$ and $D_4$ are taken large.
The figures show that the $\Delta\eta$ dependences of
higher order cumulants are sensitive to the initial condition.
In particular, it is interesting that the third- and fourth-order 
cumulants can have non-monotonic dependences on $\Delta\eta$.
These $\Delta\eta$ dependences can directly be compared with 
the experimental results \cite{Luo:2015ewa,Thaeder:2015QM}.
The comparison of the $\Delta\eta$ dependences in these 
experiments with those in the DME in Figs.~\ref{fig:dydep} 
and \ref{fig:dydep4} will lead us to deeper understanding 
on the fluctuation observables, such as the effects of 
diffusion in the hadronic stage and thermal blurring accompanied with 
the rapidity conversion.
It would also enable us to extract the initial values of cumulants.
Experimental measurement of the non-monotonic $\Delta\eta$ 
dependences in Figs.~\ref{fig:dydep} and \ref{fig:dydep4} 
is also an interesting subject.

\section{Binomial model}
\label{sec:binomial}

In this section, we address two problems associated with the 
experimental measurement of conserved-charge fluctuations which 
are not covered in Secs.~\ref{sec:e-v-e} and \ref{sec:diffusion}.
One of them is concerned with the measurement of 
net-baryon number cumulants.
As discussed in Sec.~\ref{sec:equil}, among the thermal fluctuations
of conserved charges that of net-baryon number shows the most
clear signal of the phase transitions.
The present experimental detectors, however, are not capable of 
their measurement because the detectors cannot count neutral baryons, 
in particular neutrons.
As proxies of net-baryon number cumulants, 
those of net-proton number are measured 
\cite{STAR-fluc,Adamczyk:2013dal,Luo:2015ewa}
and compared with theoretical studies on the net-baryon number 
cumulants in the literature.
The systematic error arising from this substitution has to 
be estimated carefully \cite{Kitazawa:2011wh,Kitazawa:2012at}.
The second issue is the effect of the finite efficiency and 
acceptance of detectors.
The real detectors cannot observe all particles entering them,
but lose some of them with some nonzero probability.
The probability to observe a particle is called efficiency
\footnote{This definition of efficiency is an ideal one and
different from ones used in most of experiments, in which
acceptance and efficiency cannot be separated uniquely for
various reasons such as the existence of magnetic field,
non-zero probability of simultaneous hits, and so forth.}
The detectors also have limitation in acceptance.
For example, some azimuthal angles are not covered by the detectors, 
or hidden by materials in front of the detectors, and 
the particles arriving at such azimuthal angles are not detected.
The finite efficiency and acceptance modify 
the event-by-event fluctuation
\cite{Kitazawa:2012at,Bzdak:2012ab}.
These two problems are in fact understood simultaneously.
In fact, the measurement of net-proton number in place of the 
net-baryon number is regarded as $50\%$ efficiency loss.
In this section, in order to describe these effects, 
we employ a model for probability distribution that 
we call the ``binomial model''.
The binomial model is first introduced in 
Ref.~\cite{Kitazawa:2011wh} to discuss the similarity and 
difference between the net-baryon and net-proton number 
cumulants, and then extended to investigate the effect of 
efficiency correction \cite{Kitazawa:2012at,Bzdak:2012ab}.
The purpose of this section is to review the binomial model
and deal with these problems.

\subsection{A model for efficiency}

For an illustration of the binomial model,
let us first consider the following simple problem.
We consider a probability distribution function $P(N)$
for an integer stochastic variable $N$.
To be specific, we suppose that $N$ is the number of some 
particles in each ``event'' in some experiment.
In order to determine $P(N)$ experimentally,
one would repeat the measurements of the particle number for 
individual events.
If measurements of the particles are carried out exactly 
for each event without any loss, 
one can determine the histogram corresponding 
to $P(N)$ and the corresponding cumulants.
By increasing the number of measurements,
the error of the cumulants can be suppressed arbitrarily.

Now, suppose that we are in a situation where our detector is not perfect, 
and can count particles only with a probability less than unity.
The particle number observed in each event, $n$, then would 
be smaller than the actual number $N$.
In this case, the histogram $\tilde{P}(n)$ for the detected
particle number $n$ is of course different from $P(N)$.
Accordingly, the cumulants of $\tilde{P}(n)$, i.e. those
constructed from the experimental result, are not the same as 
the real cumulants of $P(N)$, either.
The question is how to obtain the cumulants of $P(N)$
in this incomplete experiment.

This problem can be solved completely when the 
probability to detect a particle is a constant for all events and 
{\it uncorrelated with one another for individual particles} in an event.
Let us denote the probability $p$.
Then, if the actual particle number in an event is $N$,
the probability to find $n$ particles in this event is
given by the binomial distribution function $B_{p,N}(n)$ in 
Eq.~(\ref{eq:binomial}).
The distribution function $\tilde{P}(n)$ is then given by
\begin{align}
\tilde{P}(n) = \sum_N B_{p,N}(n) P(N).
\label{eq:tildeP}
\end{align}
We refer to Eq.~(\ref{eq:tildeP}) as the binomial model.

When Eq.~(\ref{eq:tildeP}) is justified, the cumulants of $P(N)$
can be written by using those of $\tilde{P}(n)$.
To understand the origin of such relations,
we note that Eq.~(\ref{eq:tildeP}) is
a linear relation connecting two distribution functions.
In fact, this equation can be rewritten using an infinite-dimensional 
matrix $M(n,N)$ as $\tilde{P}(n) = \sum_N M(n,N) P(N)$ with 
$M(n,N)=B_{p,N}(n)$ \cite{Kitazawa:2012at}.
Moreover, the matrix $M(n,N)$ is upper triangular in the sense
that $M(n,N)=0$ for $n>N$.
The inverse of $M(n,N)$ thus can be constructed iteratively.
Using the inverse matrix $M(N,n)^{-1}$, $P(N)$ can be represented as 
\begin{align}
P(N) = \sum_n M(N,n)^{-1} \tilde{P}(n).
\label{eq:P(N)MP}
\end{align}
Unfortunately, however, it is easily shown that $M(N,n)^{-1}$ has
a singular behavior, and the determination of the distribution
function $P(N)$ itself from Eq.~(\ref{eq:P(N)MP}) with the information
on $\tilde{P}(n)$ is not possible.
Nevertheless, in the binomial model Eq.~(\ref{eq:tildeP})
the cumulants of $P(N)$ can be represented by those of 
$\tilde{P}(n)$, and vice versa.
The results up to the fourth-order are given by 
\cite{Kitazawa:2012at,Kitazawa:2016awu}
\begin{align}
\langle n \rangle &= \xi_1 \langle N \rangle ,
\\
\langle n^2 \rangle_{\rm c} 
&= \xi_1^2 \langle N^2 \rangle_{\rm c} + \xi_2 \langle N \rangle ,
\\
\langle n^3 \rangle_{\rm c} 
&= \xi_1^3 \langle N^3 \rangle_{\rm c} + 3 \xi_1 \xi_2 \langle N^2 \rangle_{\rm c} 
+ \xi_3 \langle N \rangle ,
\\
\langle n^4 \rangle_{\rm c} 
&= \xi_1^4 \langle N^4 \rangle_{\rm c} + 6 \xi_1^2 \xi_2 \langle N^3 \rangle_{\rm c} 
+ ( 3\xi_2^2 + 4 \xi_1 \xi_3 ) \langle N^2 \rangle_{\rm c} + \xi_4 \langle N \rangle ,
\end{align}
and 
\begin{align}
\langle N \rangle &= \xi_1^{-1} \langle n \rangle ,
\label{eq:<N1>}
\\
\langle N^2 \rangle_{\rm c} 
&= \xi_1^{-2} \langle n^2 \rangle_{\rm c} - \xi_2 \xi_1^{-3} \langle n \rangle ,
\label{eq:<N2>}
\\
\langle N^3 \rangle_{\rm c} 
&= \xi_1^{-3} \langle n^3 \rangle_{\rm c} - 3 \xi_2 \xi_1^{-4} \langle n^2 \rangle_{\rm c} 
+ ( 3 \xi_2^{2} \xi_1^{-5} - \xi_3 \xi_1^{-4} ) \langle n \rangle ,
\label{eq:<N3>}
\\
\langle N^4 \rangle_{\rm c} 
&= \xi_1^{-4} \langle n^4 \rangle_{\rm c} - 6 \xi_2 \xi_1^{-5} \langle n^3 \rangle_{\rm c} 
+ ( 15 \xi_2^2 \xi_1^{-6} -4 \xi_3 \xi_1^{-5} ) \langle n^2 \rangle_{\rm c} 
- ( 15 \xi_2^3 \xi_1^{-7} - 10 \xi_2 \xi_3 \xi_1^{-6} + \xi_4 \xi_1^{-5} )
\langle n \rangle ,
\label{eq:<N4>}
\end{align}
where $\xi_n$ are the coefficient of the cumulants of the binomial 
distribution function defined in Eqs.~(\ref{eq:<m>_c=xi_n})
and (\ref{eq:xi_binomial}).
The derivation of these results is found in 
Ref.~\cite{Kitazawa:2016awu}, which makes use of 
the relations of cumulants summarized in 
Appendix.~\ref{app:superposition}.
The equivalent result can be obtained on the basis of the factorial 
moments \cite{Bzdak:2012ab}.
Equations~(\ref{eq:<N1>}) -- (\ref{eq:<N4>}) show that 
the cumulants of $P(N)$ can be reconstructed from the incomplete 
information obtained in experiments.

To apply the binomial model to the analysis of net-particle 
number in relativistic heavy ion collisions,
the model has to be extended to probability distribution
functions for at least two stochastic variables representing
particle and anti-particle numbers.
Suppose that the probability that $N$ particles and $\bar{N}$ 
anti-particles arrive at the detector for each event is given 
by $P(N,\bar{N})$.
We then assume that the detector finds these particles and
anti-particles with probabilities $p$ and $\bar{p}$, respectively, 
which are independent for individual particles.
The distribution function $\tilde{P}(n,\bar{n})$ that 
$n$ particles and $\bar{n}$ anti-particles are found by 
the detector in each event is then given by
\begin{align}
\tilde{P}(n,\bar{n}) = \sum_{N,\bar{N}}
B_{p,N}(n) B_{\bar{p},\bar{N}} P(N,\bar{N}).
\label{eq:binomialmodel}
\end{align}
Assuming this factorization, the cumulants of $N$ and $\bar{N}$
can again be given by those of $n$ and $\bar{n}$
\cite{Kitazawa:2012at,Bzdak:2012ab}.
The model can also be extended to multi variable cases
\cite{Luo:2014rea,Bzdak:2013pha,Kitazawa:2016awu}.
In the analysis of fluctuations by STAR collaboration, the 
efficiency correction is taken into account using the binomial 
model \cite{Adamczyk:2013dal,Adamczyk:2014fia,Luo:2015ewa}.
We note that a similar statistical model is also applied in 
\cite{Gorenstein:2011hr,Rustamov:2012bx}.

When one applies the binomial model, however, it has to be
remembered that this model is justified only when the efficiency
to measure particles is independent for individual particles.
It has been recently pointed out that the violation of this 
assumption in real detectors can significantly modify the 
reconstructed values of the cumulants especially for higher 
order ones \cite{efficiency}.

\subsection{Net-baryon vs net-proton number cumulants}

Now, let us consider the relation between the net-baryon 
and net-proton number cumulants \cite{Kitazawa:2011wh,Kitazawa:2012at}.
In this discussion, we use the fact that an (anti-)baryon arriving 
at the detector is an (anti-)proton with some probability
about $50\%$.
If the probability that an (anti-)baryon is an (anti-)proton is
uncorrelated for individual (anti-)baryons, one thus can apply 
the binomial model to relate the (anti-)baryon and (anti-)proton 
numbers.
Similarly to the previous case,
the probability ${\cal G}(N_p,N_{\bar p})$ to observe $N_p$ protons 
and $N_{\bar p}$ anti-protons in an event is related to the 
probability ${\cal F}(N_{\rm B},N_{\bar{\rm B}})$ that 
$N_{\rm B}$ baryons and $N_{\bar \rm B}$ antibaryons enters 
the detector in the event as 
\begin{align}
{\cal G}(N_p,N_{\bar p})
&= \sum_{N_{\rm B},N_{\bar{\rm B}}} 
{\cal P}(N_p,N_{\bar p};N_{\rm B},N_{\bar{\rm B}})
\nonumber \\
&= \sum_{N_{\rm B},N_{\bar{\rm B}}} 
B_{r,N_{\rm B}} (N_p) B_{{\bar r},N_{\bar{\rm B}}} (N_{\bar p})
{\cal F}(N_{\rm B},N_{\bar{\rm B}}),
\label{eq:G=FM}
\end{align}
where $r$ ($\bar{r}$) is the probability that a baryon (an anti-baryon)
arriving at the detector is a proton (anti-proton).
Using Eq.~(\ref{eq:G=FM}), one can relate the net-proton number
cumulants with those of net-baryons, and vice versa;
explicit formulas are given in Ref.~\cite{Kitazawa:2012at}.
Using these formulas, it is possible to obtain the net-baryon
number cumulants experimentally by measuring only
protons and anti-protons.

An important point of this argument is that in this case the 
assumption on the independence of the probabilities
required for the validity of the binomial model Eq.~(\ref{eq:G=FM})
can be justified from a microscopic argument for 
sufficiently large $\sqrt{s_{_{\rm NN}}}$ 
\cite{Kitazawa:2011wh,Kitazawa:2012at}.
Therefore, the use of the binomial model is well justified 
in this problem.
The key ingredient for this discussion is ${\rm N}\pi$ reactions
in the hadronic stage mediated by 
$\Delta(1232)$ resonances having the isospin $I=3/2$.
These reactions frequently take place even after chemical freezeout 
in the hadronic medium during the time evolution of the fireballs,
and in fact are the most dominant reactions of nucleons in the 
hadronic medium.
These reactions contain charge exchange reactions, which alter 
the third component of the isospin of the nucleon, or the nucleon 
isospin for short, in the reaction.
The reactions of a proton to form $\Delta$ 
are:
\begin{align}
p + \pi^+ &\to \Delta^{++} \to p + \pi^+,
\label{eq:D++} 
\\
p + \pi^0 &\to \Delta^+ \to p(n) + \pi^0(\pi^+),
\label{eq:D+}
\\
p + \pi^- &\to \Delta^0 \to p(n) + \pi^-(\pi^0).
\label{eq:D0}
\end{align}
Among these reactions, Eqs.~(\ref{eq:D+}) and (\ref{eq:D0}) 
are responsible for the change of the nucleon isospin.
The ratio of the cross sections of a proton to form 
$\Delta^{++}$, $\Delta^+$, and $\Delta^0$ is $3:1:2$, 
which is determined by the isospin SU(2) symmetry of 
the strong interaction.
The isospin symmetry also tells us that the branching 
ratios of $\Delta^+$ ($\Delta^0$) decaying into the final 
state having a proton and a neutron are $1:2$ ($2:1$).
Using these ratios, one obtains the ratio of the probabilities
that a proton in 
the hadron gas forms $\Delta^+$ or $\Delta^0$ with
a reaction with a thermal pion, and then 
decays into a proton and a neutron, respectively, $P_{p\to p}$ and 
$P_{p\to n}$, as
\begin{align}
P_{p\to p} : P_{p\to n} = 5:4,
\label{eq:P_pn}
\end{align}
provided that the hadronic medium is isospin symmetric
and that the three isospin states of pions are equally 
distributed in the medium.
Because of the isospin symmetry of the strong interaction 
one also obtains the same conclusion for neutron reactions:
\begin{align}
P_{n\to n} : P_{n\to p} = 5:4.
\label{eq:P_np}
\end{align}
Similar results are also obtained for anti-nucleons.
Equations~(\ref{eq:P_pn}) and (\ref{eq:P_np}) show that 
these reactions act to randomize the isospin 
of nucleons during the hadronic stage.
Moreover, the mean time of protons to undergo the charge
exchange reaction is $3\sim4$~fm for $T=150\simeq170$~MeV
in the hadronic gas \cite{Kitazawa:2012at}.
Because this mean time is shorter than the typical lifetime
of the hadronic stage in heavy ion collisions, all nucleons 
have chances to undergo the above reaction several times.
Because of the charge exchange reaction, the isospins of 
(anti-)baryons in the final state is randomized almost completely.
Although various effects on this conclusion is investigated
in Ref.~\cite{Kitazawa:2012at}, this conclusion is not 
altered.
This is sufficient to justify the binomial model Eq.~(\ref{eq:G=FM})
for the relation between (anti-)baryon and (anti-)proton number 
distributions.

\section{Summary}
\label{sec:concl}

In this article, we have reviewed physics of 
bulk fluctuations in relativistic heavy ion collisions.
Now, let us recall the experimental results 
in Figs.~\ref{fig:ALICEfluc} and \ref{fig:STARfluc}.
After reading this review, the readers should be able to 
understand how to interpret these experimental results.
In the experimental result of the second-order cumulant of 
net-electric charge fluctuation in Fig.~\ref{fig:ALICEfluc}, 
the second-order cumulant shows
a suppression compared with the hadronic value, and the 
suppression is more prominent for larger rapidity window $\Delta\eta$.
As discussed in Sec.~\ref{sec:equil}, the thermal fluctuations 
are suppressed when the medium undergoes a deconfinement 
phase transition.
The experimental result in Fig.~\ref{fig:ALICEfluc} thus can be
interpreted as the remnant of the small fluctuations in 
the primordial stage.
In Sec.~\ref{sec:diffusion}, we have seen that the $\Delta\eta$
dependence in this experimental result can also be 
understood reasonably in this picture.
The readers should also be able to understand the reason why the quantities 
plotted in Fig.~\ref{fig:STARfluc}, especially their deviations
from unity, contain important information.
In Sec.~\ref{sec:equil}, we have learned that these ratios should
take unity if the cumulants
are well described by hadronic degrees of freedom
in equilibrium.
Interestingly, the ratios of the cumulants in 
Fig.~\ref{fig:STARfluc} show deviations compared to 
this ``baseline'' behavior.
This deviation clearly shows that the fluctuation carries 
information of non-hadronic and/or non-thermal physics 
in relativistic heavy ion collisions.
The origin of the deviations, however, is still in debate
and is not settled when this manuscript is written.
As discussed in Secs.~\ref{sec:e-v-e}, \ref{sec:diffusion} and 
\ref{sec:binomial}, there are many subtle problems in the 
comparison between event-by-event fluctuations with theoretical
studies on thermal fluctuations.
These problems have to be revealed
in the cooperation between experimental and theoretical
researches as well as numerical analysis in lattice QCD.
The experimental analysis of rapidity window dependence discussed in 
Sec.~\ref{sec:diffusion} and its theoretical description are 
one of the important subjects.

As these examples show, bulk fluctuations are important 
observables in heavy ion collisions, which encode nontrivial
physics on the early thermodynamics of the hot medium
and diffusion processes in later stages.
The fluctuation observables are expected to become 
one of the most important quantities in the study of the QCD 
phase structure in future experimental programs with intermediate 
collision energies, $3\lesssim \sqrt{s_{_{\rm NN}}}\lesssim20$ GeV,
such as the beam-energy scan II (BES-II) program \cite{BES-II} 
at RHIC, and those planned in FAIR, NICA and J-PARC.
Careful analyses of fluctuation observables in these experiments
will provide us plenty of information on the QCD phase 
structure.

\section*{Acknowledgment}

A large part of this review is written on the basis of 
a lecture by M.~K. at Tsukuba University on Oct. 29--31, 2014. 
He thanks members of the high energy nuclear experiment group
at Tsukuba University, especially Shin-ichi Esumi and 
Hiroshi Masui for the invitation and discussions during his stay, 
which are reflected in this article.
The authors also thank for many invitations to international 
workshops on fluctuations and active discussions there, especially
Adam Bzdak, Bengt Friman, Frithjof Karsch, Volker Koch, Xiaofeng Luo, 
Tapan Nayak, Krzysztof Redlich and Nu Xu.
The authors thank Hiroshi Horii and Miki Sakaida for reading 
the manuscript.
This work is supported in part by JSPS KAKENHI Grant
Numbers 23540307, 25800148 and 26400272.

\appendix

\section{Superposition of probability distribution functions}
\label{app:superposition}

In this appendix, we consider cumulants of a probability 
distribution function which is given by a superposition 
of probability distribution functions 
\cite{Kitazawa:2012at,Kitazawa:2015ira}.

Let us consider a probability distribution function $P(x)$
for an integer stochastic variable $x$, and assume that 
$P(x)$ consists of the superposition of sub-probabilities as 
\begin{align}
P(x) = \sum_N F(N) P_N(x),
\label{eq:PFPN}
\end{align}
where $P_N(x)$ are sub-probabilities 
labeled by integer $N$. 
Each sub-probability is summed with a weight
$F(N)$ satisfying $\sum_N F(N)=1$ , which is also regarded 
as a probability.
The purpose of this appendix is to represent 
the cumulants of $P(x)$ using those of $P_N(x)$
and $F(N)$.
Although we write down the results explicitly 
up to the fourth-order in this article, 
the result can be extended to higher orders straightforwardly.

We start from the cumulant generating function of 
Eq.~(\ref{eq:PFPN}),
\begin{align}
K(\theta) 
&= \log \sum_x e^{ \theta x } P(x)
= \log \sum_N F(N) \sum_x e^{ \theta x } P_N(x)
\\
&= \log \sum_N F(N) \sum_x e^{ K_N(\theta) }
\label{eq:KNdef}
\end{align}
where $K_N(\theta) = \log \sum_x e^{ \theta x } P_N(x)$ is 
the cumulant generating function for $P_N(x)$.
Using the cumulant expansion Eq.~(\ref{eq:cumulantexpansion}), 
Eq.~(\ref{eq:KNdef}) is written as
\begin{align}
K(\theta)
=& \sum_m \frac1{m!} \sum_F [ K_N(\theta) ]_{\rm c}^m
\\
=& \sum_F K_N(\theta) + \frac12 \sum_F ( \delta K_N(\theta) )^2
+ \frac1{3!} \sum_F ( \delta K_N(\theta) )^3
\nonumber
\\
&+ \frac1{4!} \sum_F [ K_N(\theta) ]_{\rm c}^4
+ \cdots,
\label{eq:K1234}
\end{align}
where $\sum_F$ is a shorthand notation for $\sum_N F(N)$,
and $\sum_F [ K_N(\theta) ]_c^m$ is the $m$th-order cumulant
of $K_N$ for the sum over $F$, 
whose explicit forms up to the fourth-order are given on 
the far right hand side with 
\begin{align}
\sum_F (\delta K_N(\theta))^n 
&= \sum_F \left( K_N(\theta) - \sum_F K_N(\theta) \right)^n ,
\\
\sum_F [ K_N(\theta)]_{\rm c}^4
&= \sum_F (\delta K_N(\theta))^4
- 3 \left( \sum_F (\delta K_N(\theta))^2 \right)^2 .
\end{align}

Cumulants of $P(x)$ are given by derivatives of $K(\theta)$ as 
\begin{align}
\langle x^n \rangle_{\rm c} 
= \left . \frac{ \partial^n }{ \partial \theta^n } K(\theta) 
\right | _{\theta}
\equiv K^{(n)}.
\label{eq:<x^n>c}
\end{align}
All cumulants can be obtained with Eqs.~(\ref{eq:<x^n>c}) 
and (\ref{eq:K1234}).
In order to calculate the cumulants explicitly,
we first note that the normalization condition 
$\sum_x P_N(x)=1$ yields $K_N(0)=0$. 
From this property, it is immediately concluded that 
all $K_N(\theta)$ in each term on the far right hand 
side of Eq.~(\ref{eq:K1234}) must receive at least 
one differentiation so that the term gives nonzero 
contribution to Eq.~(\ref{eq:<x^n>c}).
This means that the $m$th-order term in 
Eq.~(\ref{eq:K1234}) can affect Eq.~(\ref{eq:<x^n>c})
only if $m\le n$.
Keeping this rule in mind, derivatives of 
Eq.~(\ref{eq:K1234}) with $\theta=0$ is given by
\begin{align}
K^{(1)}
=& \sum_F K_N^{(1)},
\label{eq:K1KN}
\\
K^{(2)}
=& \sum_F K_N^{(2)} + \sum_F ( \delta K_N^{(1)} )^2,
\label{eq:K2KN}
\\
K^{(3)}
=& \sum_F K_N^{(3)} + 3 \sum_F \delta K_N^{(1)} \delta K_N^{(2)} 
+ \sum_F ( \delta K_N^{(1)} )^3,
\label{eq:K3KN}
\\
K^{(4)}
=& \sum_F K_N^{(4)} + 4 \sum_F \delta K_N^{(1)} \delta K_N^{(3)} 
+ 3 \sum_F ( \delta K_N^{(2)} )^2
+ 6 \sum_F ( \delta K_N^{(1)} )^2 \delta K_N^{(2)}
\nonumber
\\ &
+ \sum_F ( \delta K_N^{(1)} )_{\rm c}^4,
\label{eq:K4KN}
\end{align}
with 
$\displaystyle{K_N^{(n)}= \left .
\frac{\partial^n K_N(\theta)}{\partial \theta^n} \right |_{\theta=0}}$
being the cumulants of the sub-probabilities $P_N(x)$.
Equations~(\ref{eq:K1KN}) - (\ref{eq:K4KN}) relate 
the cumulants $K^{(n)}$ with $K_N^{(n)}$.

The above relations are further simplified when 
the cumulants of $P_N(x)$ are at most linear with
respect to $N$, i.e.
\begin{align}
K_N^{(n)}= N \xi_{(n)} + \zeta_{(n)} ,
\label{eq:K_Nxi}
\end{align}
where $\xi_{(n)}$ and $\zeta_{(n)}$ are constants 
which do not depend on $N$.
Substituting Eq.~(\ref{eq:K_Nxi}) into 
Eqs.~(\ref{eq:K1KN}) - (\ref{eq:K4KN})
one obtains
\begin{align}
K^{(1)} 
=& \zeta_{(1)} + \xi_{(1)} \langle N \rangle_F ,
\label{eq:K1NF}
\\
K^{(2)} 
=& \zeta_{(2)} + \xi_{(2)} \langle N \rangle_F 
+ {\xi_{(1)}}^2 \langle \delta N^2 \rangle_F ,
\label{eq:K2NF}
\\
K^{(3)} 
=& \zeta_{(3)} + \xi_{(3)} \langle N \rangle_F 
+ 3 \xi_{(1)} \xi_{(2)} \langle \delta N^2 \rangle_F 
+ {\xi_{(1)}}^3 \langle \delta N^3 \rangle_F ,
\label{eq:K3NF}
\\
K^{(4)} 
=& \zeta_{(4)} + \xi_{(4)} \langle N \rangle_F 
+ ( 4 \xi_{(1)} \xi_{(3)} + 3 {\xi_{(2)}}^2 ) \langle \delta N^2 \rangle_F 
+ 6 {\xi_{(1)}}^2 \xi_{(2)} \langle \delta N^3 \rangle_F
\nonumber \\ &
+ {\xi_{(1)}}^4 \langle \delta N^4 \rangle_{{\rm c},F} ,
\label{eq:K4NF}
\end{align}
where $\langle O(N) \rangle_F = \sum_N O(N) F(N)$ 
denotes the average over $F(N)$; these averages in 
the above formulas represent cumulants of the probability $F(N)$.

Extension of these results to the case of multi variable distribution 
functions is addressed in Ref.~\cite{Kitazawa:2015ira}.


\begin{thebibliography}{99}
\itemsep -2pt 


\bibitem{Cheng:2006qk} 
  M.~Cheng, N.~H.~Christ, S.~Datta, J.~van der Heide, C.~Jung, F.~Karsch, O.~Kaczmarek and E.~Laermann {\it et al.},
  Phys.\ Rev.\ D {\bf 74}, 054507 (2006).

\bibitem{Aoki:2006we} 
  Y.~Aoki, G.~Endrodi, Z.~Fodor, S.~D.~Katz and K.~K.~Szabo,
  Nature {\bf 443}, 675 (2006).

\bibitem{Asakawa:1989bq}
  M.~Asakawa and K.~Yazaki,
  Nucl.\ Phys.\ A {\bf 504} (1989) 668.

\bibitem{Kitazawa:2002bc}
  M.~Kitazawa, T.~Koide, T.~Kunihiro and Y.~Nemoto,
  Prog.\ Theor.\ Phys.\  {\bf 108} (2002) 929
  [hep-ph/0207255].

\bibitem{Stephanov:2007fk}
  M.~A.~Stephanov,
  Proc. Sci. LAT2006 (2006) 024.
  [arXiv:hep-lat/0701002].

\bibitem{Fukushima:2010bq}
  K.~Fukushima and T.~Hatsuda,
  Rept.\ Prog.\ Phys.\  {\bf 74} (2011) 014001
  [arXiv:1005.4814 [hep-ph]].

\bibitem{Philipsen:2011zx}
  O.~Philipsen,
  Acta Phys.\ Polon.\ Supp.\  {\bf 5} (2012) 825

\bibitem{RHIC}
  I.~Arsene {\it et al.}  [BRAHMS Collaboration],
  Nucl.\ Phys.\ A {\bf 757}, 1 (2005);
  B.~B.~Back {\it et al.} [PHOBOS Collaboration],
  Nucl.\ Phys.\ A {\bf 757}, 28 (2005);
  J.~Adams {\it et al.}  [STAR Collaboration],
  Nucl.\ Phys.\ A {\bf 757}, 102 (2005);
  K.~Adcox {\it et al.}  [PHENIX Collaboration],
  Nucl.\ Phys.\ A {\bf 757}, 184 (2005).

\bibitem{LHC}
  B.~Muller, J.~Schukraft and B.~Wyslouch,
  Ann.\ Rev.\ Nucl.\ Part.\ Sci.\  {\bf 62}, 361 (2012)
  [arXiv:1202.3233 [hep-ex]].

\bibitem{Andronic:2008gu}
  A.~Andronic, P.~Braun-Munzinger and J.~Stachel,
  Phys.\ Lett.\ B {\bf 673} (2009) 142
   [Phys.\ Lett.\ B {\bf 678} (2009) 516]
  [arXiv:0812.1186 [nucl-th]].

\bibitem{Kumar:2012fb}
  L.~Kumar [STAR Collaboration],
  Nucl.\ Phys.\ A {\bf 904-905} (2013) 256c
  [arXiv:1211.1350 [nucl-ex]].

\bibitem{BraunMunzinger:2003zd}
  P.~Braun-Munzinger, K.~Redlich and J.~Stachel,
  arXiv:nucl-th/0304013.

\bibitem{Randrup:2006nr}
  J.~Randrup and J.~Cleymans,
  Phys.\ Rev.\ C {\bf 74} (2006) 047901
  [hep-ph/0607065].

\bibitem{BES-I}
  STAR Collaboration,
  ``Experimental Study of the QCD Phase Diagram \& Search for the Critical Point: Selected Arguments for the Run-10 Beam Energy Scan,''
  STAR Notes SN0493,
  https://drupal.star.bnl.gov/STAR/starnotes/public/sn0493 (2009).

\bibitem{BES-II}
  STAR Collaboration, 
  ``Studying the Phase Diagram of QCD Matter at RHIC,''
  STAR Notes SN0598,
  https://drupal.star.bnl.gov/STAR/starnotes/public/sn0598 (2014).

\bibitem{FAIR}
  R.~Rapp {\it et al.},
  Lect.\ Notes Phys.\  {\bf 814} (2011) 335.

\bibitem{NICA}
  ``Design and construction of nuclotron-based ion collider facility (NICA) conceptual design report'',
  http://nica.jinr.ru/files/NICA\_CDR.pdf (2008).

\bibitem{Stephanov:1998dy}
  M.~A.~Stephanov, K.~Rajagopal, and E.~V.~Shuryak,
  Phys.\ Rev.\ Lett.\  {\bf 81} (1998) 4816.

\bibitem{Kitazawa:2014nja}
  M.~Kitazawa,
  Nucl.\ Phys.\ A {\bf 931} (2014) 92.

\bibitem{Koch:2008ia}
  V.~Koch,
  arXiv:0810.2520 [nucl-th].

\bibitem{Asakawa:2000wh}
  M.~Asakawa, U.~W.~Heinz, and B.~M\"uller,
  Phys.\ Rev.\ Lett.\  {\bf 85} (2000) 2072.

\bibitem{Jeon:2000wg}
  S.~Jeon and V.~Koch,
  Phys.\ Rev.\ Lett.\  {\bf 85} (2000) 2076.

\bibitem{Ejiri:2005wq}
  S.~Ejiri, F.~Karsch, and K.~Redlich,
  Phys. Lett. {\bf B633} (2006) 275.

\bibitem{Stephanov:1999zu}
  M.~A.~Stephanov, K.~Rajagopal and E.~V.~Shuryak,
  Phys.\ Rev.\ D {\bf 60} (1999) 114028
  [hep-ph/9903292].

\bibitem{Hatta:2003wn}
  Y.~Hatta and M.~A.~Stephanov,
  Phys.\ Rev.\ Lett.\  {\bf 91} (2003) 102003
   [Phys.\ Rev.\ Lett.\  {\bf 91} (2003) 129901]
  [hep-ph/0302002].

\bibitem{STAR-fluc}
  M.~M.~Aggarwal {\it et al.}  [STAR Collaboration],
  Phys.\ Rev.\ Lett.\  {\bf 105} (2010) 022302.

\bibitem{ALICE}
  B.~Abelev {\it et al.}  [ALICE Collaboration],
  Phys.\ Rev.\ Lett.\ {\bf 110} (2013) 152301.

\bibitem{Adamczyk:2013dal} 
  L.~Adamczyk {\it et al.}  [STAR Collaboration],
  Phys.\ Rev.\ Lett.\  {\bf 112} (2014) 032302.

\bibitem{Anticic:2013htn}
  T.~Anticic {\it et al.},
  Phys.\ Rev.\ C {\bf 89} (2014) 5,  054902
  [arXiv:1310.3428 [nucl-ex]].

\bibitem{Adamczyk:2014fia} 
  L.~Adamczyk {\it et al.}  [STAR Collaboration],
  Phys.\ Rev.\ Lett.\  {\bf 113} (2014) 092301.
  [arXiv:1402.1558 [nucl-ex]].

\bibitem{PHENIX-fluc}
  J.~T.~Mitchell [PHENIX Collaboration],
  Nucl.\ Phys.\ A904-905 (2013) 903c;
  [arXiv:1211.6139 [nucl-ex]].

\bibitem{Luo:2015ewa}
  X.~Luo [STAR Collaboration],
  PoS CPOD {\bf 2014} (2015) 019
  [arXiv:1503.02558 [nucl-ex]].

\bibitem{Adare:2015aqk}
  A.~Adare {\it et al.} [PHENIX Collaboration],
  arXiv:1506.07834 [nucl-ex].
  
\bibitem{Ding:2015ona}
  H.~T.~Ding, F.~Karsch and S.~Mukherjee,
  Int.\ J.\ Mod.\ Phys.\ E {\bf 24} (2015) 10,  1530007
  [arXiv:1504.05274 [hep-lat]].

\bibitem{Borsanyi:2015axp}
  S.~Borsanyi,
  Proc.\ Sci.\ {\bf LATTICE2015} (2015) 015
  [arXiv:1511.06541 [hep-lat]].
  
\bibitem{Stephanov:2008qz}
  M.~A.~Stephanov,
  Phys.\ Rev.\ Lett.\  {\bf 102} (2009) 032301.

\bibitem{Asakawa:2009aj}
  M.~Asakawa, S.~Ejiri, and M.~Kitazawa,
  Phys.\ Rev.\ Lett.\  {\bf 103} (2009) 262301
  [arXiv:0904.2089 [nucl-th]].
  
\bibitem{Landauer}
  R.~Landauer, 
  Nature\ {\bf 392} (1998) 658.

\bibitem{Einstein}
  A.~Einstein, Annalen der Physik {\bf 17} (1905) 549.

\bibitem{Perrin}
  J.~Perrin, 
  Annales de chimie et de physiqe VIII {\bf 18} (1909) 5.

\bibitem{Ade:2013sjv}
  P.~A.~R.~Ade {\it et al.} [Planck Collaboration],
  Astron.\ Astrophys.\  {\bf 571} (2014) A1
  [arXiv:1303.5062 [astro-ph.CO]].

\bibitem{Baumann:2009ds}
  D.~Baumann,
  arXiv:0907.5424 [hep-th].

\bibitem{Maldacena:2002vr}
  J.~M.~Maldacena,
  JHEP {\bf 0305} (2003) 013
  [astro-ph/0210603].

\bibitem{Bartolo:2004if}
  N.~Bartolo, E.~Komatsu, S.~Matarrese and A.~Riotto,
  Phys.\ Rept.\  {\bf 402} (2004) 103
  [astro-ph/0406398].

\bibitem{Ade:2013ydc}
  P.~A.~R.~Ade {\it et al.} [Planck Collaboration],
  Astron.\ Astrophys.\  {\bf 571} (2014) A24
  [arXiv:1303.5084 [astro-ph.CO]].

\bibitem{Johnson}
  J.~B.~Johnson, Phys.\ Rev.\ {\bf 32} (1928) 97.

\bibitem{Nyquist}
  H.~Nyquist, Phys.\ Rev.\ {\bf 32} (1928) 110.

\bibitem{Schottky}
  W.~Schottky, Ann. der Phys. {\bf 57} (1918) 541.

\bibitem{Jehl}
  X.~Jehl, {\it et al.}, 
  Nature {\bf 405} (2000) 50.

\bibitem{FQHE}
  L.~Saminadayar, D.~C.~Glattli, Y.~Jin, and B.~Etienne,
  Phys.\ Rev.\ Lett.\ {\bf 79}, 2526 (1997).

\bibitem{Gustavssona}
  See, for example,
  L.~S.~Levitov and G.~B.~Lesovik, JETP Lett., {\bf 58} (1993) 230;
  B.~Reulet, J.~Senzier and D.~E.~Prober, Phys.\ Rev.\ Lett.\ {\bf 91} (2003) 196601;
  S.~Gustavssona, {\it et al.}, Surf. Sci. Rep. {\bf 64} (2009) 191.



\bibitem{Pratt:2012dz}
  S.~Pratt,
  Phys.\ Rev.\ Lett.\  {\bf 108} (2012) 212301.

\bibitem{Bass:2000az} 
  S.~A.~Bass, P.~Danielewicz, and S.~Pratt,
  Phys.\ Rev.\ Lett.\  {\bf 85} (2000) 2689.

\bibitem{Jeon:2001ue} 
  S.~Jeon and S.~Pratt,
  Phys.\ Rev.\ C {\bf 65} (2002) 044902.

\bibitem{Ling:2013ksb}
  B.~Ling, T.~Springer and M.~Stephanov,
  Phys.\ Rev.\ C {\bf 89} (2014) 6,  064901.



\bibitem{Pearson}
  K.~Pearson,
  ``On the General Theory of Skew Correlation and Non-linear Regression,''
  ( Dulau and Co., London, 1905).

\bibitem{GPY}
  D.~J.~Gross, R.~D.~Pisarski, and L.~G.~Yaffe,
  Rev. Mod. Phys. {\bf 53} (1981) 43.

\bibitem{Bonati:2013tt} 
  C.~Bonati, M.~D'Elia, H.~Panagopoulos and E.~Vicari,
  Phys.\ Rev.\ Lett.\  {\bf 110}, no. 25, 252003 (2013)
  [arXiv:1301.7640 [hep-lat]].

\bibitem{Kitano:2015fla} 
  R.~Kitano and N.~Yamada,
  arXiv:1506.00370 [hep-ph].

\bibitem{Borsanyi:2015cka} 
  S.~Borsanyi {\it et al.},
  arXiv:1508.06917 [hep-lat].

\bibitem{Luo:2011tp}
  X.~Luo,
  J.\ Phys.\ G {\bf 39} (2012) 025008
  [arXiv:1109.0593 [physics.data-an]].

\bibitem{Morita:2013tu} 
  K.~Morita, B.~Friman, K.~Redlich, and V.~Skokov,
  Phys.\ Rev.\ C {\bf 88} (2013) 034903
  [arXiv:1301.2873 [hep-ph]].
  
\bibitem{Morita:2012kt}
  K.~Morita, V.~Skokov, B.~Friman and K.~Redlich,
  Eur.\ Phys.\ J.\ C {\bf 74} (2014) 2706
  [arXiv:1211.4703 [hep-ph]].

\bibitem{Morita:2014fda}
  K.~Morita, B.~Friman and K.~Redlich,
  Phys.\ Lett.\ B {\bf 741} (2015) 178
  [arXiv:1402.5982 [hep-ph]].

\bibitem{Kitazawa:2016awu}
  M.~Kitazawa,
  arXiv:1602.01234 [nucl-th].

\bibitem{Negele}
  J.~W.~Negele and H.~Orland, ``Quantum Many-particle Systems''
  (Perseus, 1998).

\bibitem{Bzdak:2012ab} 
  A.~Bzdak and V.~Koch,
  Phys.\ Rev.\ C {\bf 86} (2012) 044904.

\bibitem{Kitazawa:2013bta} 
  M.~Kitazawa, M.~Asakawa, and H.~Ono,
  Phys.\ Lett.\ B {\bf 728} (2014) 386
  [arXiv:1307.2978].

\bibitem{Luo:2014rea}
  X.~Luo,
  Phys.\ Rev.\ C {\bf 91} (2015) 3,  034907
  [arXiv:1410.3914 [physics.data-an]].

\bibitem{Kitazawa:2015ira}
  M.~Kitazawa,
  Nucl.\ Phys.\ A {\bf 942} (2015) 65
  [arXiv:1505.04349 [nucl-th]].



\bibitem{Yagi}
  K.~Yagi, T.~Hatsuda, and Y.~Miake,
  ``Quark-Gluon Plasma'' (Cambridge University Press, 2005).

\bibitem{Kapusta}
  J.~I.~Kapusta and C.~Gale,
  ``Finite-Temperature Field Theory'' (Cambridge University Press, 2006).

\bibitem{Gardiner}
  Crispin Gardiner, 
  ``Stochastic Methods: A Handbook for the Natural and Social Sciences''
  (Springer Series in Synergetics, 2009).

\bibitem{Gavai:2010zn}
  R.~V.~Gavai and S.~Gupta,
  Phys.\ Lett.\  B {\bf 696}, 459 (2011)
  [arXiv:1001.3796 [hep-lat]].

\bibitem{Schmidt:2010xm} 
  C.~Schmidt,
  Prog.\ Theor.\ Phys.\ Suppl.\  {\bf 186}, 563 (2010)
  [arXiv:1007.5164 [hep-lat]].

\bibitem{Mukherjee:2011td} 
  S.~Mukherjee,
  J.\ Phys.\ G G {\bf 38}, 124022 (2011)
  [arXiv:1107.0765 [nucl-th]].

\bibitem{Borsanyi:2011sw} 
  S.~Borsanyi, {\it et al.}, 
  JHEP {\bf 1201}, 138 (2012)
  [arXiv:1112.4416 [hep-lat]].

\bibitem{Nagata:2012pc}
  K.~Nagata and A.~Nakamura,
  JHEP {\bf 1204} (2012) 092
  [arXiv:1201.2765 [hep-lat]].

\bibitem{Bazavov:2012jq} 
  A.~Bazavov {\it et al.}  [HotQCD Collaboration],
  Phys.\ Rev.\ D {\bf 86} (2012) 034509.

\bibitem{Bazavov:2012vg} 
  A.~Bazavov, {\it et al.}, 
  Phys.\ Rev.\ Lett.\  {\bf 109} (2012) 192302.

\bibitem{Bazavov:2013dta} 
  A.~Bazavov, {\it et al.}, 
  Phys.\ Rev.\ Lett.\  {\bf 111} (2013) 082301.

\bibitem{Nakamura:2013ska} 
  A.~Nakamura and K.~Nagata,
  arXiv:1305.0760 [hep-ph].
  
\bibitem{Borsanyi:2013hza}
  S.~Borsanyi, Z.~Fodor, S.~D.~Katz, S.~Krieg, C.~Ratti and K.~K.~Szabo,
  Phys.\ Rev.\ Lett.\  {\bf 111} (2013) 062005
  [arXiv:1305.5161 [hep-lat]].

\bibitem{Bellwied:2013cta} 
  R.~Bellwied, {\it et al.}, 
  Phys.\ Rev.\ Lett.\  {\bf 111} (2013) 202302.

\bibitem{Bazavov:2013uja}
  A.~Bazavov {\it et al.},
  Phys.\ Rev.\ D {\bf 88} (2013) 9,  094021
  [arXiv:1309.2317 [hep-lat]].

\bibitem{Borsanyi:2014ewa} 
  S.~Borsanyi,{\it et al.}, 
  Phys.\ Rev.\ Lett.\  {\bf 113} (2014) 052301.

\bibitem{Bazavov:2014yba}
  A.~Bazavov, 
  {\it et al.},
  Phys.\ Lett.\ B {\bf 737} (2014) 210.

\bibitem{Bazavov:2014xya}
  A.~Bazavov {\it et al.},
  Phys.\ Rev.\ Lett.\  {\bf 113} (2014) 7,  072001
  [arXiv:1404.6511 [hep-lat]].

\bibitem{Gupta:2014qka}
  S.~Gupta, N.~Karthik and P.~Majumdar,
  Phys.\ Rev.\ D {\bf 90} (2014) 3,  0340010.

\bibitem{Nakamura:2015jra}
  A.~Nakamura, S.~Oka and Y.~Taniguchi,
  arXiv:1504.04471 [hep-lat].

\bibitem{Bazavov:2015zja}
  A.~Bazavov {\it et al.},
  arXiv:1509.05786 [hep-lat].

\bibitem{PDG}
  The Review of Particle Physics,
  K.~Nakamura, {\it et al.} (Particle Data Group), 
  J. Phys. G {\bf 37}, 075021 (2010).

\bibitem{Cleymans:1998fq}
  J.~Cleymans and K.~Redlich,
  Phys.\ Rev.\ Lett.\  {\bf 81}, 5284 (1998)
  [arXiv:nucl-th/9808030].

\bibitem{Landau1}
  L.D.~Landau and E.M.~Lifshitz, 
  ``Statistical Physics: Part 1'', 
  (Pergamon, Oxford, 1980).

\bibitem{Landau2}
  L.D.~Landau and E.M.~Lifshitz, 
  ``Statistical Physics: Part 2'', 
  (Pergamon, Oxford, 1980).

\bibitem{Koch:2005vg}
  V.~Koch, A.~Majumder and J.~Randrup,
  Phys.\ Rev.\ Lett.\  {\bf 95} (2005) 182301
  [nucl-th/0505052].

\bibitem{Karsch:2010ck}
  F.~Karsch and K.~Redlich,
  Phys.\ Lett.\ B {\bf 695} (2011) 136
  [arXiv:1007.2581 [hep-ph]].

\bibitem{Kitazawa:2011wh} 
  M.~Kitazawa and M.~Asakawa,
  Phys.\ Rev.\ C {\bf 85} (2012) 021901
  [arXiv:1107.2755 [nucl-th]];

\bibitem{Kitazawa:2012at} 
  M.~Kitazawa and M.~Asakawa,
  Phys.\ Rev.\ C {\bf 86} (2012) 024904
  [Erratum-ibid.\ C {\bf 86} (2012) 069902]
  [arXiv:1205.3292 [nucl-th]].

\bibitem{Mogliacci:2013mca}
  S.~Mogliacci, J.~O.~Andersen, M.~Strickland, N.~Su and A.~Vuorinen,
  JHEP {\bf 1312} (2013) 055
  [arXiv:1307.8098 [hep-ph]].

\bibitem{Haque:2013qta}
  N.~Haque, M.~G.~Mustafa and M.~Strickland,
  JHEP {\bf 1307} (2013) 184
  [arXiv:1302.3228 [hep-ph]].

\bibitem{Haque:2013sja}
  N.~Haque, J.~O.~Andersen, M.~G.~Mustafa, M.~Strickland and N.~Su,
  Phys.\ Rev.\ D {\bf 89} (2014) 6,  061701
  [arXiv:1309.3968 [hep-ph]].

\bibitem{Son:2004iv}
  D.~T.~Son and M.~A.~Stephanov,
  Phys.\ Rev.\ D {\bf 70} (2004) 056001
  [hep-ph/0401052].

\bibitem{Minami:2011un}
  Y.~Minami,
  Phys.\ Rev.\ D {\bf 83} (2011) 094019
  [arXiv:1102.5485 [hep-ph]].

\bibitem{Hohenberg:1977ym}
  P.~C.~Hohenberg and B.~I.~Halperin,
  Rev.\ Mod.\ Phys.\  {\bf 49} (1977) 435.

\bibitem{Hatta:2002sj}
  Y.~Hatta and T.~Ikeda,
  Phys.\ Rev.\ D {\bf 67} (2003) 014028
  [hep-ph/0210284].

\bibitem{Fujii:2003bz}
  H.~Fujii,
  Phys.\ Rev.\ D {\bf 67} (2003) 094018
  [hep-ph/0302167].

\bibitem{Fujii:2004jt}
  H.~Fujii and M.~Ohtani,
  Phys.\ Rev.\ D {\bf 70} (2004) 014016
  [hep-ph/0402263].

\bibitem{Kunihiro:1991qu}
  T.~Kunihiro,
  Phys.\ Lett.\ B {\bf 271} (1991) 395.

\bibitem{Sasaki:2006ws}
  C.~Sasaki, B.~Friman and K.~Redlich,
  Phys.\ Rev.\ D {\bf 75} (2007) 054026
  [hep-ph/0611143].

\bibitem{Sasaki:2006ww}
  C.~Sasaki, B.~Friman and K.~Redlich,
  Phys.\ Rev.\ D {\bf 75} (2007) 074013
  [hep-ph/0611147].

\bibitem{Fukushima:2008wg}
  K.~Fukushima,
  Phys.\ Rev.\ D {\bf 77} (2008) 114028
   [Phys.\ Rev.\ D {\bf 78} (2008) 039902]
  [arXiv:0803.3318 [hep-ph]].

\bibitem{Schaefer:2009ui}
  B.~J.~Schaefer, M.~Wagner and J.~Wambach,
  Phys.\ Rev.\ D {\bf 81} (2010) 074013
  doi:10.1103/PhysRevD.81.074013
  [arXiv:0910.5628 [hep-ph]].

\bibitem{Fu:2009wy}
  W.~j.~Fu, Y.~x.~Liu and Y.~L.~Wu,
  Phys.\ Rev.\ D {\bf 81} (2010) 014028
  doi:10.1103/PhysRevD.81.014028
  [arXiv:0910.5783 [hep-ph]].

\bibitem{Skokov:2010uh}
  V.~Skokov, B.~Friman and K.~Redlich,
  Phys.\ Rev.\ C {\bf 83} (2011) 054904
  doi:10.1103/PhysRevC.83.054904
  [arXiv:1008.4570 [hep-ph]].

\bibitem{Ichihara:2015kba}
  T.~Ichihara, K.~Morita and A.~Ohnishi,
  arXiv:1507.04527 [hep-lat].

\bibitem{Friman:2011pf} 
  B.~Friman, F.~Karsch, K.~Redlich, and V.~Skokov,
  Eur.\ Phys.\ J.\ C {\bf 71} (2011) 1694.

\bibitem{Stephanov:2011pb} 
  M.~A.~Stephanov,
  Phys.\ Rev.\ Lett.\  {\bf 107} (2011) 052301.

\bibitem{Gupta:2011wh} 
  S.~Gupta, {\it et al.}, 
  Science {\bf 332} (2011) 1525.

\bibitem{Athanasiou:2010kw}
  C.~Athanasiou, K.~Rajagopal and M.~Stephanov,
  Phys.\ Rev.\ D {\bf 82} (2010) 074008
  [arXiv:1006.4636 [hep-ph]].

\bibitem{BraunMunzinger:2011dn} 
  P.~Braun-Munzinger, B.~Friman, F.~Karsch, K.~Redlich and V.~Skokov,
  Phys.\ Rev.\ C {\bf 84}, 064911 (2011)
  [arXiv:1107.4267 [hep-ph]].

\bibitem{BraunMunzinger:2011ta}
  P.~Braun-Munzinger, B.~Friman, F.~Karsch, K.~Redlich and V.~Skokov,
  Nucl.\ Phys.\  A {\bf 880}, 48 (2012)
  [arXiv:1111.5063 [hep-ph]].

\bibitem{Fukushima:2014lfa}
  K.~Fukushima,
  Phys.\ Rev.\ C {\bf 91} (2015) 4,  044910
  [arXiv:1409.0698 [hep-ph]].

\bibitem{Bluhm:2014wha}
  M.~Bluhm, P.~Alba, W.~Alberico, R.~Bellwied, V.~Mantovani Sarti, M.~Nahrgang and C.~Ratti,
  Nucl.\ Phys.\ A {\bf 931} (2014) 814
  [arXiv:1408.4734 [hep-ph]].

\bibitem{Albright:2015uua}
  M.~Albright, J.~Kapusta and C.~Young,
  Phys.\ Rev.\ C {\bf 92} (2015) 4,  044904
  [arXiv:1506.03408 [nucl-th]].
  
\bibitem{Karsch:2012wm}
  F.~Karsch,
  Central Eur.\ J.\ Phys.\  {\bf 10} (2012) 1234
  [arXiv:1202.4173 [hep-lat]].


\bibitem{Asakawa:QM15}
  M.~Asakawa, M.~Kitazawa, Y.~Ohnishi and M.~Sakaida,
  talk given in Quark Matter 2015.

\bibitem{Ohnishi}
  Y.~Ohnishi, M.~Asakawa, M.~Kitazawa, in preparation.

\bibitem{Abelev:2013vea}
  B.~Abelev {\it et al.} [ALICE Collaboration],
  Phys.\ Rev.\ C {\bf 88} (2013) 044910
  [arXiv:1303.0737 [hep-ex]].

\bibitem{Alba:2014eba}
  P.~Alba, W.~Alberico, R.~Bellwied, M.~Bluhm, V.~Mantovani Sarti, M.~Nahrgang and C.~Ratti,
  Phys.\ Lett.\ B {\bf 738} (2014) 305
  [arXiv:1403.4903 [hep-ph]].

\bibitem{Morita:2014nra}
  K.~Morita and K.~Redlich,
  PTEP {\bf 2015} (2015) 4,  043D03
  [arXiv:1409.8001 [hep-ph]].

\bibitem{Karsch:2015zna}
  F.~Karsch, K.~Morita and K.~Redlich,
  arXiv:1508.02614 [hep-ph].

\bibitem{Bleicher:2000ek} 
  M.~Bleicher, S.~Jeon, and V.~Koch,
  Phys.\ Rev.\ C {\bf 62}, 061902 (2000).

\bibitem{Begun:2004gs}
  V.~V.~Begun, M.~Gazdzicki, M.~I.~Gorenstein and O.~S.~Zozulya,
  Phys.\ Rev.\ C {\bf 70} (2004) 034901
  [nucl-th/0404056].

\bibitem{Begun:2006uu}
  V.~V.~Begun, M.~Gazdzicki, M.~I.~Gorenstein, M.~Hauer, V.~P.~Konchakovski and B.~Lungwitz,
  Phys.\ Rev.\ C {\bf 76} (2007) 024902
  [nucl-th/0611075].

\bibitem{Bzdak:2012an}
  A.~Bzdak, V.~Koch and V.~Skokov,
  Phys.\ Rev.\ C {\bf 87} (2013) 1,  014901
  [arXiv:1203.4529 [hep-ph]].

\bibitem{Sakaida:2014pya}
  M.~Sakaida, M.~Asakawa and M.~Kitazawa,
  Phys.\ Rev.\ C {\bf 90} (2014) 6,  064911.

\bibitem{Cao:2015cba}
  S.~Cao, G.~Y.~Qin and S.~A.~Bass,
  Phys.\ Rev.\ C {\bf 92} (2015) 5,  054909
  doi:10.1103/PhysRevC.92.054909
  [arXiv:1505.01869 [nucl-th]].

\bibitem{Niemi:2015qia}
  H.~Niemi, K.~J.~Eskola and R.~Paatelainen,
  arXiv:1505.02677 [hep-ph].

\bibitem{Skokov:2012ds}
  V.~Skokov, B.~Friman and K.~Redlich,
  Phys.\ Rev.\ C {\bf 88} (2013) 034911
  [arXiv:1205.4756 [hep-ph]].

\bibitem{Alba:2015iva}
  P.~Alba, R.~Bellwied, M.~Bluhm, V.~M.~Sarti, M.~Nahrgang and C.~Ratti,
  arXiv:1504.03262 [hep-ph].

\bibitem{Gorenstein:2011vq}
  M.~I.~Gorenstein and M.~Gazdzicki,
  Phys.\ Rev.\ C {\bf 84} (2011) 014904
  [arXiv:1101.4865 [nucl-th]].

\bibitem{Begun:2012rf}
  V.~V.~Begun, M.~Gazdzicki and M.~I.~Gorenstein,
  Phys.\ Rev.\ C {\bf 88} (2013) 2,  024902
  [arXiv:1208.4107 [nucl-th]].

\bibitem{Sangaline:2015bma}
  E.~Sangaline,
  arXiv:1505.00261 [nucl-th].

\bibitem{OAK}
  H.~Ono, M.~Asakawa and M.~Kitazawa,
  Phys.\ Rev.\ C {\bf 87} (2013) 041901 R.

\bibitem{Nahrgang:2014fza}
  M.~Nahrgang, M.~Bluhm, P.~Alba, R.~Bellwied and C.~Ratti,
  arXiv:1402.1238 [hep-ph].

\bibitem{Feckova:2015qza}
  Z.~Feckova, J.~Steinheimer, B.~Tomasik and M.~Bleicher,
  arXiv:1510.05519 [nucl-th].





\bibitem{Kapusta:2011gt}
  J.~I.~Kapusta, B.~Muller and M.~Stephanov,
  Phys.\ Rev.\ C {\bf 85} (2012) 054906
  [arXiv:1112.6405 [nucl-th]].

\bibitem{Shuryak:2000pd} 
  E.~V.~Shuryak and M.~A.~Stephanov,
  Phys.\ Rev.\ C {\bf 63} (2001) 064903.

\bibitem{Ono:master}
  H.~Ono, Master thethis, Osaka University (2013).

\bibitem{Haussler:2007un}
  S.~Haussler, S.~Scherer and M.~Bleicher,
  Phys.\ Lett.\ B {\bf 660} (2008) 197
  [hep-ph/0702188 [HEP-PH]].

\bibitem{Aziz:2004qu} 
  M.~A.~Aziz and S.~Gavin,
  Phys.\ Rev.\ C {\bf 70} (2004) 034905.

\bibitem{Young:2014pka}
  C.~Young, J.~I.~Kapusta, C.~Gale, S.~Jeon and B.~Schenke,
  Phys.\ Rev.\ C {\bf 91} (2015) 4,  044901
  [arXiv:1407.1077 [nucl-th]].

\bibitem{Adcox:2002mm}
  K.~Adcox {\it et al.} [PHENIX Collaboration],
  Phys.\ Rev.\ Lett.\  {\bf 89} (2002) 082301
  [nucl-ex/0203014].

\bibitem{Adams:2003st}
  J.~Adams {\it et al.} [STAR Collaboration],
  Phys.\ Rev.\ C {\bf 68} (2003) 044905
  [nucl-ex/0307007].

\bibitem{Abelev:2008jg}
  B.~I.~Abelev {\it et al.} [STAR Collaboration],
  Phys.\ Rev.\ C {\bf 79} (2009) 024906
  [arXiv:0807.3269 [nucl-ex]].

\bibitem{Berdnikov:1999ph}
  B.~Berdnikov and K.~Rajagopal,
  Phys.\ Rev.\ D {\bf 61} (2000) 105017
  [hep-ph/9912274].

\bibitem{Nonaka:2004pg}
  C.~Nonaka and M.~Asakawa,
  Phys.\ Rev.\ C {\bf 71} (2005) 044904
  [nucl-th/0410078].

\bibitem{Mukherjee:2015swa}
  S.~Mukherjee, R.~Venugopalan and Y.~Yin,
  Phys.\ Rev.\ C {\bf 92} (2015) 3,  034912
  [arXiv:1506.00645 [hep-ph]].

\bibitem{Stephanov:2009ra}
  M.~A.~Stephanov,
  Phys.\ Rev.\ D {\bf 81} (2010) 054012
  [arXiv:0911.1772 [hep-ph]].

\bibitem{Nahrgang:2011mg}
  M.~Nahrgang, S.~Leupold, C.~Herold and M.~Bleicher,
  Phys.\ Rev.\ C {\bf 84} (2011) 024912
  [arXiv:1105.0622 [nucl-th]].

\bibitem{Nahrgang:2011vn}
  M.~Nahrgang, C.~Herold, S.~Leupold, I.~Mishustin and M.~Bleicher,
  J.\ Phys.\ G {\bf 40} (2013) 055108
  [arXiv:1105.1962 [nucl-th]].

\bibitem{Herold:2014zoa}
  C.~Herold, M.~Nahrgang, Y.~Yan and C.~Kobdaj,
  J.\ Phys.\ G {\bf 41} (2014) 11,  115106.

\bibitem{Sasaki:2007db}
  C.~Sasaki, B.~Friman and K.~Redlich,
  Phys.\ Rev.\ Lett.\  {\bf 99} (2007) 232301
  [hep-ph/0702254 [HEP-PH]].

\bibitem{Steinheimer:2012gc}
  J.~Steinheimer and J.~Randrup,
  Phys.\ Rev.\ Lett.\  {\bf 109} (2012) 212301
  [arXiv:1209.2462 [nucl-th]].

\bibitem{Steinheimer:2013xxa}
  J.~Steinheimer, J.~Randrup and V.~Koch,
  Phys.\ Rev.\ C {\bf 89} (2014) 3,  034901
  [arXiv:1311.0999 [nucl-th]].

\bibitem{Thaeder:2015QM}
  J.~THAEDER, for STAR Collaboration, 
  presentation at Quark Matter 2015, 
  Kobe, Japan, 27 Sep. -- 3 Oct., 2015.




\bibitem{Bzdak:2013pha}
  A.~Bzdak and V.~Koch,
  Phys.\ Rev.\ C {\bf 91} (2015) 2,  027901
  [arXiv:1312.4574 [nucl-th]].

\bibitem{Gorenstein:2011hr}
  M.~I.~Gorenstein,
  Phys.\ Rev.\ C {\bf 84} (2011) 024902
  [arXiv:1106.4473 [nucl-th]].

\bibitem{Rustamov:2012bx}
  A.~Rustamov and M.~I.~Gorenstein,
  Phys.\ Rev.\ C {\bf 86} (2012) 044906
  [arXiv:1204.6632 [nucl-th]].

\bibitem{efficiency}
  R.~Holzmann, talk given in ``HIC for FAIR Workshop on Fluctuation and Correlation Measured in Nuclear Collisions 2015,'' Jul. 29-31, FIAS, Frankfurt, Germany, and ``EMMI Workshop on Fluctuations in Strongly Interacting Hot and Dense Matter: Theory and Experiment,'' Nov. 2-6, 2015, GSI, Darmstadt, Germany; private communication with R.~Holzmann, A.~Bzdak and V.~Koch.








\end{thebibliography}
\end{document}